\documentclass{article}
\usepackage{jinstpub}
\usepackage{natbib} 
\usepackage{amsmath} 
\usepackage{subcaption} 
\usepackage{siunitx} 
\usepackage{physics} 
\usepackage[version=4]{mhchem} 
\usepackage{graphicx} 
\usepackage{caption} 
\usepackage{siunitx} 
\usepackage{makecell} 
\usepackage{lineno} 
\usepackage{placeins}
\usepackage{float}
\usepackage{xcolor}
\usepackage{url}
\usepackage{multirow}

\usepackage{booktabs} 
\usepackage{threeparttable} 

\title{Calibration Methods of Silicon Photomultiplier for JUNO-TAO Central Detector}

\renewcommand{\thefootnote}{*}

\author[1,2]{Jiayang Xu \footnotemark}
\author[1]{Yichen Li \footnotemark}
\author[1]{Guofu Cao}
\author[1]{Liang Zhan}
\author[3]{Zelin Chen}

\footnotetext{Corresponding author.}

\affiliation[1]{Institute of High Energy Physics, Chinese Academy of Sciences, No.19B Yuquan Road, Shĳingshan District, Beĳing 100049, China}
\affiliation[2]{University of Chinese Academy of Sciences, No.1 Yanqihu East Rd, Huairou District, Beĳing 101408, China}
\affiliation[3]{Nanjing  University, No.22 Hankou Road, Gulou District, Nanjing, Jiangsu 210093, China}

\emailAdd{xujy@ihep.ac.cn}
\emailAdd{liyichen@ihep.ac.cn}

\abstract{The Taishan Antineutrino Observatory (TAO or JUNO-TAO) is a satellite observatory for the Jiangmen Underground Neutrino Observatory (JUNO), located 44 meters away from the No.1 reactor of the Taishan Nuclear Power Plant. TAO can measure the reactor antineutrino energy spectrum with excellent energy resolution (better than 2\% at 1 MeV) using state-of-the-art Silicon Photomultipliers (SiPMs) operated at low temperature. To achieve this goal, the SiPMs (together with their readout electronics) must be well calibrated. This paper presents the channel-level calibration methods for the dark count rate (DCR), relative photon detection efficiency (PDE), time offset, gain, and internal optical crosstalk (IOCT) of the SiPMs based on charge and time information of the collected events. For the tile-level calibration of the external optical crosstalk (EOCT), in terms of its rate and emission angle distribution, a novel method is proposed by switching on and off different groups of SiPM tiles with an LED placed in the detector. Using one million simulated events, the expected calibration biases are evaluated for all the aforementioned parameters: relative PDE ($\sim$3\%), IOCT (1.4\%), DCR (-0.4\%), EOCT rate (<0.1\%), gain (<0.1\%), time offset (<0.2 ns). The emission angle distribution of the EOCT photons could be measured with a bias of less than 4\% in the main angular range. With this calibration accuracy, the overall impacts of SiPM parameter uncertainties and calibration biases on reconstructed vertex uncertainty and energy resolution are limited, with relative degradation below 2\% and 3\%, respectively. It verifies the validity of the calibration method for the JUNO-TAO detector.}

\keywords{Calibration, SiPM, External optical crosstalk, TAO, Reactor neutrino}

\begin{document}
\flushbottom

\maketitle

\section{Introduction}
\label{1}

Before the precise measurements of the reactor neutrino energy spectrum by experiments such as Daya Bay \cite{Ref1}, RENO \cite{Ref2}, and Double Chooz \cite{Ref3}, the primary theoretical models for calculating the expected reactor neutrino energy spectrum were the Summation model \cite{Ref4} based on nuclear database data \cite{Ref33, Ref34, Ref35, Ref36, Ref37, Ref38} and the $\beta$ conversion method such as Huber-Muller model \cite{Ref5, Ref6} based on the ILL High Flux Reactor measurement data \cite{Ref7, Ref8, Ref9}.  However, according to the measurement results from the Daya Bay, RENO, and Double Chooz experiments, both the Summation model and the Huber-Muller model exhibit systematic deviations from the experimental results in terms of the predicted energy spectrum flux and the shape of the energy spectrum at 5 MeV. Although Estienne and Fallot have improved the Summation model \cite{Ref10}; and Kopeikin, Skorokhvatov, Titov \cite{Ref11}; Hayen, Kostensalo, Severijns, Suhonen \cite{Ref12}; Giunti, Li, Ternes, Xin \cite{Ref13} have refined the Huber-Muller model, these models currently cannot simultaneously explain the systematic deviations in both flux and spectrum shape. Moreover, none of the models can reliably predict the fine structures in the reactor antineutrino spectrum \cite{Ref46}. If the JUNO experiment directly uses these predicted spectra as inputs, it will introduce significant uncertainties in the energy spectrum, which will affect the sensitivity of the neutrino mass ordering measurement \cite{Ref14}. To reduce the model dependence of the reactor neutrino energy spectrum, JUNO has deployed a satellite detector, the Taishan Antineutrino Observatory (TAO or JUNO-TAO) at a baseline of 44 meters away from Unit 1 of the Taishan Nuclear Power Plant \cite{Ref15}.

The TAO central detector (CD) consists of a ton-scale gadolinium-doped liquid scintillator coupled with silicon photomultipliers (SiPMs) array with the photosensitive area of $\sim$10 $m^{2}$ and the photon detection efficiency (PDE) about 50\%. The TAO CD operates at -50\si{\degreeCelsius} to suppress the dark noise of SiPMs. This design enables the detector to achieve an energy resolution better than 2\% at 1 MeV, and reaching sub-percent levels in the key energy region of the reactor spectrum. The inverse beta decay event rate induced by reactor neutrino is approximately 1000 per day within the fiducial volume \cite{Ref15}. In addition to providing spectral input to JUNO, TAO can measure the fine structure of the reactor neutrino spectrum, provide a reference spectrum for the nuclear database, search for sterile neutrinos, and monitor reactor operations by measuring the uranium-to-plutonium ratio.

To achieve the aforementioned physical goals, it is crucial to attain the designed energy resolution for the TAO experiment. However, the parasitic effects of SiPMs can directly impact the energy resolution. For instance, according to the TAO Conceptual Design Report, dark noise and optical crosstalk (OCT) contribute to a degradation of 0.75\% and 0.5\% on the energy resolution at 1 MeV \cite{Ref15}. In addition to affecting the energy resolution, these two effects can cause deviations of 1 mm and 2 mm in the vertex reconstruction, as reported by Shi \cite{Ref16}. Therefore, accurate calibration of all relevant SiPM parameters for each readout channel is of crucial importance.

According to the study in Ref. \cite{Ref17}, the OCT of the HPK (Hamamatsu Photonics K.K.) SiPM used in TAO is primarily composed of external optical crosstalk (EOCT). In laboratory settings, the EOCT rate of SiPMs can be measured by placing two SiPMs face-to-face and detecting the probability of coincident signals from both SiPMs \cite{Ref17, Ref18}. However, in the final detector configuration (such as the TAO CD), SiPMs are not arranged face-to-face. In recent work, Gallacher et al. proposed a time correlation method to measure the EOCT of SiPMs \cite{Ref19} in the LoLX experiment, where 96 SiPMs were operated at a temperature of 165 K. The dark count rate (DCR) in the LoLX experiment was significantly lower than that in the TAO experiment. Based on the results from the TAO SiPM mass test \cite{Ref32}, the DCR of the TAO SiPMs at -50\si{\degreeCelsius} is approximately 50 Hz/\si{\mm\squared}. In the TAO CD, 4024 SiPM tiles are deployed, where the area of each SiPM tile is $50.7 \times 50.7$ \si{\mm\squared}. Therefore, the total DCR of all SiPMs is approximately 500 MHz, the probability of accidental dark noise coincidence events within the 48 ns time correlation window (the presence of multiple dark noise events in the time window) approaches 100\%, which demonstrates that the time correlation method in the TAO detector will inevitably pick up dark noise events rather than genuine optical crosstalk signals. Therefore, a new EOCT calibration method must be developed.

This paper investigates and presents channel-level and tile-level calibration methods for SiPMs installed in the TAO CD, including calibration of the DCR, relative PDE, internal optical crosstalk (IOCT) rate, gain, and time offset. Additionally, a method based on selectively turning on and off different groups of SiPM tiles utilizing the fine control capability of the SiPM high-voltage system and an LED calibration source is proposed to calibrate the EOCT rate and emission angle distribution. This study uses simulated data generated by the TAO offline software.

This paper is structured as follows: Section \ref{2} introduces the TAO detector; Section \ref{2n} presents the simulated model; Section \ref{3} describes the calibration of the DCR, relative PDE, and time offset based on hit time information; Section \ref{4} details the calibration of gain and IOCT rate based on charge information; Section \ref{5} introduces a novel method for EOCT rate and emission angle distribution calibration; Section \ref{7} discusses the SiPM parameter uncertainties induced by temperature fluctuation and the impact of uncertainties originating from both temperature fluctuation and the calibration method on the reconstructed vertex and energy; Section \ref{6} summarizes the research work presented in this paper.

\section{TAO Detector and SiPM}
\label{2}

\begin{figure}[htbp]
  \centering
  \includegraphics[width=0.75\textwidth]{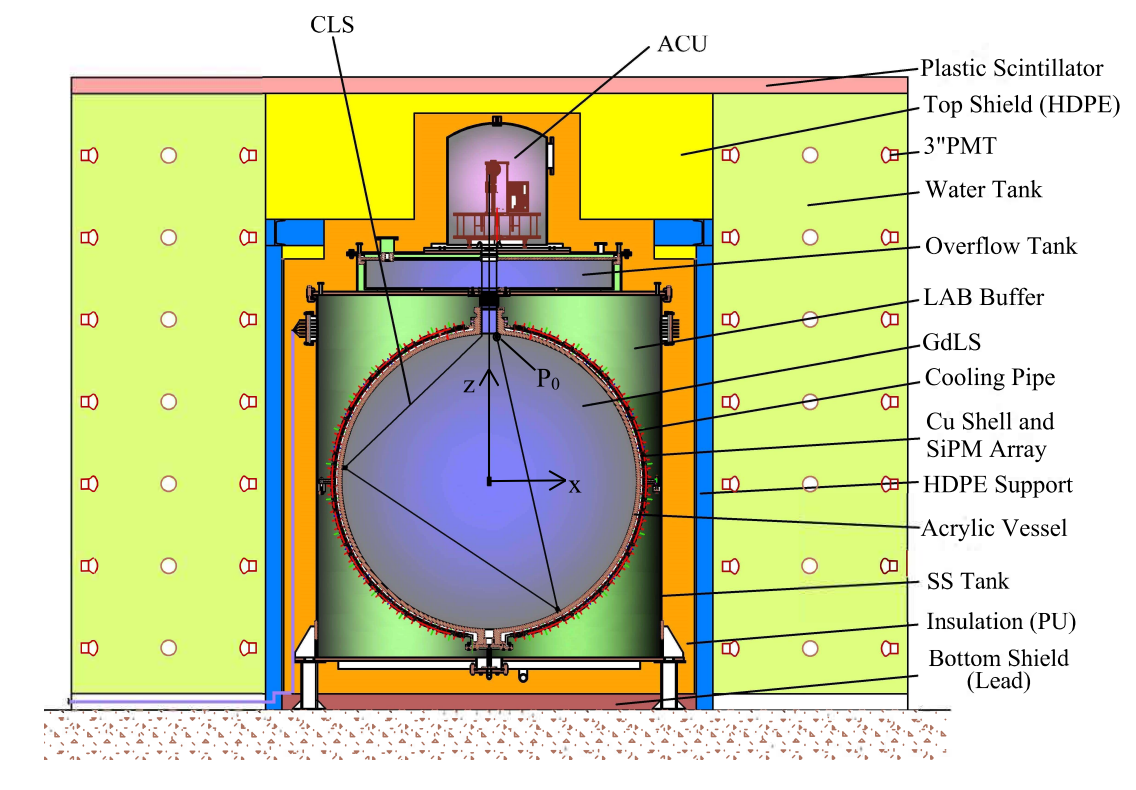}
  \caption{Design of the TAO detector.}
  \label{pic2.1}
\end{figure}

The TAO detector structure is shown in Figure \ref{pic2.1}. It consists of three sub-detectors: the central detector, water tanks and the top veto tracker.

The central detector contains 2.8 tonnes of gadolinium-doped liquid scintillator \cite{Ref20} enclosed within an acrylic sphere. The exterior of the acrylic sphere is covered by a copper shell, which inner surface is equipped with 4024 HPK SiPM tiles, each measuring $50.7 \times 50.7$ \si{\mm\squared}. The copper shell is housed inside a stainless steel tank filled with linear alkylbenzene (LAB), which serves as a buffer liquid to shield external radioactive background, maintain temperature stability, and provide optical coupling between the acrylic container and the SiPM tiles.  Both the tank and the copper shell have refrigeration tubes that cool the central detector to -50\si{\degreeCelsius} to reduce the SiPM DCR to below 100 Hz/\si{\mm\squared}. At the top of the CD  is the calibration system \cite{Ref21}, which includes an Automatic Calibration Unit and a Cable Loop System. The Automatic Calibration Unit can deploy calibration sources (including LED source, $^{68}{\rm Ge}$ source and combined $\gamma$ sources) along the central axis of the gadolinium-doped liquid scintillator. It is used to calibrate the CD's non-linearity and perform channel-level calibration of SiPMs. The Cable Loop System consists of the stainless steel cable passing through specific anchor points with a small segment plated with $^{137}{\rm Cs}$, which is used to calibrate the non-uniformity of the central detector.

Surrounding the central detector are three water tanks. Each tank is filled with 50 tonnes of pure water, which can shield external radioactive background and produces water Cherenkov light signals when muons pass through. The tanks are equipped with 300 3-inch PMTs, which detect the water Cherenkov signals to veto muons.

Above the central detector is the top veto tracker \cite{Ref47,Ref48}, which consists of four layers of plastic scintillators and is designed for muon veto.

\begin{figure}[htbp]
  \centering
  \includegraphics[width=0.75\textwidth]{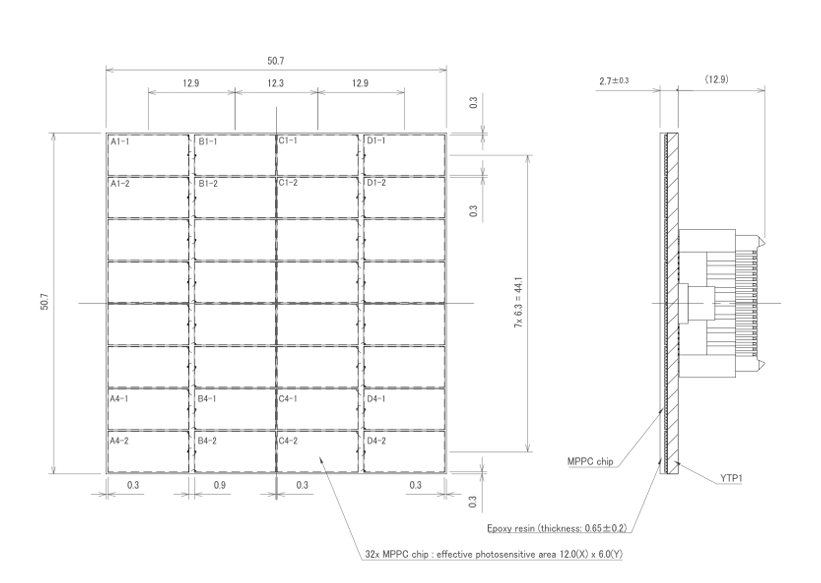}
  \caption{Dimensional outline of the front and side view of the S16088 SiPM tile.}
  \label{pic2x.1}
\end{figure}

We employ HPK MPPC S16088 SiPM in the TAO CD with their device design illustrated in Figure \ref{pic2x.1} \cite{Ref58}, which integrates thirty-two $6 \times 12$ \si{\mm\squared} chips into a single tile. Each chip incorporates 12782 APDs. The SiPM tile has a $0.65 \pm 0.20\ \mathrm{mm}$ thick epoxy resin coating on its surface, which protects the sensitive devices on the tile. In the TAO CD, each tile is fitted with 16 chips on both its left and right sides, with each side corresponding to one readout channel.

\section{TAO Central Detector Simulated Model}
\label{2n}

This paper adopts simulated data produced with the TAO offline software to investigate calibration methods for SiPM parameters, and this section details the corresponding simulation model.

The TAO offline software is a dedicated framework for experimental data processing, physics data analysis, and simulated data production for the TAO experiment. As a core offline component of the experiment, it is built on the SNiPER framework \cite{Ref23}. For the simulated data production in this work, three core modules are employed: physics event generation, detector simulation, and electronics simulation.

\subsection{Physics Event Generator}
\label{2n.1}

The first step of the full Monte Carlo (MC) simulation chain for the TAO is to produce primary particles using the physics event generator. The generator used in the TAO is directly inherited from the well-validated offline framework of the JUNO experiment. As documented in the JUNO simulation paper \cite{Ref24}, the JUNO framework supports three primary types of physics event generators: the particle gun, HepEvt, and GENIE. 

In this work, we adopt the particle gun, a dedicated tool that generates particles with well-defined species, momenta, and positions. We generate one million particle events for each SiPM parameter, and the particle events information corresponding to each SiPM parameter is summarized in Table \ref{t2.1}.

\begin{table}[htbp]
  \centering
  \begin{threeparttable}
  \caption{Summary of the simulated data used for the calibration of the SiPM parameters}
  \label{t2.1}
  \begin{tabular}{llll}
    \toprule
    SiPM Parameter & Particle & Position & Trigger Type \\
    \midrule
    Dark Count Rate & Arbitrary\tnote{1} & Arbitrary & Self Trigger \\
    Gain & Arbitrary & Arbitrary & Self Trigger \\
    Time Offset & LED & CD Center & External Trigger \\
    Relative PDE & $^{68}\text{Ge}$ & CD Center & Self Trigger \\
    Internal Crosstalk & Arbitrary & Arbitrary & Self Trigger \\
    External Crosstalk & LED & CD Center & External Trigger \\
    \bottomrule
  \end{tabular}
  \begin{tablenotes}
    \item[1] We use electrons uniformly distributed within the liquid scintillator with energies following a uniform distribution in the range of 1–10 MeV to represent arbitrary events.
  \end{tablenotes}
  \end{threeparttable}
\end{table}

\subsection{Detector Simulation}
\label{2n.2}

Following the physics event generation, detector simulation is performed as the next step of the simulation chain. The detector simulation module in the TAO offline software is built based on the Geant4 (version:10.04.p02) \cite{Ref41} toolkit. Firstly, the TAO detector geometry is constructed in Geant4 according to the detector structure of the TAO shown in Figure \ref{pic2.1}. Subsequently, the physics processes of particles propagating inside the detector are defined. All physics processes are consistent with those implemented in the detector simulation of the JUNO \cite{Ref24}, except for two dedicated optical processes: the scintillation light emission induced by energy deposition of particles in the gadolinium-doped liquid scintillator, and the physical process of optical photon detection by SiPMs.

For the simulation of optical processes in the liquid scintillator, a quenching effect occurs when charged particles deposit energy in the liquid scintillator. This quenching effect is described by Birks' law \cite{Ref54}, and the Birks constant of the liquid scintillator adopted in the TAO is determined from dedicated experimental measurements \cite{Ref55}. The remaining deposited energy, after excluding the quenched component, drives the scintillation light emission of the liquid scintillator. The wavelength distribution of the emitted scintillation photons is determined by the 2,5-diphenyloxazole (PPO) and  p-bis-(o-methylstyryl)-benzene (bis-MSB), and the wavelength spectrum implemented in the TAO detector simulation is fixed based on the experimental measurements in Ref. \cite{Ref50}. The optical attenuation length of the TAO liquid scintillator is jointly determined by the LAB, PPO and bis-MSB, and the value adopted in the TAO simulation is experimentally determined as specified in Ref. \cite{Ref50}. In addition, the light yield and scintillation time profile of the TAO liquid scintillator are experimentally characterized and determined as described in Ref. \cite{Ref20}.

For the simulation of optical processes for SiPMs, when optical photons hit the surface of a SiPM, they can be detected or reflected by the SiPM. The former is determined by the PDE of the SiPM, while the latter is governed by the optical reflectivity of the SiPM. Both parameters implemented in the TAO detector simulation are derived from dedicated experimental measurements: the PDE is taken from Ref. \cite{Ref26}, and the optical reflectivity is adopted from Ref. \cite{Ref51}.

Finally, the SiPMs detect both the light produced by energy deposition of particles in the liquid scintillator and the Cherenkov light generated via Cherenkov radiation, and store the corresponding detection data as SiPM hit information.








\subsection{Electronics Simulation}
\label{2n.3}

\begin{figure}[htbp]
  \centering
  \includegraphics[width=0.75\textwidth]{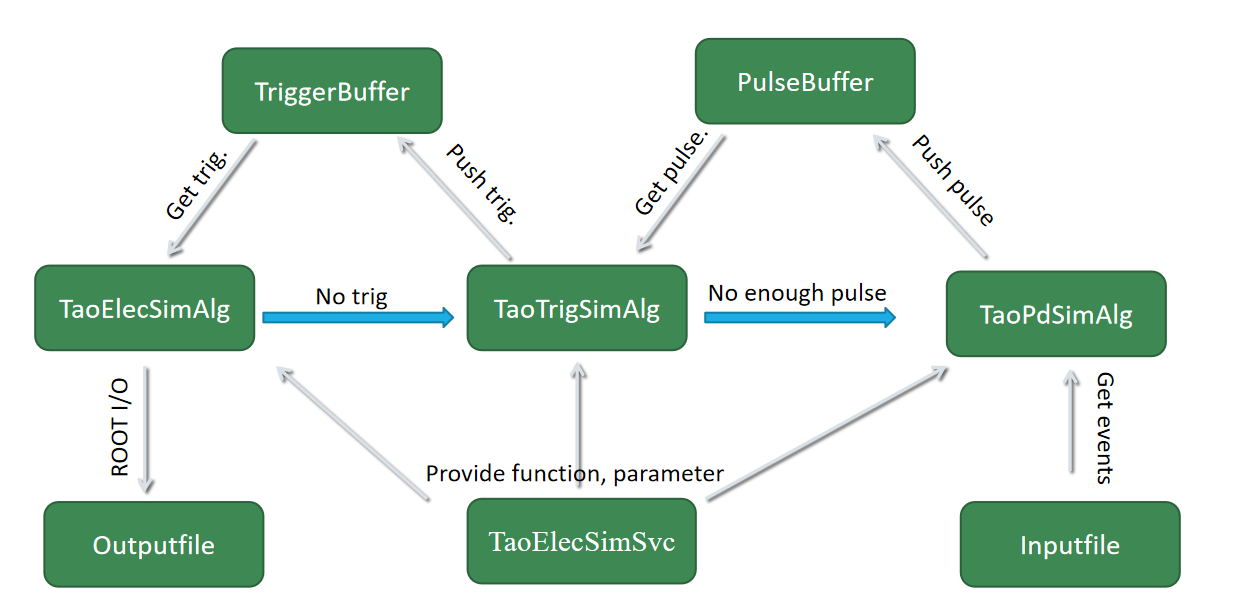}
  \caption{TAO electronics simulation workflow.}
  \label{pic2.1n}
\end{figure}

Following the generation of SiPM hit information via detector simulation, electronics simulation is performed as the subsequent stage of the full simulation chain. The TAO electronics simulation workflow is illustrated in Figure \ref{pic2.1n}, which takes the SiPM hit information produced by detector simulation as its input. The hit information is fed into the TaoPdSimAlg algorithm, which converts the SiPM hit information from each detector simulation event into electrical pulse data. Each electrical pulse is characterized by three core parameters: relative amplitude, timestamp, and type identifier. The relative amplitude is determined by the single photoelectron (PE) gain and single PE charge resolution of the SiPM. For the timestamp, the event-level timestamp of each electronics simulation event is obtained by sampling based on the physics event rate; the timestamp of each electrical pulse is then calculated by summing the event-level timestamp, the corresponding SiPM hit time, and the contributions from the electronics time offset and time smearing effects. The type identifier is used to indicate the specific SiPM channel or electronics process from which the electrical pulse originates. Thereafter, additional electrical pulses induced by the intrinsic SiPM effects, including dark noise, OCT, and afterpulse (AP), are generated based on the corresponding dedicated models. Details of the SiPM effect models are elaborated in Section \ref{2n.3.1}.

Following the generation of all electrical pulses, trigger decision and event packaging are performed within the TaoTrigSimAlg algorithm. The TAO electronic system supports two distinct trigger modes: self-trigger mode and external trigger mode. For the self-trigger mode, trigger decision is implemented based on a 300 ns trigger time window. This time window slides along the time axis, when the number of channels in which electrical pulses are detected within this time window exceeds the trigger threshold  (set to 1000 in the TAO electronics simulation), a valid trigger will be generated, and the start time of the  trigger time window is defined as the trigger time. For the external trigger mode, the internal threshold-based trigger decision logic is bypassed, and a forced trigger is generated for each individual detector simulation event. The timestamp sampled from the corresponding detector simulation event is directly adopted as the electronics trigger time. For both trigger modes, a fixed readout window spanning from 100 ns before the trigger time to 900 ns after the trigger time is applied, and all electrical pulses falling within this window are packaged into a single electronic event.

Following the generation of electronics event, the event is fed into the TaoElecSimAlg algorithm for waveform simulation, which converts each electrical pulse into an electrical pulse waveform. The amplitude of the waveform is determined by the relative amplitude of the electrical pulse, and white noise with a signal-to-noise ratio of 10 is added to the waveform. Subsequently, SiPM waveform identification and charge-time calculation are performed. A valid SiPM waveform is identified when three consecutive sampling points exceed the preset threshold (set at 50\% of the single PE waveform amplitude). The charge corresponding to the waveform is obtained via waveform integration, and the timing of the waveform is determined by linear fitting of the waveform rising edge, where the intersection of the fitted rising edge and the baseline is defined as the waveform timing. After the charge-time calculation is completed, three parameters of each waveform are obtained: the charge, the time relative to the trigger time, and the associated electronics channel number. Finally, the SiPM parameters of each channel are calibrated based on the timing and charge information.

In the TAO simulation, the true values of each SiPM parameter are known. Accordingly, we quantify the calibration performance using two core metrics: the calibration bias and the standard deviation. The calibration bias is defined as the mean difference between the calibrated value and the corresponding simulated truth, averaged over all channels. The calibration standard deviation is calculated as the standard deviation of the residuals (calibrated value minus the corresponding simulated truth) across all channels.

\subsection{SiPM Simulation Modeling}
\label{2n.3.1}

\subsubsection{Dark Noise}
\label{2n.3.1.1}

Dark noise refers to the effect in which a SiPM spontaneously generates avalanche current in the absence of photon hits. It arises from two distinct physical processes: one is thermal current, which is generated when valence band electrons in the SiPM are excited into the conduction band by thermal energy, and trigger an electron avalanche to produce avalanche current during this process; the other is tunneling current, which occurs when valence band electrons in the SiPM have a probability of tunneling into the conduction band via the quantum tunneling effect under the action of an electric field, and subsequently induce avalanche current \cite{Ref25}.

For both thermal current and tunneling current, the corresponding dark noise events are uniformly distributed in time, and the generation of different dark noise events is statistically independent. Therefore, the number of dark noise events generated within a given time interval follows a Poisson distribution. Therefore, in the TAO electronics simulation, dark noise pulses are generated via Poisson sampling, while the corresponding timestamps of the dark noise pulses are generated by sampling from a uniform distribution. The DCR is configured to 20 Hz/\si{\mm\squared} in simulation, while the most probable value of the DCR from the current TAO SiPM mass test result is approximately 50 Hz/\si{\mm\squared} \cite{Ref32}, with a formal acceptance criterion specifying that the DCR of all TAO SiPM tiles must be less than 100 Hz/\si{\mm\squared} at -50\si{\degreeCelsius}. Since there is no difference in order of magnitude, it does not affect the calibration method study. 

\subsubsection{Time Offset}
\label{2n.3.1.2}

In the TAO experiment, there exist differences in the length of signal cables for SiPM tiles at different positions, and potential time asynchrony between the ADCs used in the TAO readout system. These effects lead to relative time offsets between different electronics channels. In the TAO electronics simulation, the time offset is implemented by adding distinct time offset constants to the timestamps of electrical pulses corresponding to different channels. In this study, the time offset is configured to 0 ns for the upper hemisphere channels and 20 ns for the lower hemisphere channels.

\subsubsection{Photon Detection Efficiency }
\label{2n.3.1.3}

The SiPM PDE is defined as the ratio of the number of photons detected by the SiPM to the number of photons incident on the SiPM surface. The SiPM PDE is generally characterized by three factors \cite{Ref27, Ref28}:
\begin{equation}
\label{eq2n.1}
\text{PDE}(\lambda,V) = \text{QE}(\lambda) \cdot \text{Prob}_{\rm GM}(V) \cdot \text{FF}
\end{equation}
where QE is the Quantum Efficiency, which varies with the wavelength of incident light; $Prob_{\rm GM}$ is the probability of triggering an electron avalanche effect, which is related to the internal electric field strength of the SiPM and thus to the operating voltage of the SiPM; and FF (Fill Factor) is a geometric factor used to evaluate the impact of the SiPM geometry on the photon detection efficiency. For the TAO simulation, the detector geometry is constructed from engineering drawings, and the geometric effect is automatically set by Geant4.

The PDE used in the TAO detector simulation is set to 50\%, which is generally consistent with the measured results from TAO SiPM mass test \cite{Ref32}. This PDE value is configured on the optical surface of the Geant4 sensitive detector corresponding to the SiPMs in the TAO detector simulation.

\subsubsection{Gain}
\label{2n.3.1.4}

Gain is defined as the total charge of the waveform, which is generated after a photon hits the SiPM, is detected, produces an avalanche current, and is amplified and shaped by the electronics system.

Due to various stochastic processes in the SiPM and the electronics, there is a fluctuation in the amplification factor that converts a single photoelectron into a single photoelectron waveform, which is the charge resolution of the SiPM and the electronics. In the TAO electronics simulation, a 15\% Gaussian smearing is added to the relative amplitude of the electrical pulse to represent the 15\% charge resolution of the TAO SiPM and electronics reported in the TAO Conceptual Design Report \cite{Ref15}.

\subsubsection{Optical Crosstalk}
\label{2n.3.1.5}

OCT of the SiPM arises from the fact that avalanche electrons have a finite probability of emitting photons during acceleration in the electric field of the SiPM. These photons have a certain probability of leaving the Avalanche Photodiode (APD) where they are generated, impinging on other APDs, and triggering new avalanche signals \cite{Ref44, Ref30}.

If the APD hit by the OCT photon and the APD that generates the photon belong to the same electronics channel, this effect is defined as IOCT. If the APD hit by the OCT photon and the APD that generates the photon belong to different electronics channels, this effect is defined as EOCT.

According to the study in Ref. \cite{Ref31}, the number of OCT events generated by a signal follows a cascaded Poisson distribution. For a single avalanche signal, the total number of the primary signal and the OCT events it induces follows a Borel distribution:
\begin{equation}
\label{eq2n.2}
\text{Borel}(k)=\frac{(\lambda_{\rm OCT} \cdot k)^{k-1} \cdot e^{-k\cdot \lambda_{\rm OCT}}}{k!}(k=1,2,...)
\end{equation}
where $k$ represents the total number of SiPM electrical pulses induced by one primary avalanche signal and all subsequent OCT events triggered by the primary signal, and $\lambda_{\rm OCT}$ is the Poisson parameter characterizing the intrinsic OCT probability per avalanche event.

The physical meaning behind this mathematical formulation can be intuitively interpreted as a multi-generation cascaded avalanche process: a primary avalanche signal produces a set of first-generation OCT events through a Poisson process with rate parameter $\lambda_{\rm OCT}$. Each of these first-generation OCT events will trigger an independent avalanche, which can further generate second-generation OCT events via an identical Poisson process with the same rate parameter $\lambda_{\rm OCT}$. This iterative cascade process continues until no additional OCT events are generated in a given generation. Mathematically, the total count of the primary signal and all cascaded OCT events across all generations converges to the Borel distribution given above.

More generally, for avalanche signals generated by a Poisson process with a Poisson coefficient of $\mu$, the sum of the number of avalanche signals and the number of OCT events generated by the avalanche signals mathematically follows a generalized Poisson distribution:
\begin{equation}
\label{eq4.4}
\text{GP}(k,\mu,\lambda) = \frac{\mu \cdot (\mu + \lambda \cdot k )^{k-1} \cdot e^{-\mu-\lambda \cdot k}}{k!}
\end{equation}

In the TAO electronics simulation, IOCT and EOCT are simulated separately. For IOCT, the corresponding number of IOCT pulses is generated via Borel distribution sampling. IOCT will be induced by every electrical pulse generated from a SiPM photon hit, a dark noise electrical pulse, and an EOCT electrical pulse. According to the study in Ref. \cite{Ref17}, the IOCT of HPK SiPMs in LAB is very low, with a Poisson coefficient $\lambda_{\rm IOCT}$ less than 0.1\%, which makes this effect negligible and no corresponding calibration required. The TAO SiPM batch tests measured the in-air IOCT probability of the SiPMs, which value is approximately 15\%. We directly adopted this value in the TAO electronics simulation, to investigate how to calibrate this parameter under the scenario where the IOCT effect is non-negligible.

The simulation of EOCT is implemented by performing Poisson sampling on each generation of electrical pulses. Each electrical pulse, including those induced by SiPM photon hits, dark noise, IOCT, is subjected to Poisson sampling with a coefficient of $\lambda_{\rm EOCT}$ to generate first-generation EOCT pulses. These newly generated EOCT pulses, together with the IOCT produced by them, are further processed with Poisson sampling of the same coefficient $\lambda_{\rm EOCT}$ to produce a new generation of EOCT pulses. Such a cascaded process proceeds iteratively until no additional EOCT pulses are generated. In the TAO electronics simulation, the $\lambda_{\rm EOCT}$ is adopted from the measurement of EOCT for HPK SiPMs reported in Ref. \cite{Ref17}, with a value of 0.2447.

The electronics channel ID hit by EOCT is obtained by simulating a specific photon emission distribution on the SiPM channel surface and observing the photon hit responses of other SiPM channels. As calculated in Ref. \cite{Ref59} based on a thin-film optical model and Fresnel equations, most EOCT photons emitted from the monocrystalline silicon have their propagation directions confined within $30^\circ$ relative to the surface normal as they travel through the epoxy resin coating on the SiPM tile surface into the LAB medium. Since the actual angular distribution depends to some extent on the surface fabrication process of the SiPM, and the emission angle distribution of EOCT photons for the TAO SiPM has not been experimentally measured, we adopt a simplified assumption: the emission angle of EOCT photons follows a Gaussian distribution with a mean of $0^\circ$ and a standard deviation of $30^\circ$, to characterize the small-angle distribution feature of EOCT photons.

Taking the SiPM with SiPM ID = 0 as an example, one million optical photons are simulated using a particle gun on the surface of the corresponding SiPM. The angle between the photon propagation direction and the surface normal is sampled from a Gaussian distribution with a mean of $0^\circ$ and a standard deviation of $30^\circ$. If the sampled angle larger than $90^\circ$, the sampling is repeated until angle less than or equal to $90^\circ$.

\begin{figure}[!htbp]
\centering
\includegraphics[width=.75\textwidth]{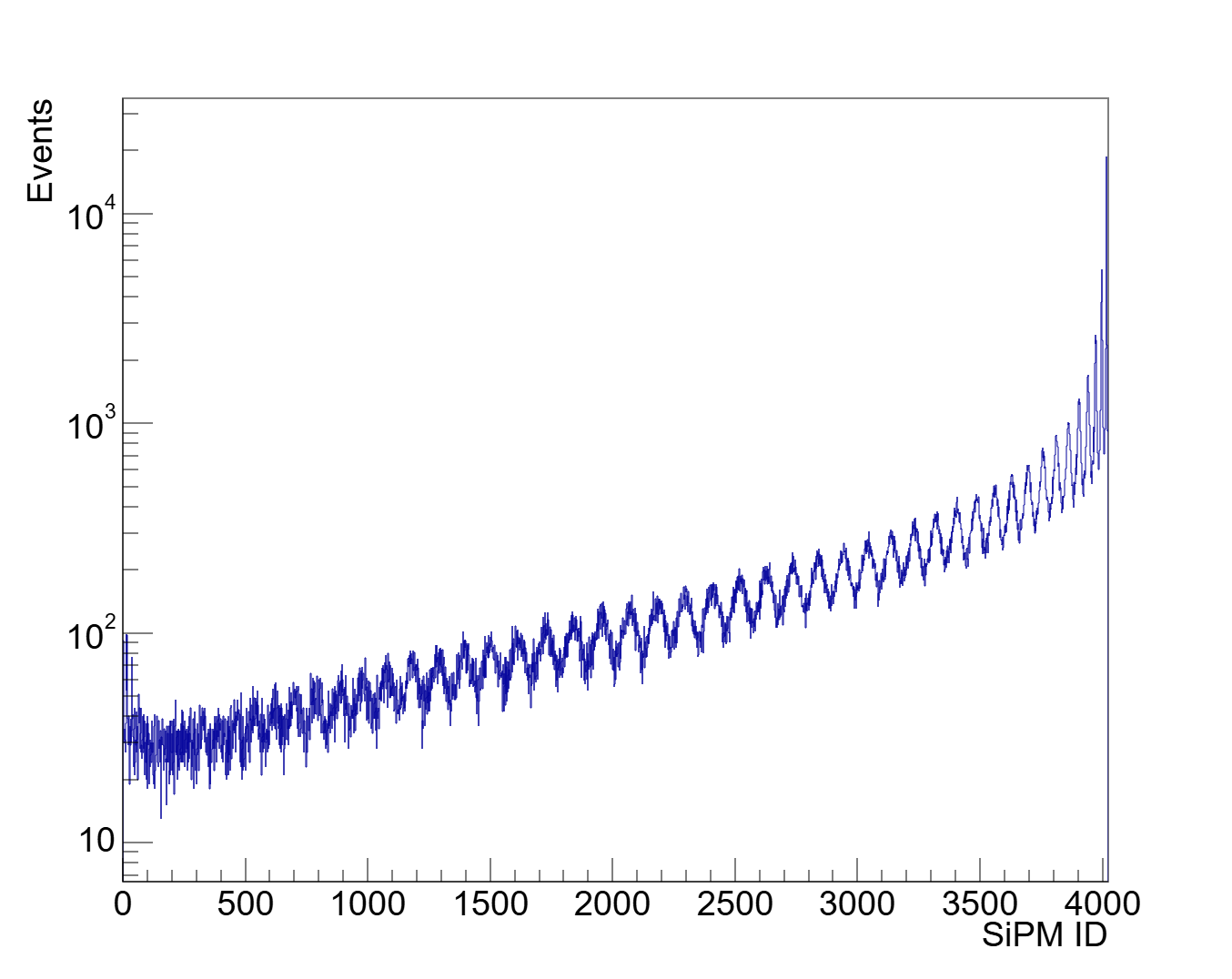}
\caption{Hit SiPM ID distribution from EOCT hits generated by the SiPM ID = 0 channel.}
\label{pic2.3n}
\end{figure}

Following the above simulation procedure, the distribution of SiPM ID hit by EOCT generated from the SiPM with ID = 0 is shown in Figure \ref{pic2.3n}. Since the photons generated by EOCT are emitted predominantly along the normal direction of the SiPM tile surface, the SiPM facing the EOCT-emitting SiPM within each layer of the SiPMs receive the largest number of EOCT photons. This results in a multi-peak structure in the SiPM ID distribution, where each peak corresponds to the EOCT response received by the SiPMs in each layer. The electronics channel ID hit by EOCT is then obtained by sampling from this histogram.

\begin{figure}[!htbp]
\centering
\includegraphics[width=.75\textwidth]{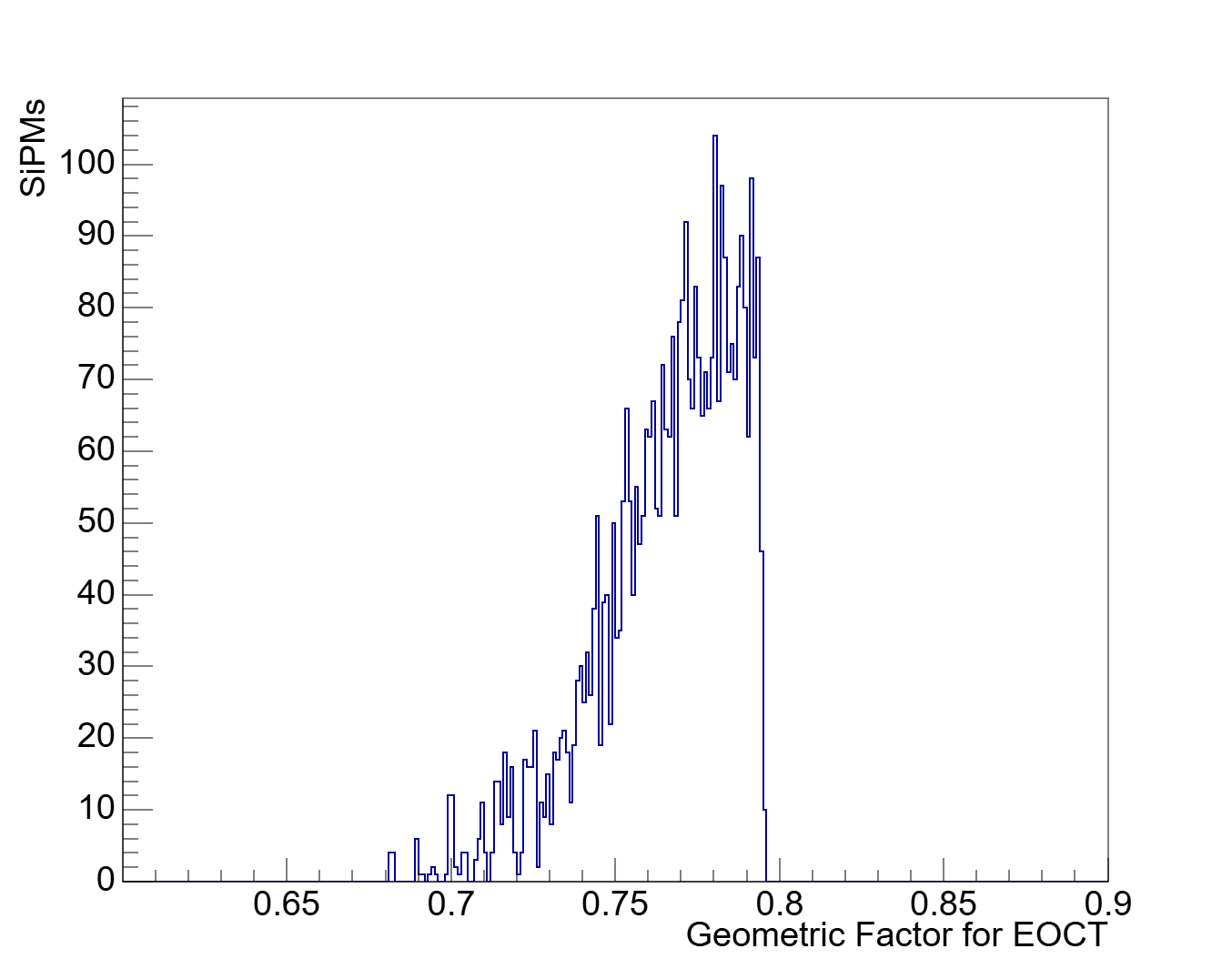}
\caption{Geometric factor distribution for EOCT.}
\label{pic2.4n}
\end{figure}

It is worth noting that since the coverage of TAO SiPMs is not 100\%, the total number of photon hits is less than one million. The ratio of the total number of photon hits to one million is defined as the geometric factor for EOCT. The actual Poisson coefficient for EOCT is 0.2447 multiplied by the geometric factor corresponding to the SiPM. The geometric factor varies for SiPMs at different positions. The distribution of the geometric factor for all 4024 SiPMs is shown in Figure \ref{pic2.4n}; the mean value is 0.7429.

\subsubsection{Afterpulse}
\label{2n.3.1.6}

During the avalanche discharge process of the APD, there is a probability that electron-hole pairs are trapped by lattice defects. The trapped carriers will be re-released, which in turn triggers a new avalanche process; this phenomenon is AP \cite{Ref30}. The AP in SiPMs can be divided into two categories. One is delayed AP, which has a long time interval from the main avalanche signal (> 1 $\mu\text{s}$). For the TAO SiPM, the measured delayed AP is negligible \cite{Ref26}. The other is non-delayed AP. Due to the short time interval between this type of AP and the main signal, the APD is still in the avalanche recovery phase at this time, so the gain of the AP is lower than that of the primary avalanche pulse.

Based on the results of the TAO SiPM mass test \cite{Ref32}, the AP probability of the TAO SiPM is very low (<1\%). In the TAO simulation, we have implemented a 1\% AP effect. The time interval $\Delta t$ between the AP and the primary pulse, as well as the relative gain of the AP $\text{Gain}_{AP}$ with respect to the primary pulse, both follow an exponential distribution \cite{Ref56}, as shown in Equation \ref{eq2n.3} and Equation \ref{eq2n.4}, respectively.
\begin{equation}
\label{eq2n.3}
f(\Delta t)=e^{-\frac{\Delta t}{\tau_{AP}}}
\end{equation}
\begin{equation}
\label{eq2n.4}
\text{Gain}_{AP} = 1-e^{-\frac{\Delta t}{\tau_{rec}}}
\end{equation}
where $\tau_{\text{AP}}$ is the AP time constant, and $\tau_{\text{rec}}$ is the avalanche recovery time constant. Since these two time constants were not measured in the TAO SiPM mass test, we adopted the values reported in Ref. \cite{Ref52}, which are 20 ns and 13 ns, respectively.


\section{Calibration Based on Hit Time Information}
\label{3}

\subsection{Dark Count Rate}
\label{3.1}

\begin{figure}[htbp]
\centering
\includegraphics[width=.75\textwidth]{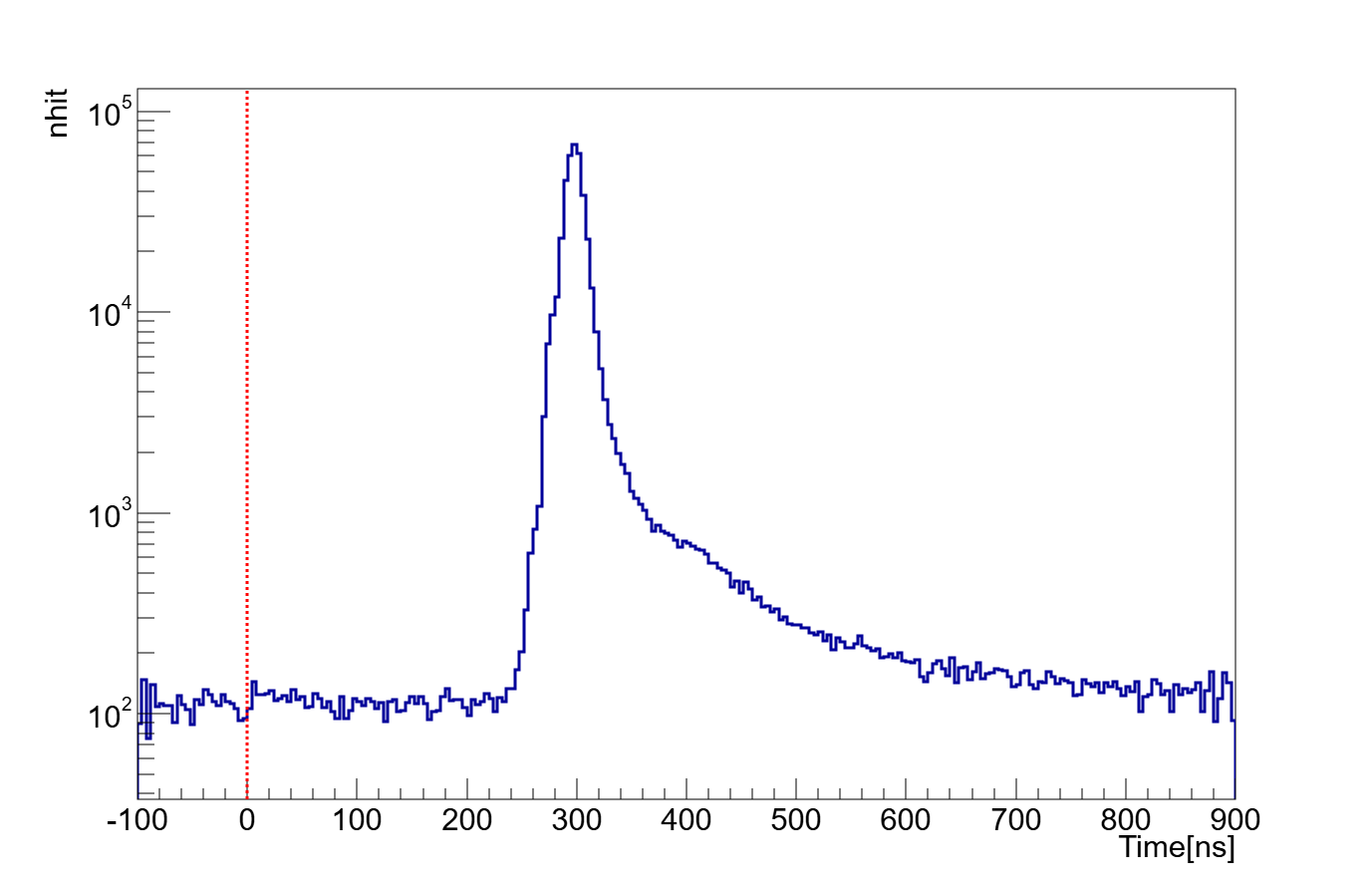}
\caption{Hit time distribution from self-triggered events. The trigger time corresponds to 0 ns, which is indicated by the red line in the figure. The hits from physics events are concentrated around 300 ns, as determined by the 300 ns width of the trigger time window. The plateau preceding the hits of physics events corresponds to the dark noise plateau. } 
\label{pic3.1}
\end{figure}

The DCR can be estimated by counting the recorded hits in the plateau region before the main signal hits from physics events. This plateau region corresponds to the pre-trigger time, which is the area to the left of the red line (indicating the trigger time) in Figure \ref{pic3.1}. The DCR can be calculated using Equation \ref{eq3.1}, where $n$ is the total number of the counted hits in the plateau, $N$ is the number of physics events, $t$ is the duration of the plateau before the trigger timestamp (from -100 ns to 0 ns), and $S$ is the sensitive area of a single SiPM channel.

\begin{equation}
\label{eq3.1}
\text{DCR} = \frac{n}{N \cdot t \cdot S}
\end{equation}

However, the hits observed in the plateau range may also include minor contributions from hits generated by the EOCT of dark noise from other SiPM channels, as well as hits from delayed APs. For the TAO SiPM, the delayed AP has been measured to be very weak~\cite{Ref26}. Therefore, only EOCT contamination needs to be considered in the DCR calibration.

\begin{figure}[htbp]
\centering
\includegraphics[width=.45\textwidth]{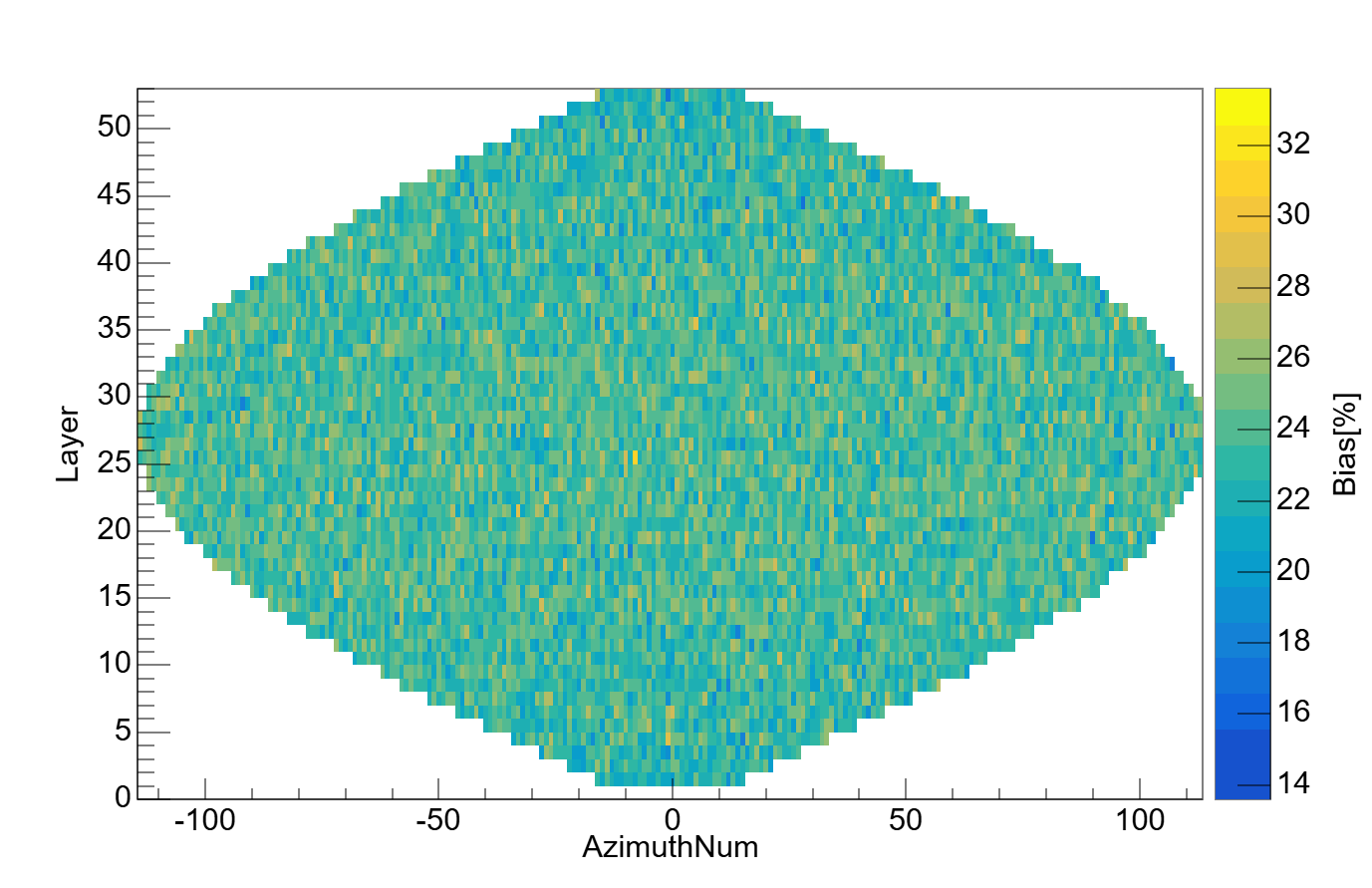}
\qquad
\includegraphics[width=.45\textwidth]{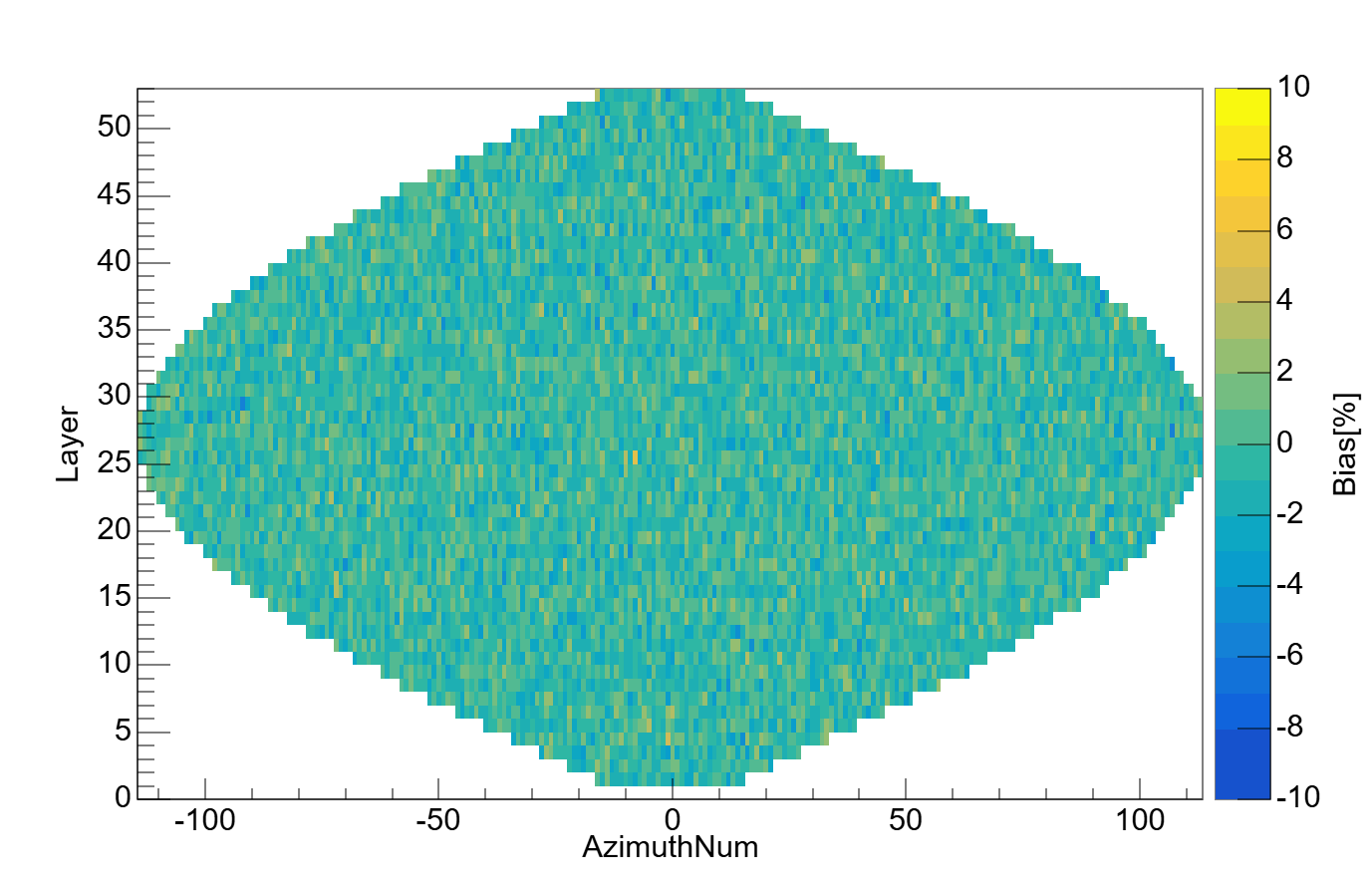}
\\
\includegraphics[width=.45\textwidth]{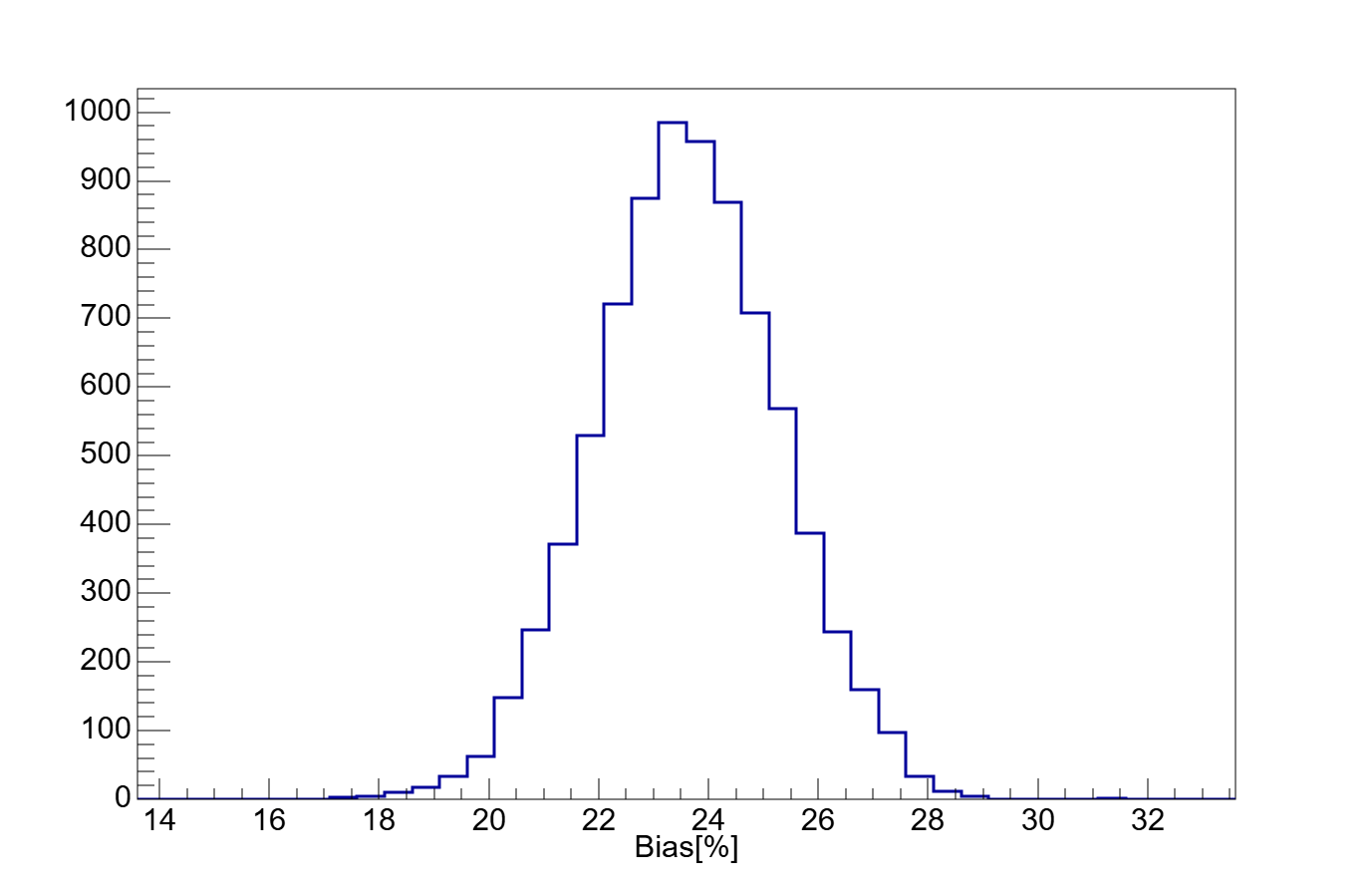}
\qquad
\includegraphics[width=.45\textwidth]{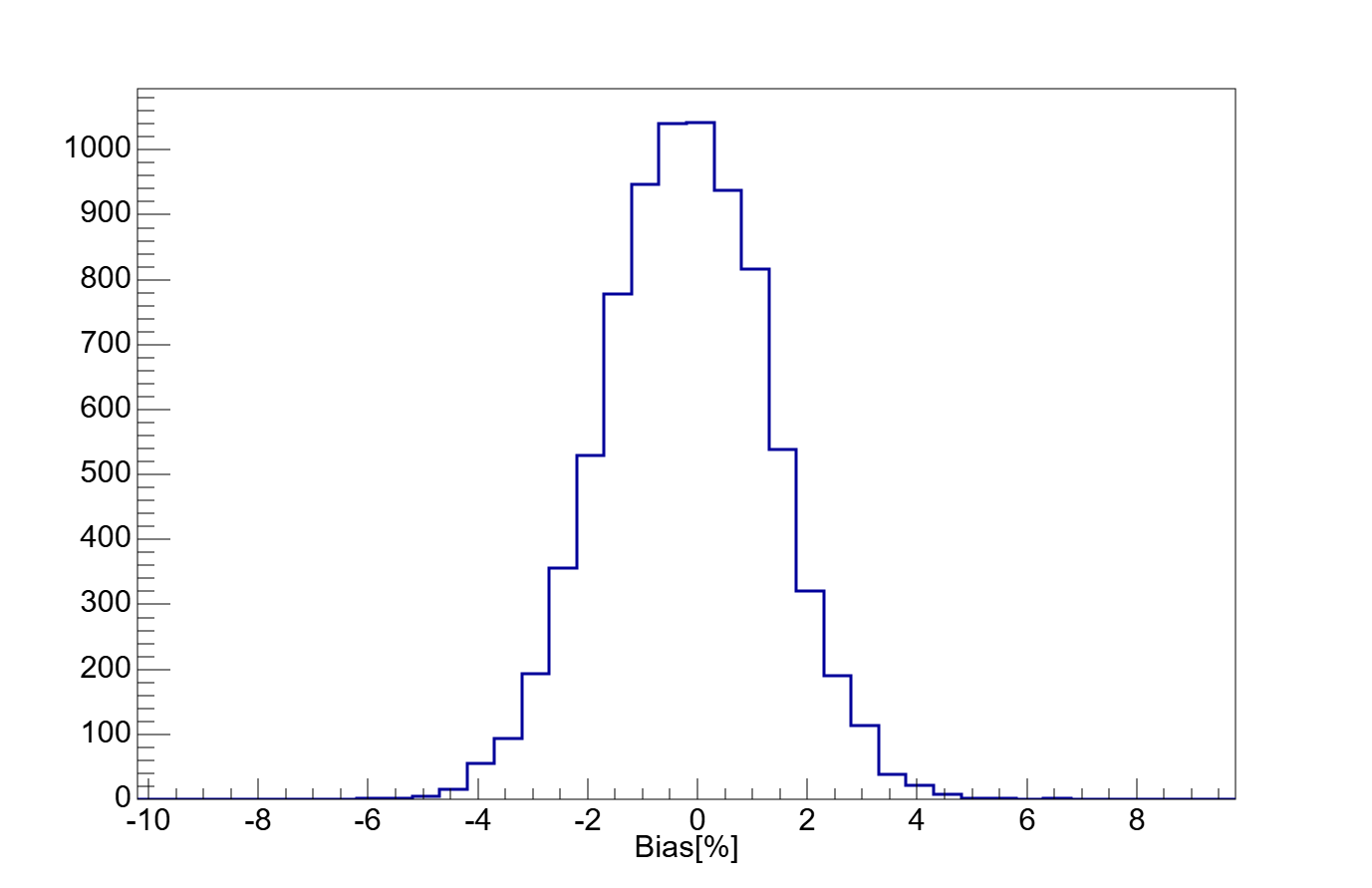}
\caption{DCR calibration results. The left two panels: calibration results with EOCT included. The right two panels: calibration results with EOCT excluded. The upper two panels: comparison between the calibration results and the simulation truth in each channel. The lower two panels: 1D histogram filled with each bin value from the upper 2D histogram. }
\label{pic3.2}
\end{figure}

The DCR calibration results are shown in the left part of Figure \ref{pic3.2}. Since the SiPM tiles are mounted in circles around the central $z$-axis of the detector and layer by layer along the central $z$-axis, the channels have azimuth number and layer number as their coordinates. A significant bias of 23.6\% with the standard deviation of 1.7\% is observed. However, when the EOCT effect is disabled, the bias is substantially reduced to -0.2\% as presented in the right part of Figure \ref{pic3.2}. This clearly demonstrates that the EOCT effect introduces a considerable bias in the calibration results. A correction method will be presented in Section \ref{5.3}.

\subsection{Time Offset}
\label{3.3}

The hit time of the SiPM obtained via the electronic readout satisfies Equation \ref{eq3.4}, where $t_R$ is the hit time from the electronic readout. It is calculated as $t_R = t_A - t_T$, where $t_A$ is the absolute timestamp and $t_T$ is the trigger timestamp. The absolute timestamp consists of several components: $t_P$ is the time when the particle reaches and deposits energy in the liquid scintillator; $t_{\rm LS}$ is the time when the liquid scintillator emits scintillation photons; $t_{\rm tof}$ is the time of flight of scintillation photons to the SiPM; the electronic processing time, which is divided into two parts: $t_{\rm same}$ and $t_{\rm timeoffset}$. Here, $t_{\rm same}$ represents the identical portion of the processing time across all electronic channels, while the differing portions constitute the time offset ($t_{\rm timeoffset}$) in different channels.

\begin{equation}
\label{eq3.4}
\begin{split}
t_R &= t_A - t_T \\
t_A &= t_P + t_{\rm LS} +t_{\rm tof} + t_{\rm same} + t_{\rm timeoffset} \\
t_{\rm timeoffset} &= t_R + t_T - t_P - t_{\rm LS} - t_{\rm tof} - t_{\rm same}
\end{split}
\end{equation}

Since the TAO experiment only requires correcting the relative time offsets between electronic channels \cite{Ref15} to ensure that the timestamps of all channels are synchronized, this study deploys an LED source at the center of the liquid scintillator to calibrate the relative time offset between different channels. The LED source acquires events using an external trigger method, ensuring that the trigger time $t_T$ is identical across all channels. Therefore, $t_T$ does not affect the calibration of relative time offset. The $t_P$ and $t_{\rm LS}$ are statistically identical across all channels, so the $t_P$ and $t_{\rm LS}$ do not influence relative time offset calibration. Because the LED is deployed at the center of the liquid scintillator, the distance from the liquid scintillator center to each SiPM channel is the same, meaning the $t_{\rm tof}$ is identical across all channels. Consequently, the relative time offsets between different channels can be derived from the relative $t_R$ differences for each channel. 

\begin{figure}[htbp]
\centering
\includegraphics[width=.75\textwidth]{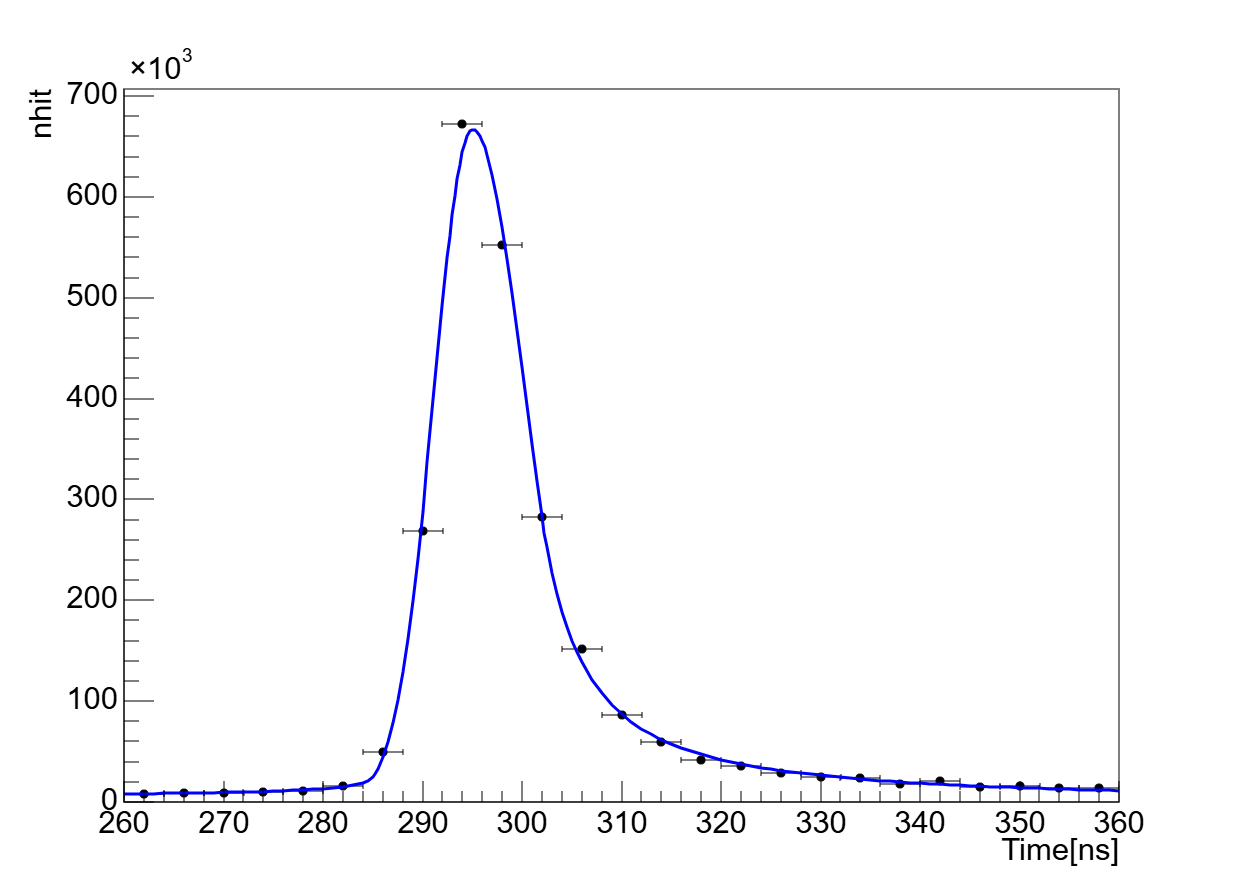}
\caption{Double-sided crystal ball function fit to the time spectrum. The mean value of the fitted function corresponds to the $t_R$ value.}
\label{pic3.7.1.1}
\end{figure}

As shown in Figure \ref{pic3.7.1.1}, we use the double-sided Crystal Ball function to fit the distribution of $t_R$. The mean value of the fitted function corresponds to the $t_R$ value for the channel. When a reference channel is selected, the time offset of $\rm channel_{\rm i}$ relative to the reference channel ($t^{\rm chid=i}_{\rm timeoffset}$) is the readout time of $\rm channel_{\rm i}$ ($t^{\rm chid=i}_{R}$) minus the readout time of the reference channel ($t^{\rm RefCh}_{R}$).

\begin{equation}
\label{eq3.5}
t^{\rm chid=i}_{\rm timeoffset} = t^{\rm chid=i}_{R} - t^{\rm RefCh}_{R}
\end{equation}

\begin{figure}[htbp]
\centering
\includegraphics[width=0.577\textwidth]{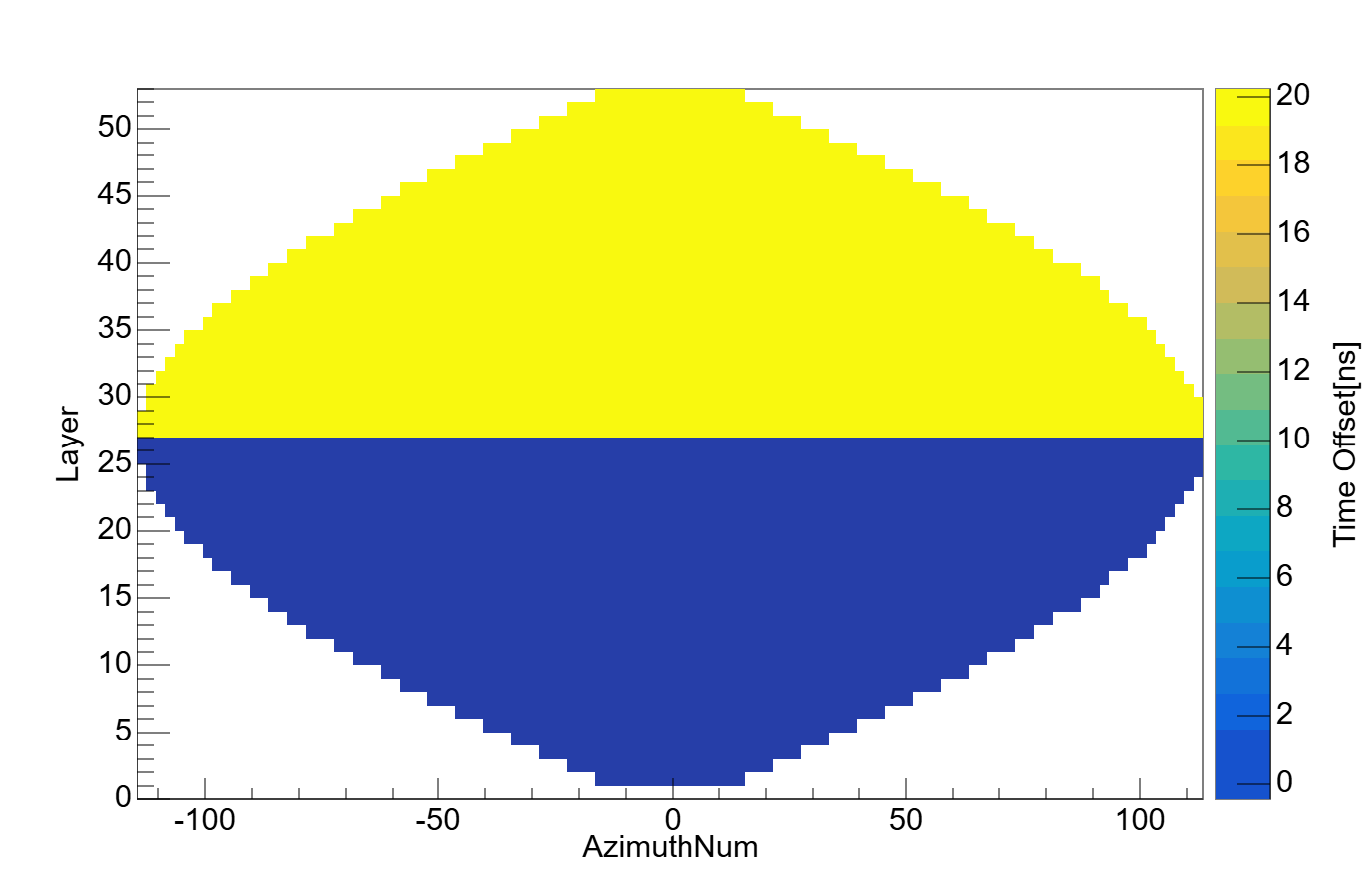}
\\
\includegraphics[width=0.577\textwidth]{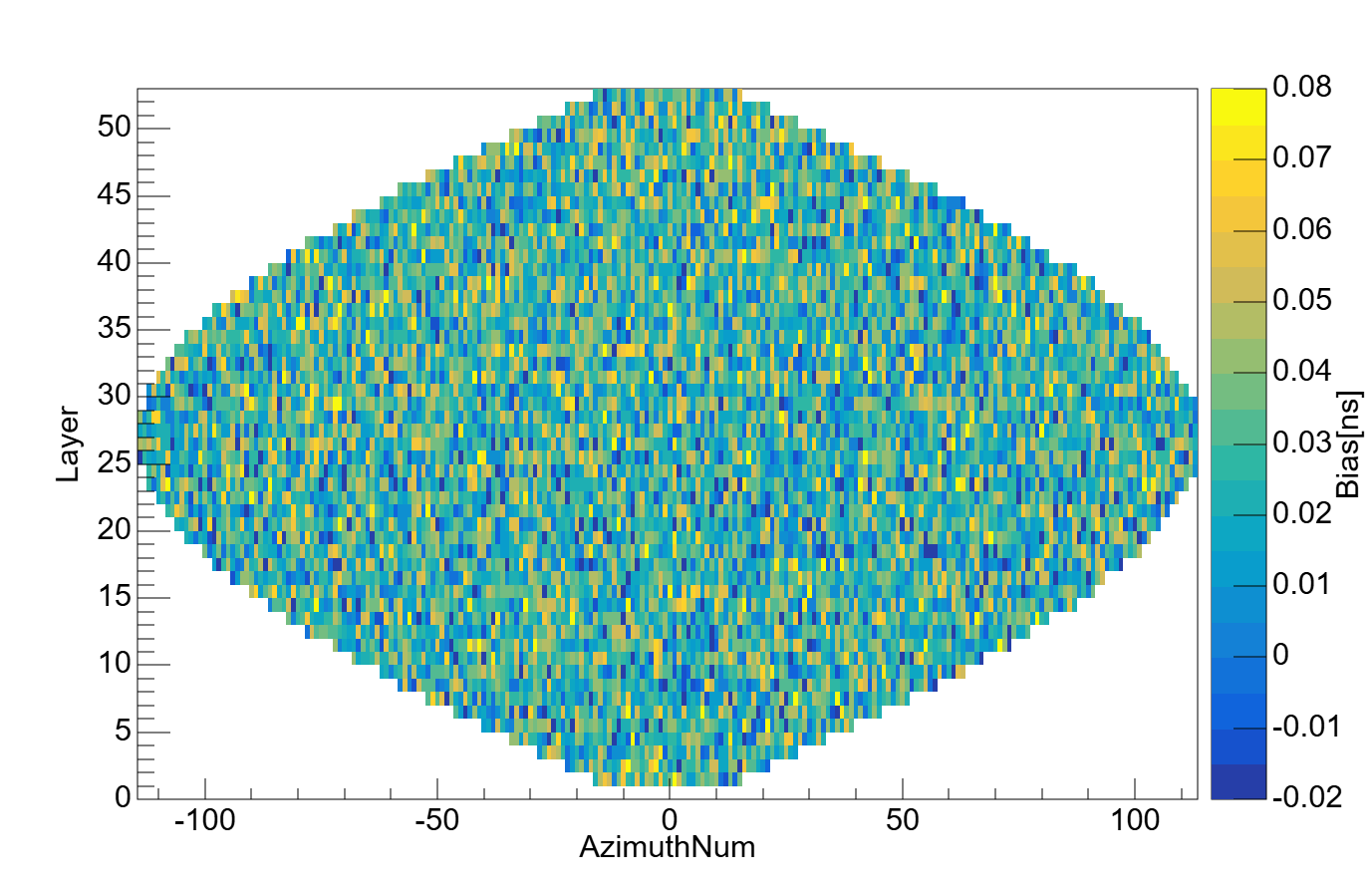}
\\
\includegraphics[width=0.577\textwidth]{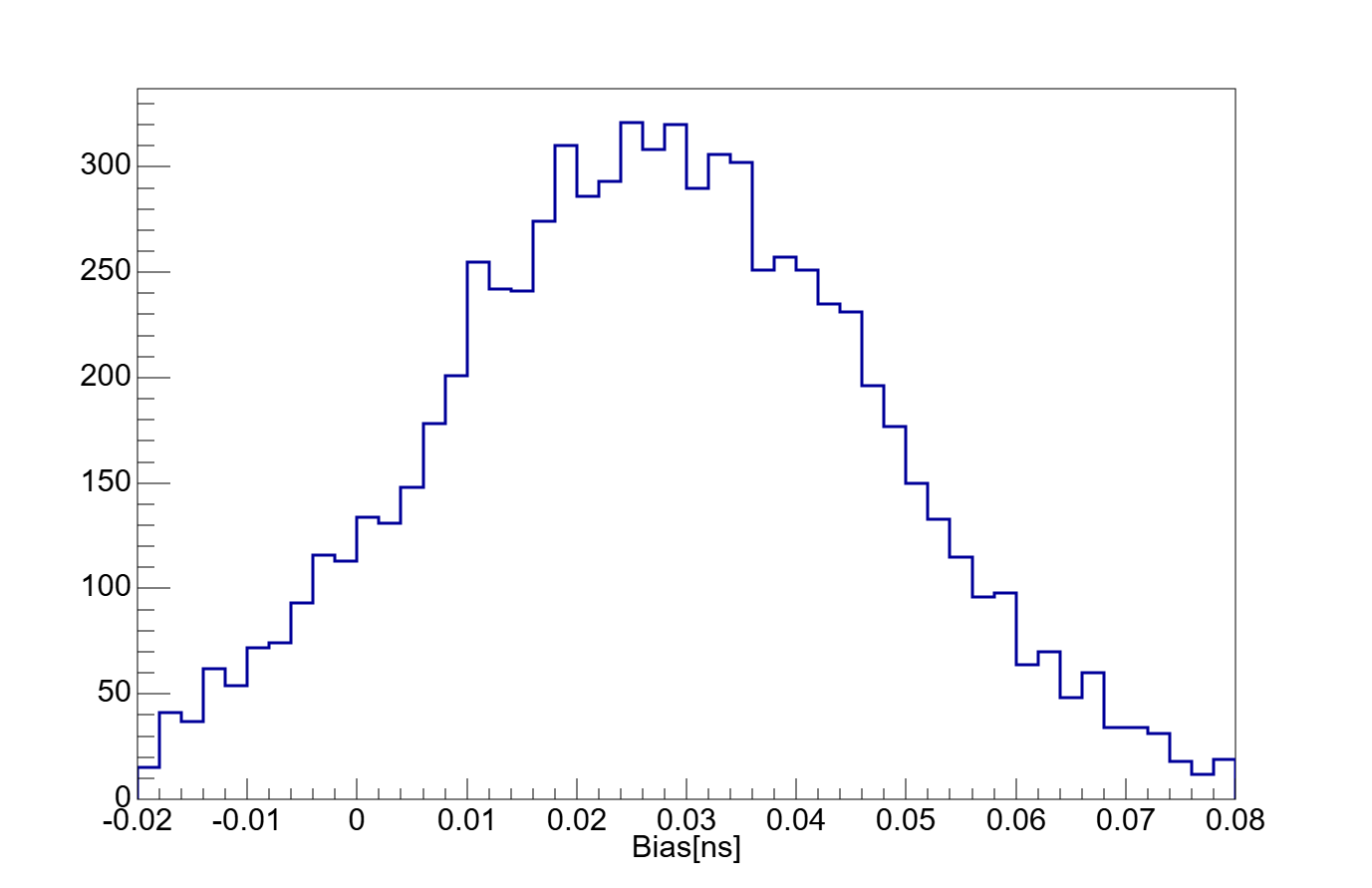}
\caption{Time offset calibration results. The upper panel: time offset calibration results. The middle panel: comparison between the calibration results and the simulation truth in each channel. Lower panel: 1D histogram filled with each bin value from the middle 2D histogram.}
\label{pic3.7.1}
\end{figure}

 The time offset calibration results are shown in Figure \ref{pic3.7.1}; the bias is 0.027 ns and the standard deviation is 0.031 ns.

 In the actual TAO CD, the calibration source's position may be offset from its deployed position, and this offset contributes to a bias in the time of flight. Specifically, a 1 cm offset results in a 0.13 ns bias between the SiPM channels located at the south and north poles of the CD. Considering that the time resolution of the TAO electronics exceeds 1 ns \cite{Ref15}, both the calibration bias and the calibration source's positional offset from the liquid scintillator center are acceptable.


\subsection{Relative Photon Detection Efficiency}
\label{3.2}

As shown in Equation \ref{eq3.2},  the PDE of a SiPM is defined as the ratio of the number of photons detected by the SiPM ($N_{\rm Det}$) to the number of photons incident on the SiPM surface ($N_{\text{photon}}$):
\begin{equation}
\label{eq3.2}
\text{PDE}=\frac{N_{\rm Det}}{N_{\text{photon}}}
\end{equation}
Since the number of photons incident on the SiPM surface depends on the particle type, particle energy, light yield of the liquid scintillator, and light attenuation length of the liquid scintillator, calibrating the absolute PDE of each SiPM is impractical. However, calibrating the relative PDE is feasible. The TAO calibration system employs three types of calibration sources: an LED source, a $^{68}{\rm Ge}$ source and a combined $\gamma$ source \cite{Ref21}. The light field produced by LED source may be not isotropic, and the combined $\gamma$ source consists of multiple $\gamma$ sources with different energies. In contrast, the $^{68}{\rm Ge}$ source is a positron annihilation $\gamma$ source. Due to the presence of the source shell, positrons annihilate inside the shell, producing a pair of $\gamma$ emitted in opposite directions, which enhances the uniformity of the light field. Therefore, the $^{68}{\rm Ge}$ source is used for relative PDE calibration.

Deploying the $^{68}{\rm Ge}$ source at the center of the liquid scintillator, the number of photons detected by each channel follows a Poisson distribution. The mean number of photons detected by the SiPM is the parameter of the Poisson distribution, which satisfies:
\begin{equation}
\label{eq3.3}
N_{\rm Det} = -\text{ln}(\frac{N_{\text{PE=0}}}{N_{\text{Event}}})
\end{equation}
where $N_{\text{PE=0}}$ is the number of events without $^{68}{\rm Ge}$ photon hits in the channel, and $N_{\text{Event}}$ is the number of events. Since the $^{68}{\rm Ge}$ source is deployed at the center of the liquid scintillator and the light emission from the liquid scintillator is isotropic, the average number of incident photons on the surface of each channel is the same. According to Equation \ref{eq3.2}, the relative PDE of each channel corresponds to the relative $N_{Det}$ value.

However, the dark noise effect can also generate hits. If the DCR varies across different readout channels, the number of hits contributed by dark noise will differ between channels, which in turn leads to variations in $N_{\text{PE=0}}$, thereby introducing a bias in the calibration results of the relative PDE. For this reason, a dedicated correction for the dark noise effect is required.

The hit time distribution is shown in Figure \ref{pic3.1}. After applying the time offset correction, we count hits from 0 ns to 160 ns, which are purely contributed by dark noise. Then, also count the hits from 240 ns to 400 ns, which include contributions from both dark noise and $^{68}{\rm Ge}$. The probability of a dark noise occurring within the 160 ns time window, denoted as $P_{\rm DN}$, is defined as the ratio of the number of events containing dark noise $N_{\rm DN}$  to the $N_{\text{Event}}$. The probability of observing either dark noise or $^{68}{\rm Ge}$ (or both), denoted as $P_{\rm Ge, DN}$, is defined as the ratio of the number of events containing dark noise, or $^{68}{\rm Ge}$, or both occurring $N_{\rm Ge, DN}$ to the $N_{\text{Event}}$. 

\begin{equation}
\label{eq3.3.1.1}
\begin{split}
P_{\rm DN} &= \frac{N_{\rm DN}}{N_{\text{Event}}} \\
P_{\rm Ge, DN} &= \frac{N_{\rm Ge, DN}}{N_{\text{Event}}} \\
\end{split}
\end{equation}

Dark noise and $^{68}{\rm Ge}$ are statistically independent. Therefore, the probability of the union of these two events $P_{\rm Ge, DN}$ equals the sum of the probabilities of three mutually exclusive scenarios: dark noise occurs without $^{68}{\rm Ge}$, $^{68}{\rm Ge}$ occurs without dark noise, and both dark noise and $^{68}{\rm Ge}$ occur simultaneously. 
\begin{equation}
\label{eq3.3.1.2}
\begin{split}
P_{\rm Ge, DN} =  P_{\rm DN} \cdot (1 - P_{\rm Ge}) + P_{\rm Ge} \cdot (1 - P_{\rm DN}) + P_{\rm Ge} \cdot P_{\rm DN} 
\end{split}
\end{equation}
Meanwhile, the probability of no $^{68}{\rm Ge}$ occurring within the 240 ns to 400 ns time window, denoted as $1-P_{\rm Ge}$, is defined as the ratio of the $N_{\text{PE=0}}$ to the $N_{\text{Event}}$. 
\begin{equation}
\label{eq3.3.1.3}
\begin{split}
 N_{\text{PE=0}} &=  (1 - P_{\rm Ge}) \cdot N_{\text{Event}}
\end{split}
\end{equation}
where $P_{\rm Ge}$ is derived from Equation \ref{eq3.3.1.1} and Equation \ref{eq3.3.1.2}, and is given by:
\begin{equation}
\label{eq3.3.1.4}
\begin{split}
 P_{\rm Ge} &= \frac{P_{\rm Ge, DN} - P_{\rm DN}}{1 - P_{\rm DN}}
\end{split}
\end{equation}

\begin{figure}[htbp]
\centering
\includegraphics[width=0.577\textwidth]{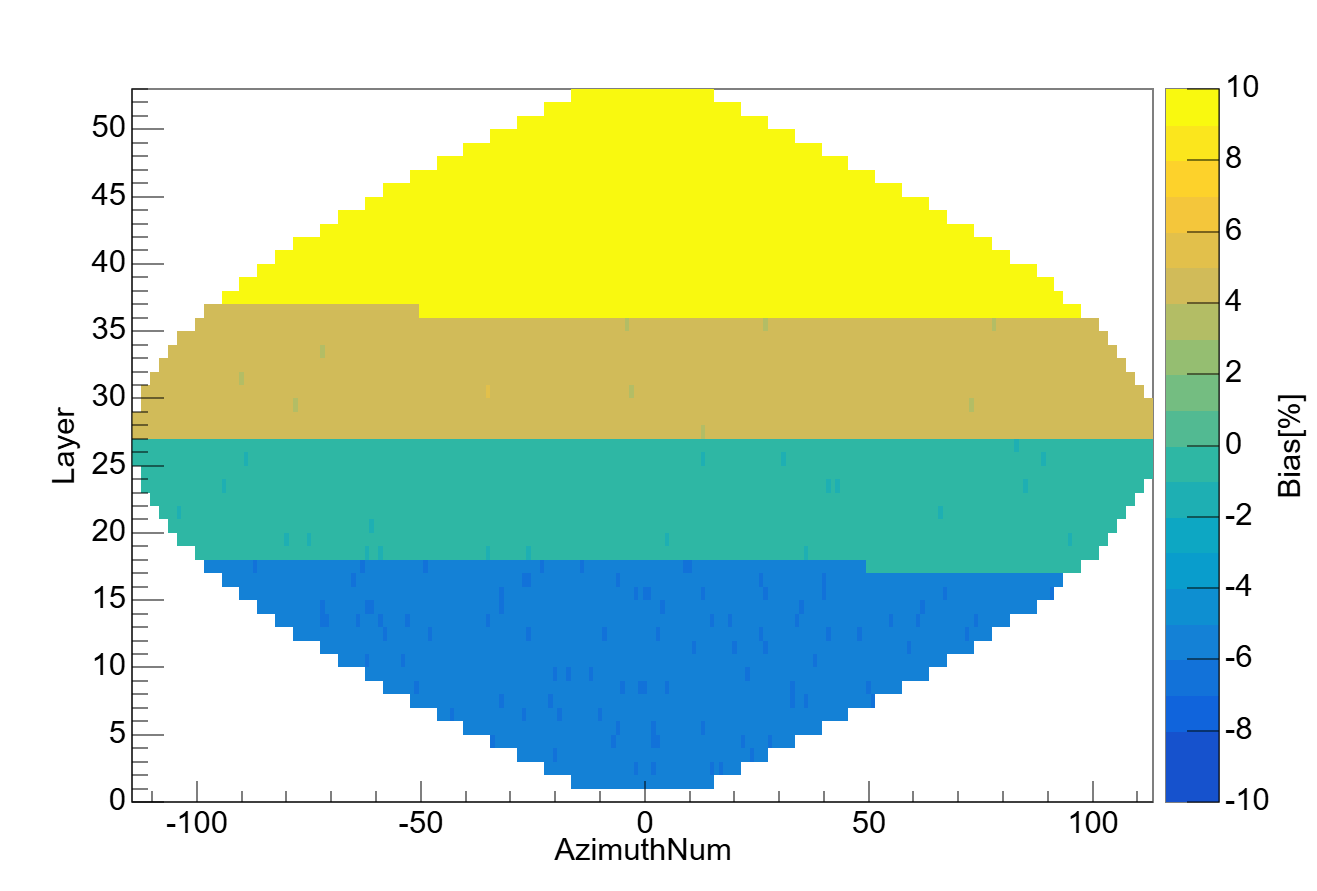}
\\
\includegraphics[width=0.577\textwidth]{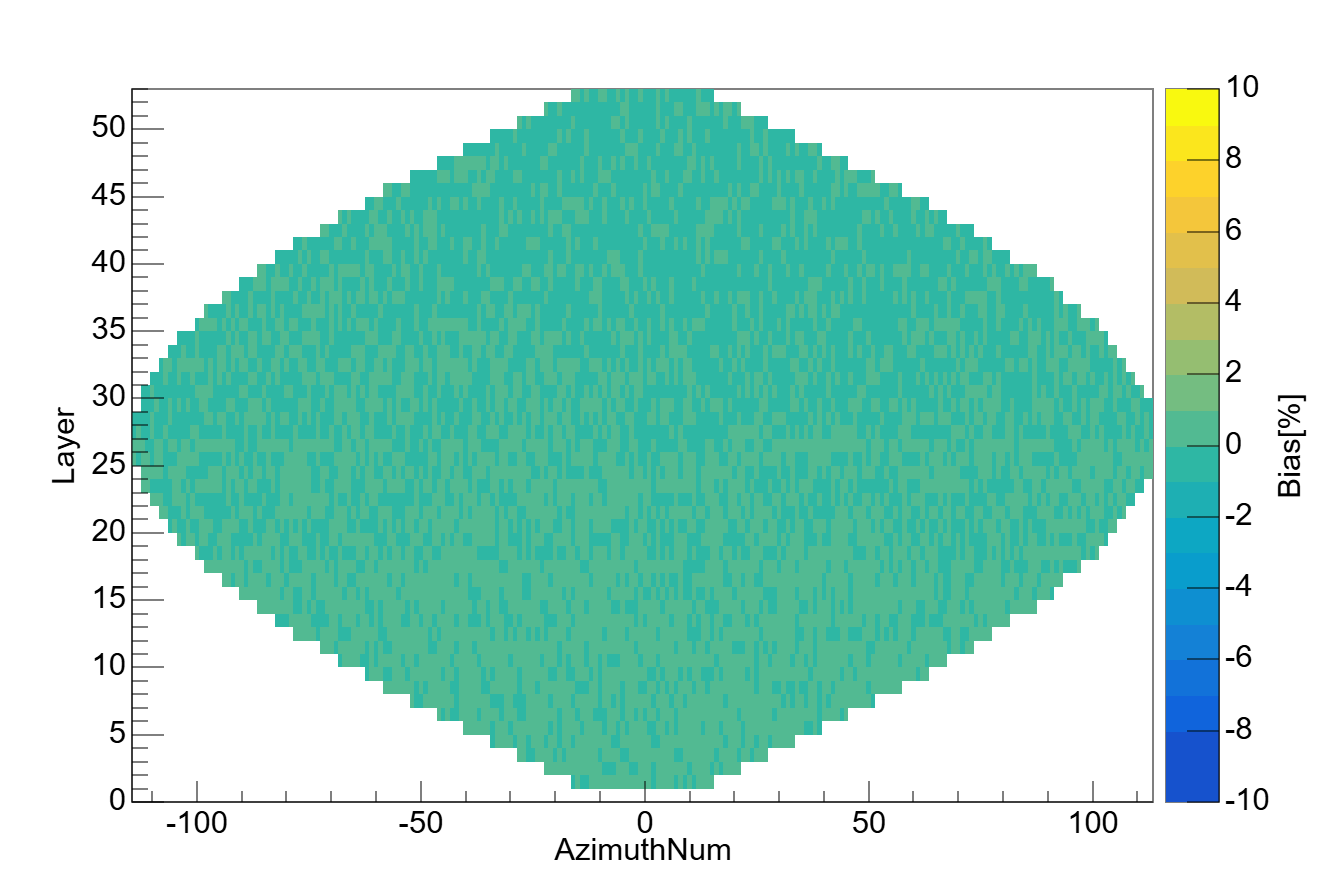}
\\
\includegraphics[width=0.577\textwidth]{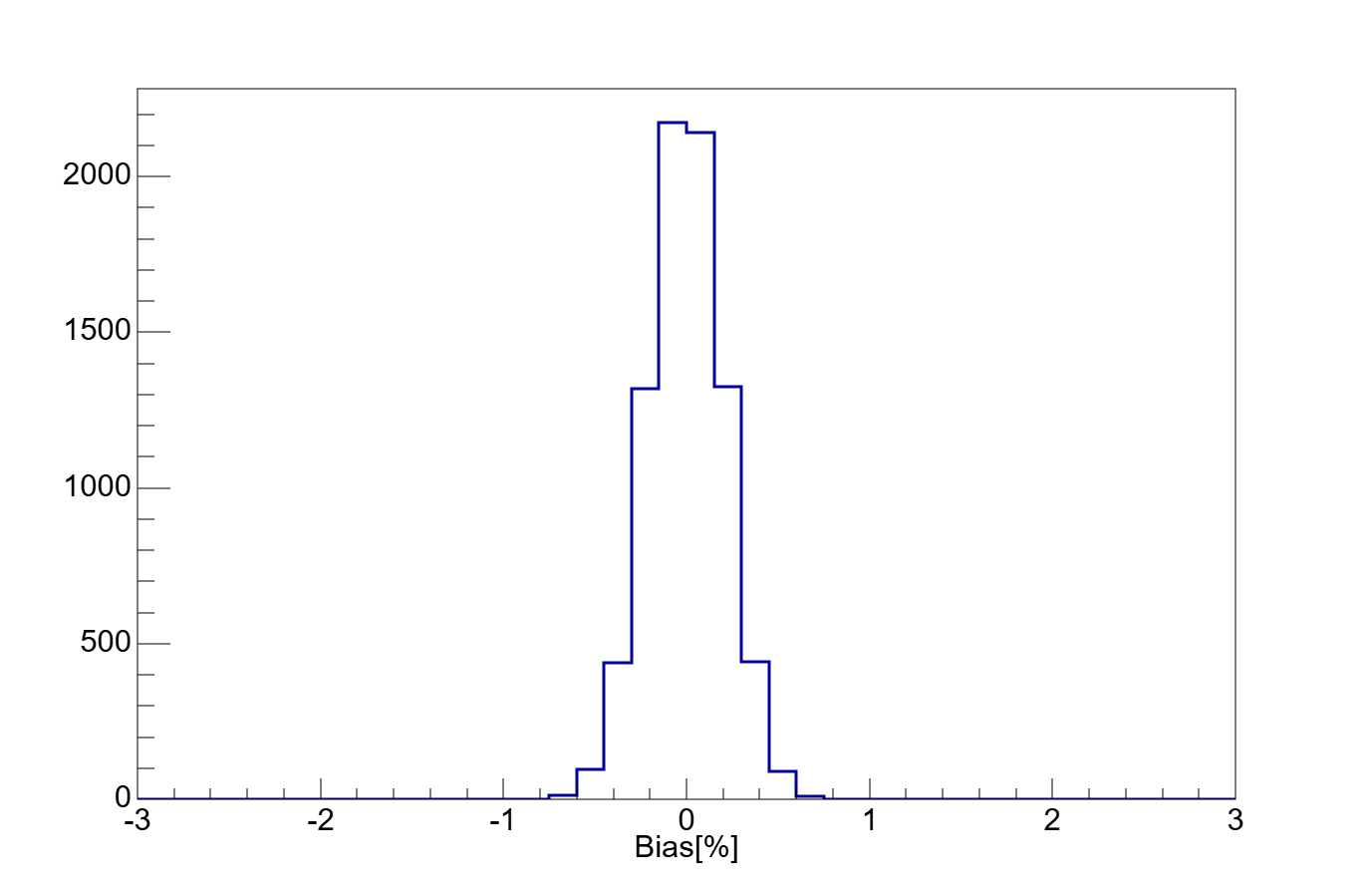}
\caption{Relative PDE calibration results before and after DCR correction with the SiPM tile surface optical reflection effect disabled. The upper panel: comparison between the calibration results and the simulation truth in each channel before DCR correction with the SiPM tile surface optical reflection effect disabled. The middle panel: comparison between the calibration results and the simulation truth in each channel after DCR correction with the SiPM tile surface optical reflection effect disabled. Lower panel: 1D histogram filled with each bin value from the middle 2D histogram.}
\label{pic3.4.1}
\end{figure}

In this simulation, we set four different DCR values: 20, 40, 60, and 80 Hz/\si{\mm\squared}. The results before and after applying this correction method are shown in Figure \ref{pic3.4.1}. It can be observed that prior to DCR correction, channels with different DCR introduce bias in the inter-channel relative PDE calibration results. After correcting for the dark noise effect using Equation \ref{eq3.3.1.3}, this bias is effectively eliminated. After DCR correction, the bias is 0.011\% and the standard deviation is 0.17\%.

The relative PDE calibration results presented earlier in this section were obtained with the SiPM tile surface optical reflection effect disabled. The SiPM tile coverage of the TAO detector is not 100\%: specifically, there is no SiPM tile coverage at the north and south poles, as these regions are reserved for liquid scintillator filling and the deployment of the calibration source. As a result, the light field is no longer uniform when the SiPM tile surface optical reflection effect is taken into account.

\begin{figure}[htbp]
\centering
\includegraphics[width=0.45\textwidth]{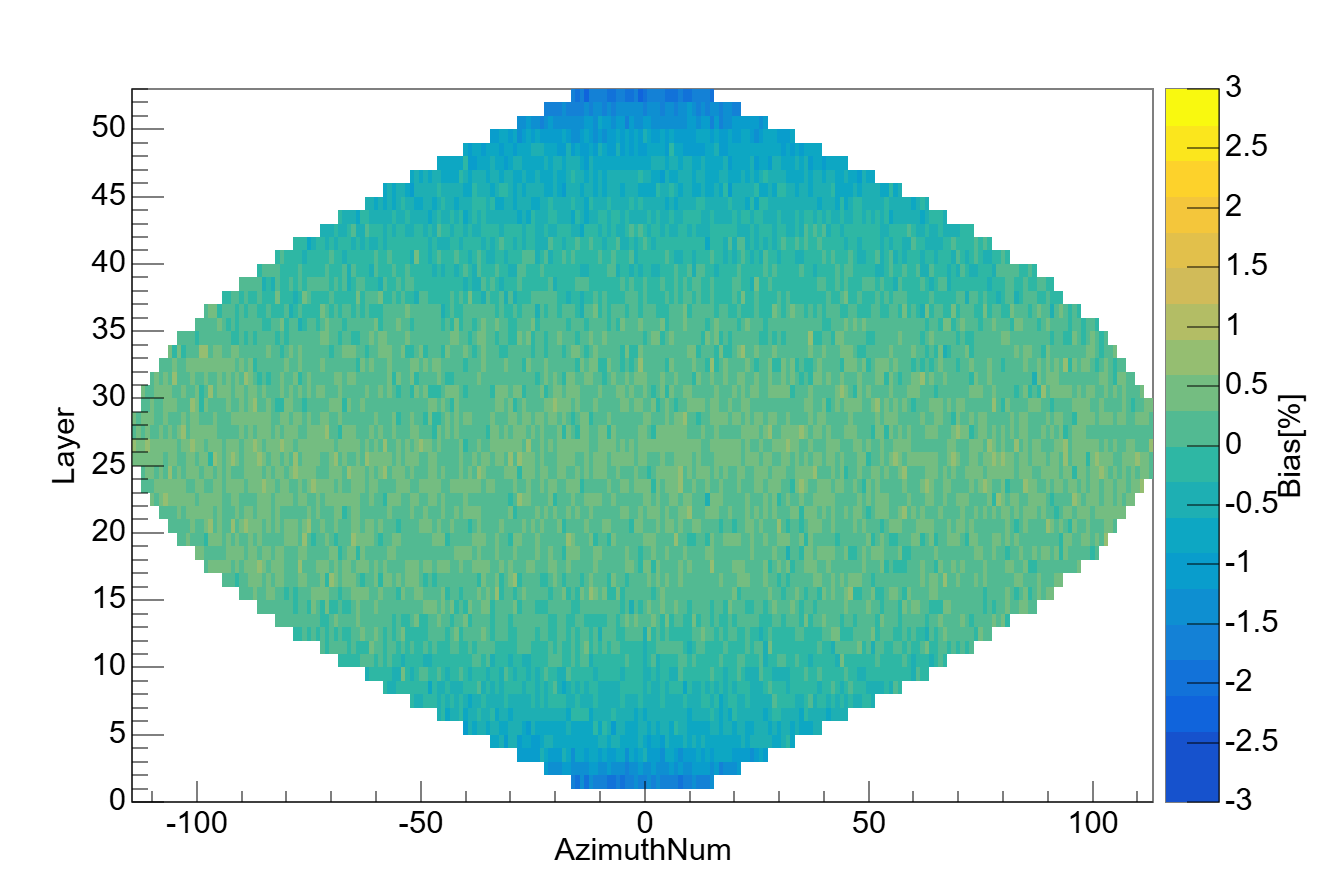}
\qquad
\includegraphics[width=0.45\textwidth]{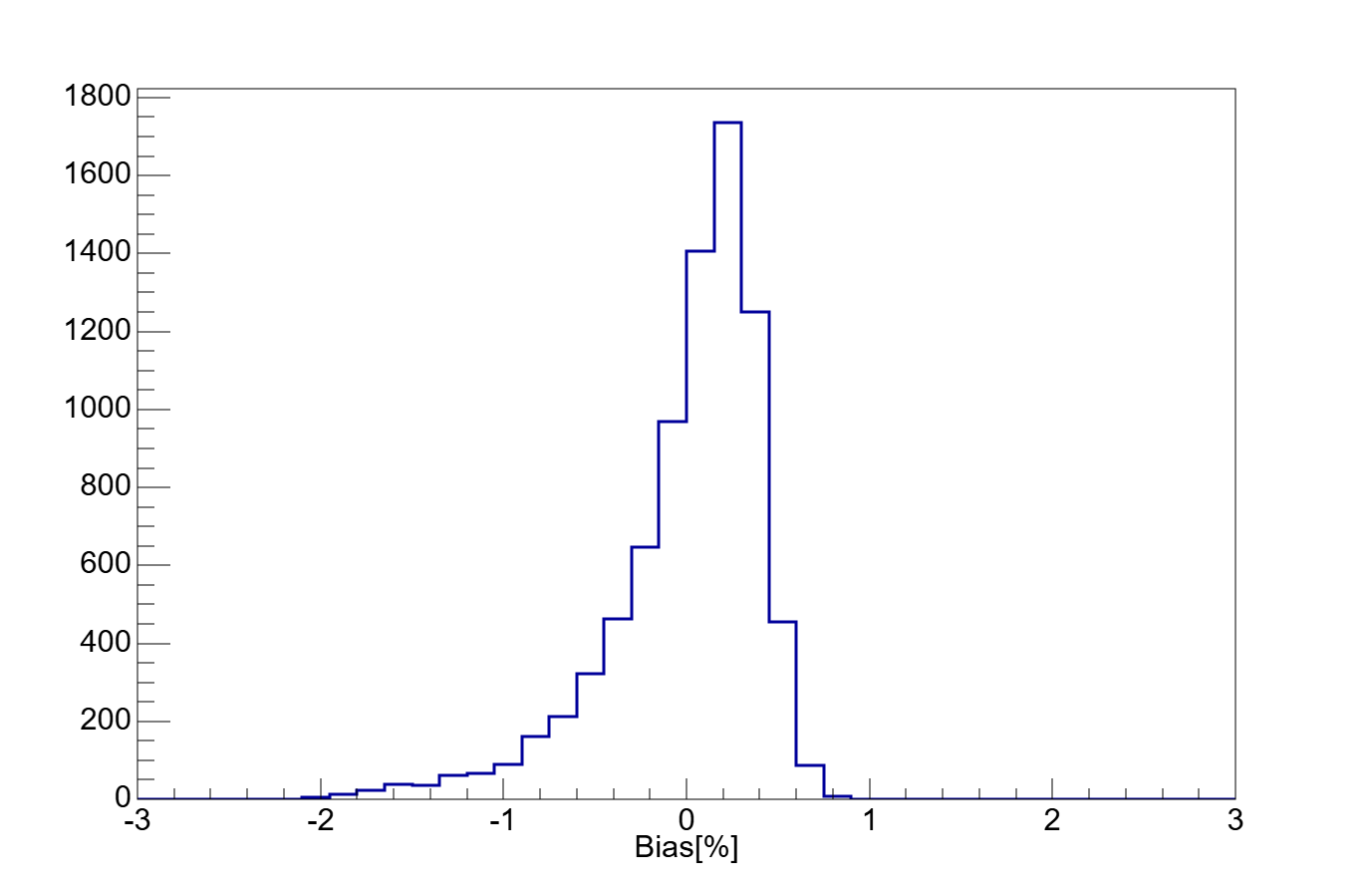}
\caption{Relative PDE calibration results with the SiPM tile surface optical reflection effect enabled. The left panel: comparison between the calibration results and the simulation truth in each channel with the SiPM tile surface optical reflection effect enabled in simulation. The right panel: 1D histogram filled with each bin value from the left 2D histogram.}
\label{pic3.4}
\end{figure}

Figure \ref{pic3.4} shows the relative PDE calibration results after DCR correction, obtained from simulations with the SiPM tile surface optical reflection effect enabled. It can be observed that bias exists in the calibrated relative PDE for channels in different layers, with an approximately 3\% bias between the pole layer and the equatorial layer.

Currently, this work does not provide a solution to the bias in relative PDE calibration caused by the degradation of light field uniformity induced by SiPM tile surface optical reflection effect. This bias will be corrected based on the results of the detector optical model and simulation in the future.

\section{Calibration Based on Charge Information}
\label{4}

\subsection{Gain}
\label{4.1}

\begin{figure}[htbp]
\centering
\includegraphics[width=.7\textwidth]{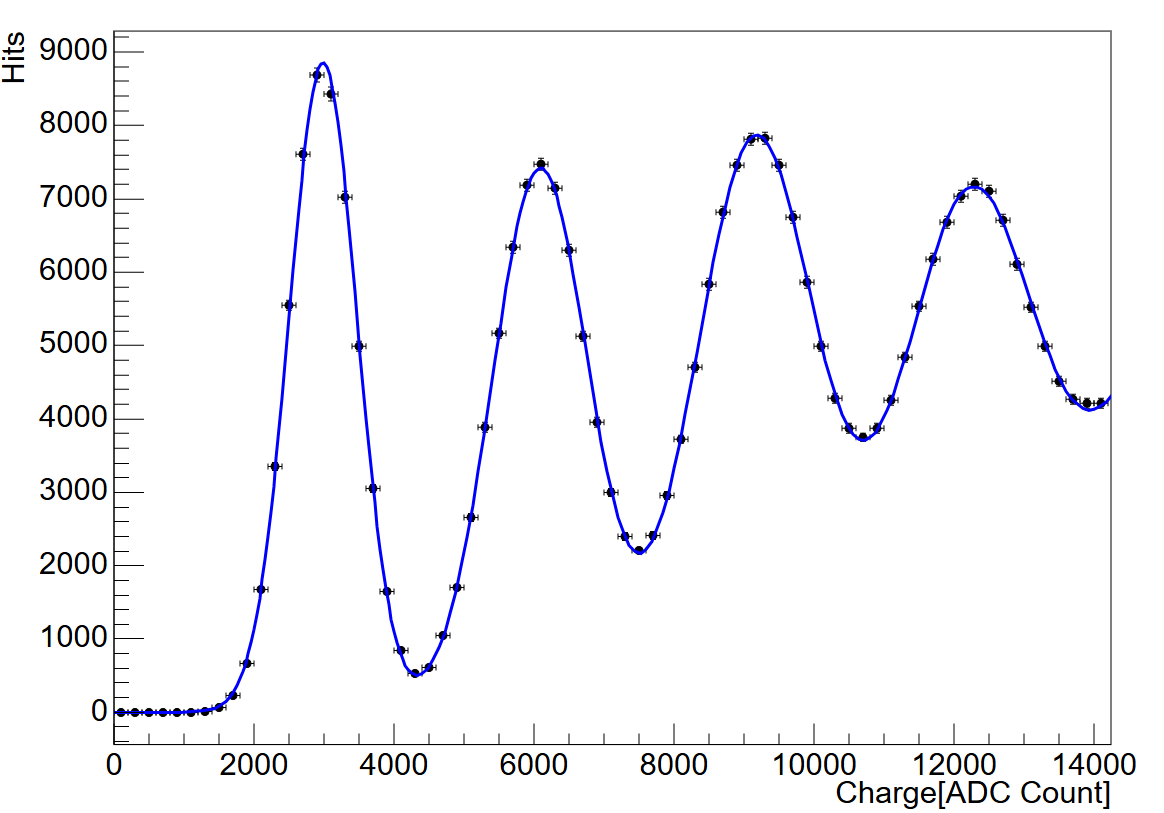}
\caption{Multi-Gaussian function fit to charge distribution. The horizontal axis is measured in ADC counts, where 1.8 volts correspond to $2^{16}$ ADC counts. Each peak represents the charge response for a specific number of PE: the first peak corresponds to 1PE, the second to 2 PEs, and so on. The separation between adjacent PE peaks indicates the gain.}
\label{pic4.1}
\end{figure}

The charge generated by a single PE waveform follows a Gaussian distribution, and the individual photons detected by the APDs in the SiPM chip are independent. Consequently, the total charge resulting from the simultaneous detection of multiple PEs also follows a Gaussian distribution, and the overall distribution is a multi-Gaussian distribution (Equation \ref{eq4.0} where f is the multi-Gaussian function, $N_{i}$ is the number of hits corresponding to each PE, $\mu_{i}$ is the mean value, and $\sigma_{i}$ is the standard deviation). Different Gaussian peaks correspond to the charge responses for different numbers of PEs. As shown in Figure \ref{pic4.1}, by fitting the charge spectrum with a multi-Gaussian function, the gain is determined from the separation between adjacent peaks.

\begin{equation}
\label{eq4.0}
\begin{split}
f = \sum_{i=1}^{n} N_{i} \cdot \text{Gaus}(\mu_{i},\sigma_{i})
\end{split}
\end{equation}

\begin{figure}[htbp]
\centering
\includegraphics[width=.45\textwidth]{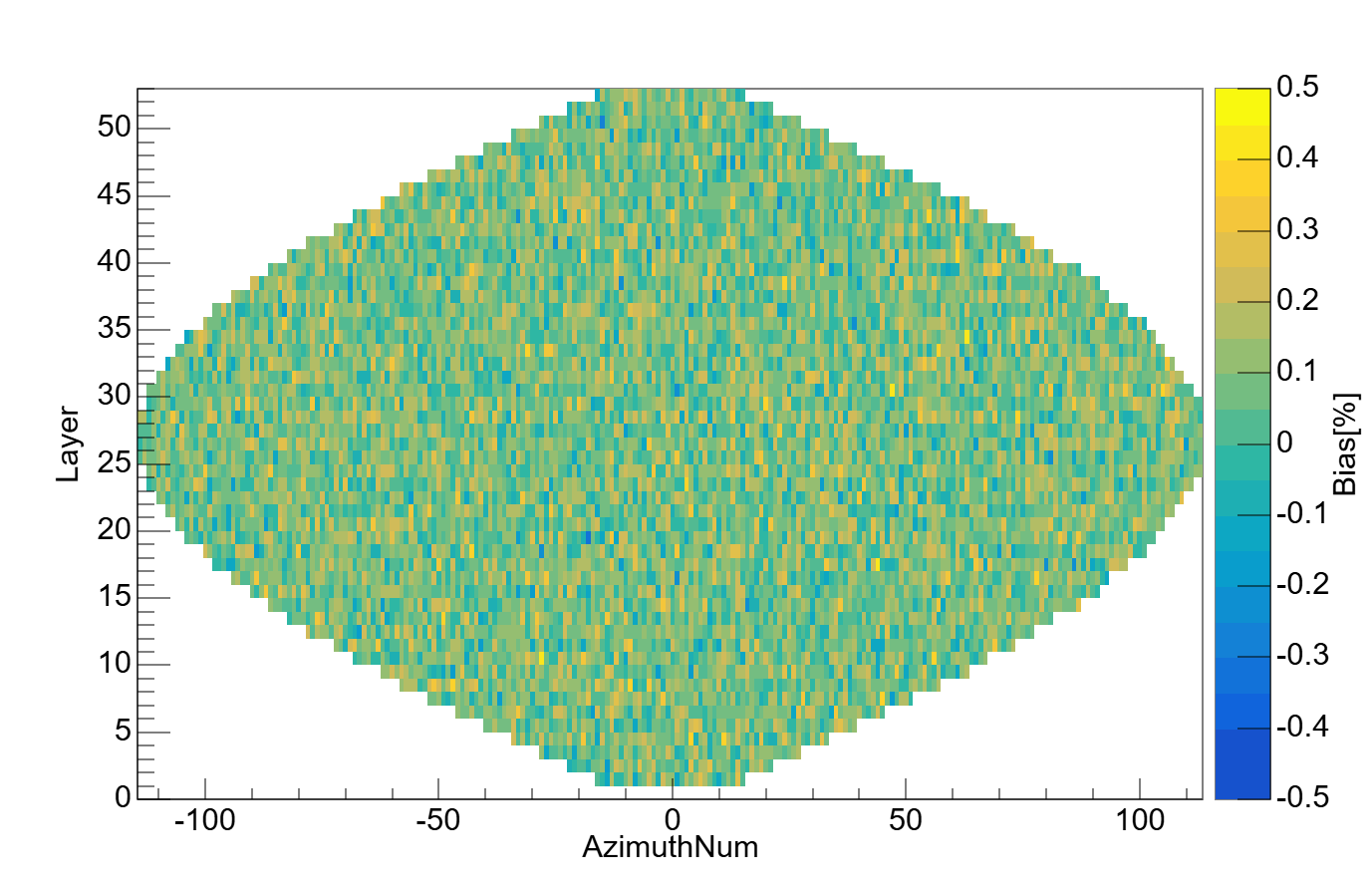}
\qquad
\includegraphics[width=.45\textwidth]{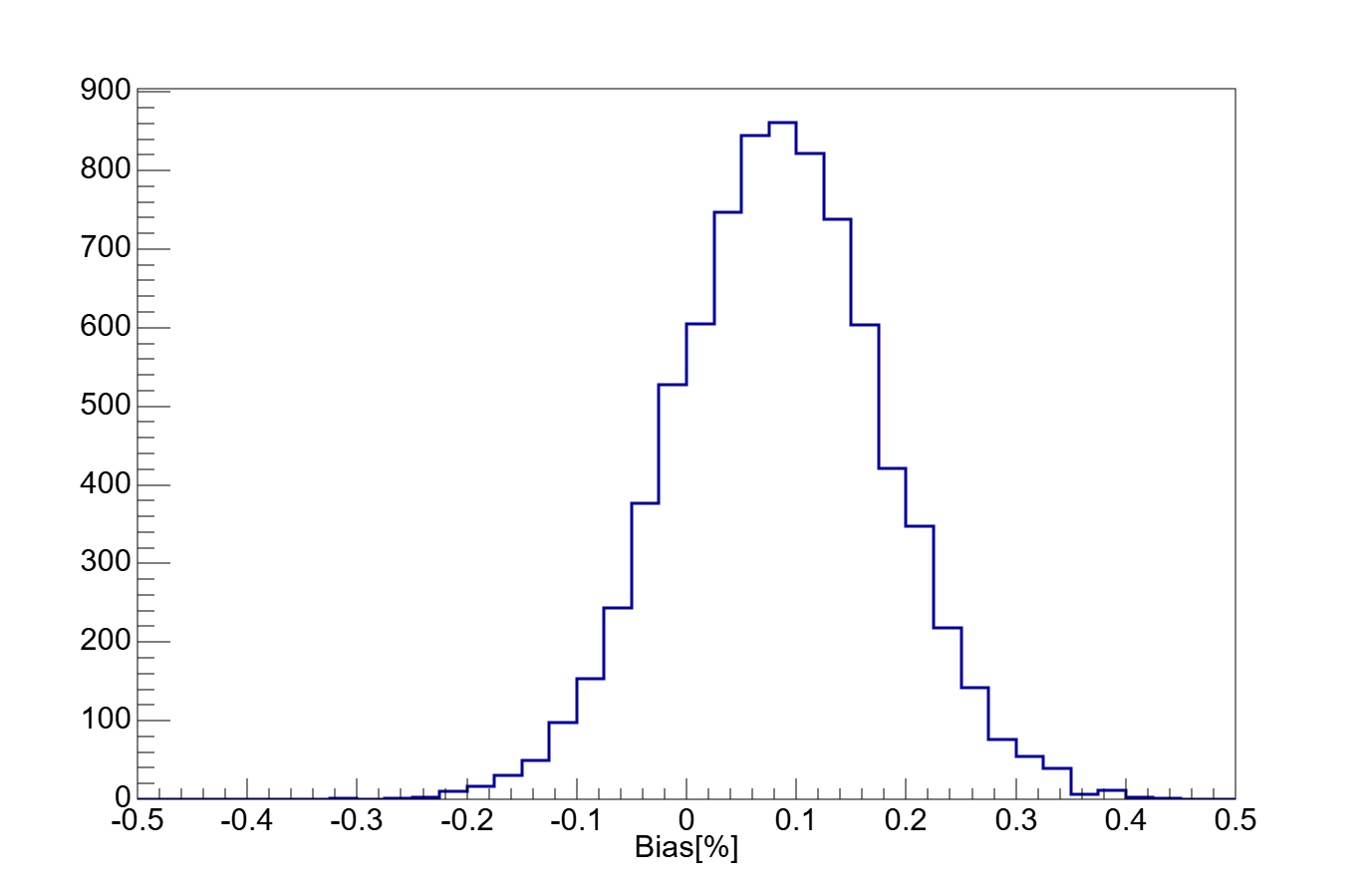}
\caption{Gain calibration results. The left panel: comparison between the calibration results and the simulation truth in each channel. The right panel: 1D histogram filled with each bin value from the left 2D histogram.}
\label{pic4.2}
\end{figure}

The gain calibration results are shown in Figure \ref{pic4.2}. The unit of gain is ADC counts, where 1.8 volts corresponds to $2^{16}$ ADC counts. The bias is 0.084\% and the standard deviation is 0.094\%.

\subsection{Internal Optical Crosstalk}
\label{4.2}

There are two methods for calibrating the IOCT in the TAO CD: multiple PEs hit analysis and generalized Poisson fitting. For multiple PEs hit analysis, the calibrated quantity $P_{\rm IOCT}$ represents the probability that a single electrical pulse generates IOCT pulses. In generalized Poisson fitting, the calibrated quantity $\lambda_{\rm IOCT}$ is the Poisson parameter of the IOCT. According to the mathematical properties of the Generalized Poisson distribution \cite{Ref31}, the conversion relationship between these two quantities is given by Equation \ref{eq4.2.1}.

\begin{equation}
\label{eq4.2.1}
P_{\rm IOCT} = 1-e^{-\lambda_{\rm IOCT}}
\end{equation}

\subsubsection{Multiple Photoelectrons Hit Analysis}
\label{4.2.1}

\begin{figure}[htbp]
\centering
\includegraphics[width=.75\textwidth]{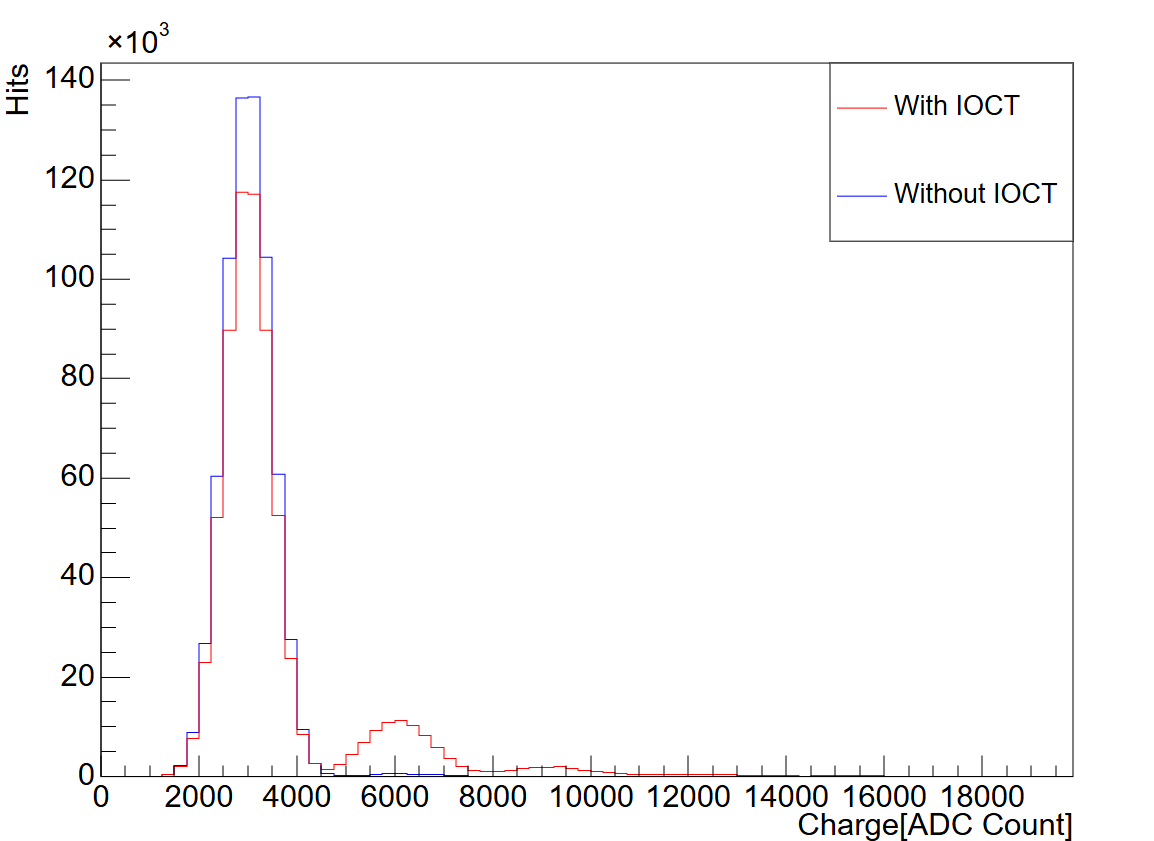}
\caption{Charge spectrum distribution with and without IOCT. The red line represents a case with IOCT ($\lambda_{\rm IOCT}$ = 0.15), and the blue line represents a case without IOCT. It can be observed that the multiple PEs component is mainly contributed by IOCT.}
\label{pic4.3}
\end{figure}

Hits occurring before the trigger time are generated by dark noise pulses and EOCT pulses induced by dark noise, which are referred to as equivalent dark noise in this paper. These equivalent dark noise pulses can induce IOCT.  When IOCT is induced by these pulses, a multiple PEs hit composed of both the equivalent dark noise and IOCT signals is invariably generated. In contrast, the probability of forming a multiple PEs hit from equivalent dark noise alone is minimal. As shown in Figure \ref{pic4.3}, it can be observed that multiple PEs hits are primarily contributed by the IOCT effect.

Therefore, the multiple PEs region of the charge spectrum acquired in a dark environment is dominated by IOCT. Hence, the IOCT rate can be quantified using Equation \ref{eq4.2}, where $P_{\rm IOCT}$ is the probability of IOCT occurrence, $N_{\rm multPE}$ is the number of multiple PEs hits and $N_{\rm total}$ is the total number of hits.

\begin{equation}
\label{eq4.2}
P_{\rm IOCT} = \frac{N_{\rm multPE}}{N_{\rm total}}
\end{equation}

\begin{figure}[htbp]
\centering
\includegraphics[width=.45\textwidth]{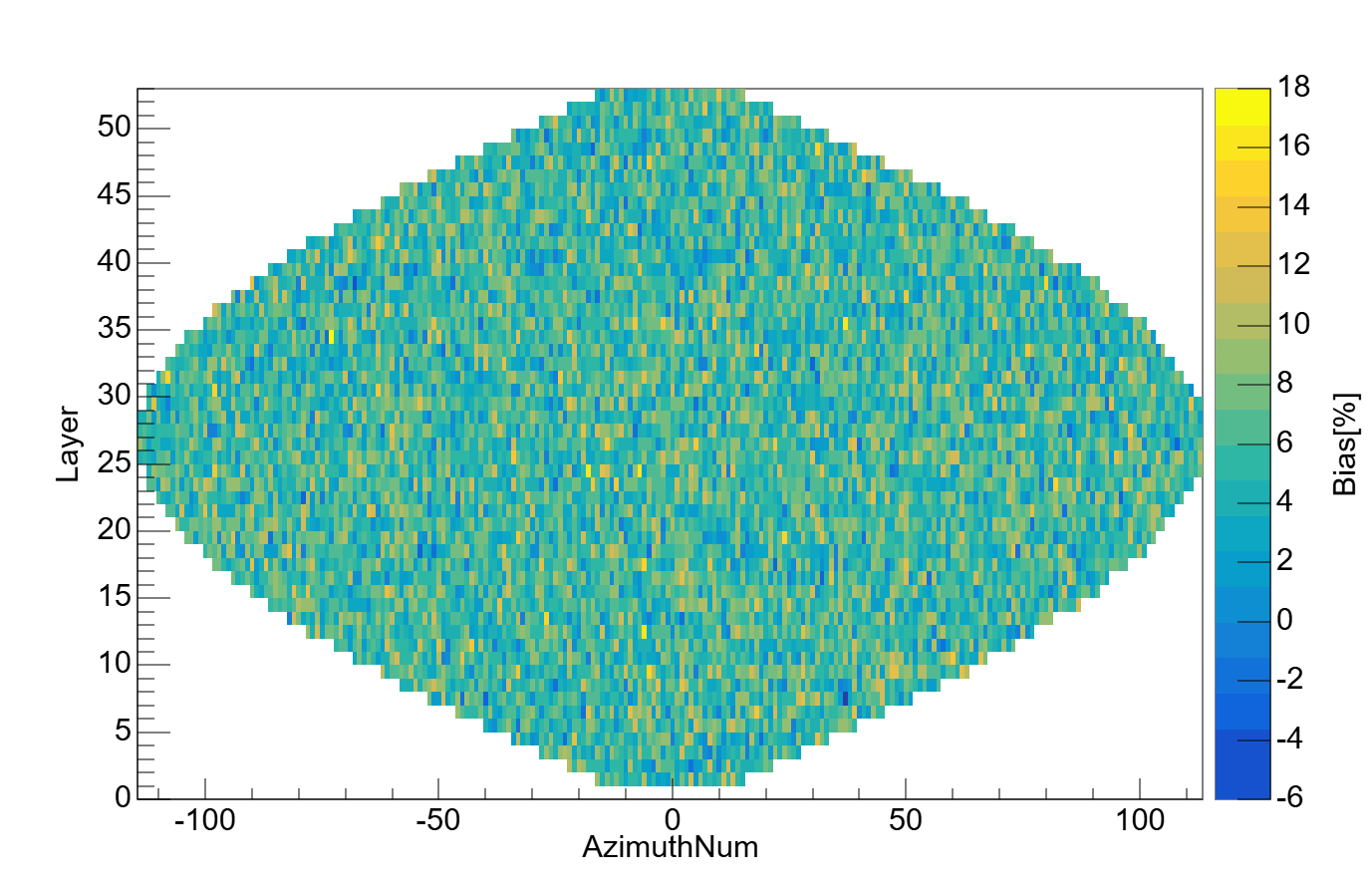}
\qquad
\includegraphics[width=.45\textwidth]{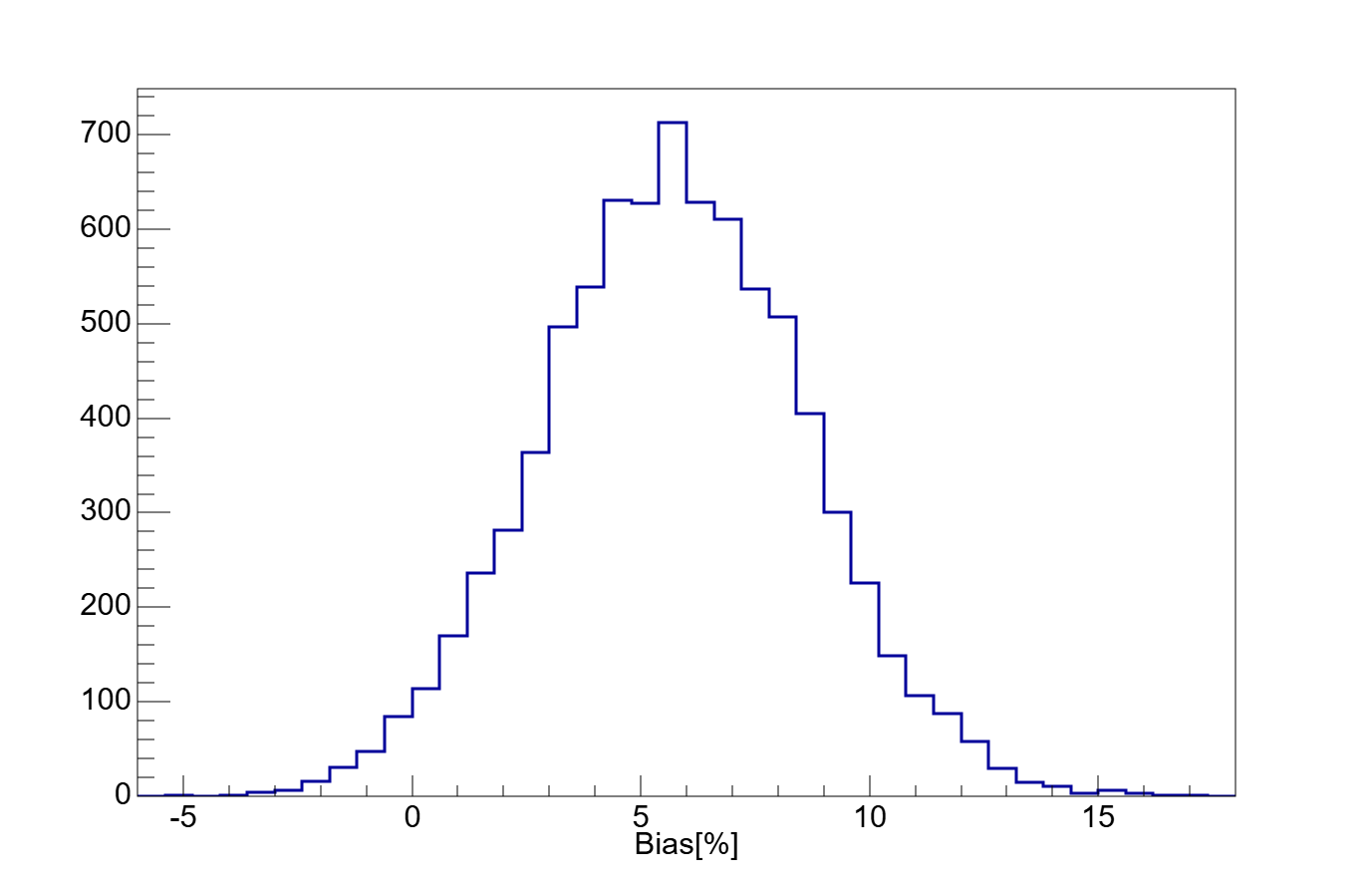}
\caption{IOCT calibration results based on multiple PEs hit analysis. The left panel: comparison between the calibration results and the simulation truth in each channel. The right panel: 1D histogram filled with each bin value from the left 2D histogram.}
\label{pic4.4}
\end{figure}

The IOCT calibration results are shown in Figure \ref{pic4.4}; the bias is 5.70\% and the standard deviation is 2.73\%. The bias results from multiple PEs contributions caused by effects other than IOCT. By enabling and disabling various SiPM and electronics effects in the TAO simulation, we found that the multiple PEs hits excluding those from IOCT are dominated by two components: the multiple PEs hits generated by the equivalent dark noise itself, and contributions from the AP. 

For the multiple PEs equivalent dark noise component, simulation samples with the same statistical size are generated in the TAO simulation with both IOCT and AP effects disabled. We find that the multiple PEs contribution from equivalent dark noise accounts for a bias of 4.57\%. For the AP component, samples of identical statistical size are produced with the IOCT effect disabled and only 1-PE equivalent dark noise included in the TAO simulation. We find that the AP effect accounts for a bias of 1.42\%. (The discrepancy between the sum of the biases from the AP effect and multiple PEs equivalent dark noise and the total IOCT calibration bias arises because the three values are derived from three independent statistical sample sets. Statistical fluctuations across these separate datasets lead to a mismatch between the sum of the individual bias components and the total IOCT calibration bias.) It is clear that the IOCT calibration bias is dominated by the contribution of multiple PEs equivalent dark noise, which necessitates a dedicated correction scheme to mitigate and eliminate this bias.

\subsubsection{Generalized Poisson Fitting}
\label{4.2.2}

To eliminate the influence of multiple PEs equivalent dark noise on the IOCT calibration, the charge spectrum can be fitted using a generalized Poisson distribution. The total distribution of the number of pulses from the $\mu$ Poisson primary pulses and IOCT follows the generalized Poisson distribution. Since IOCT pulses have no delay relative to the primary pulses, the distribution of the number of pulses corresponds to the number of PEs.

\begin{figure}[htbp]
\centering
\includegraphics[width=.75\textwidth]{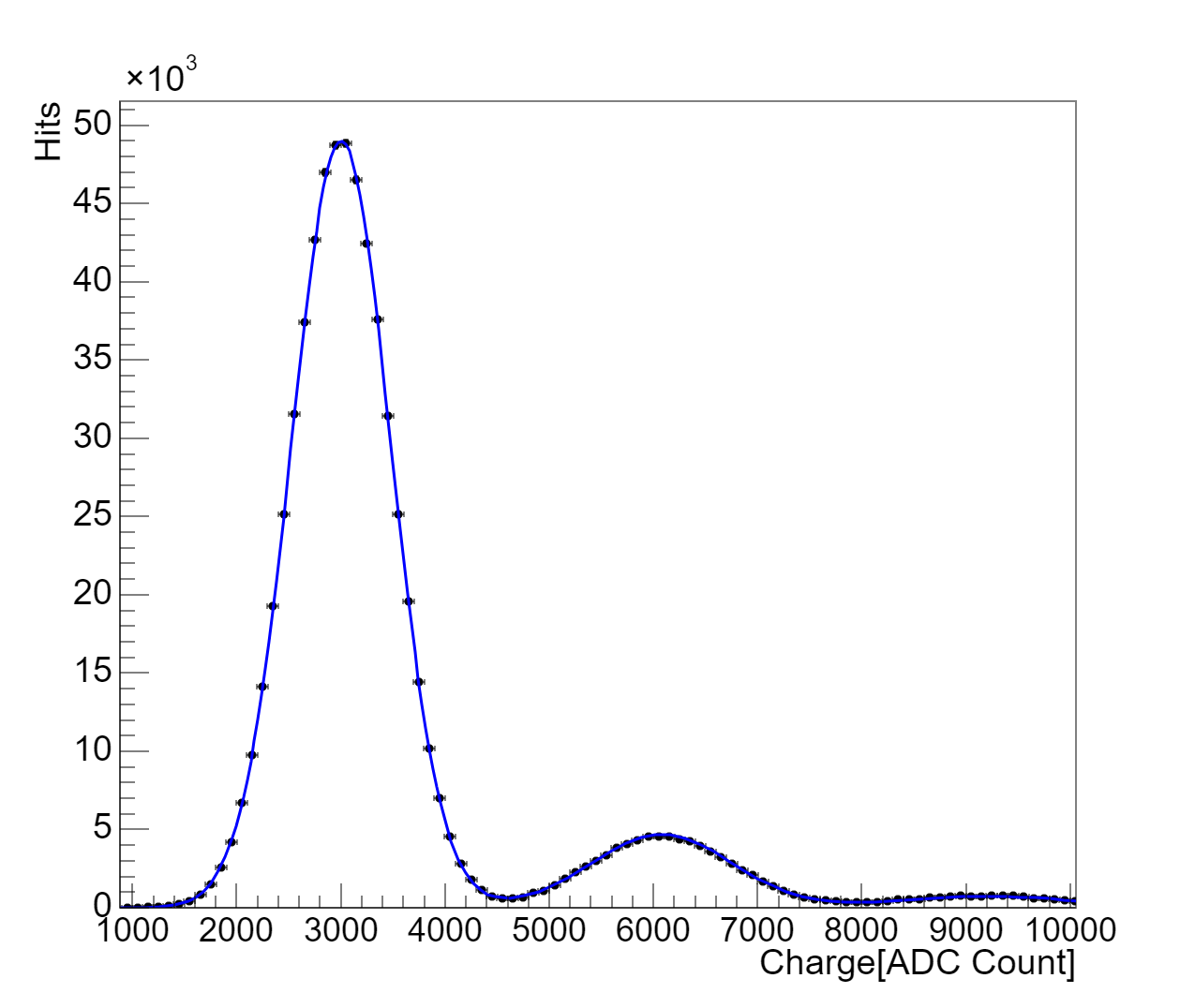}
\caption{Generalized Poisson distribution fit to the charge spectrum.}
\label{pic4.5.1}
\end{figure}

For the equivalent dark noise signal, $\mu = \text{DCR}^{*} \cdot t_{W} \cdot S $, where $\text{DCR}^{*}$ is equivalent DCR, $t_{W}$ is the average width of a single PE waveform, and S is the sensitive area of the SiPM in one channel. By replacing N in Equation \ref{eq4.0} with GP in Equation \ref{eq4.4} and fitting this function to the charge spectrum (as shown in Figure \ref{pic4.5.1}), the fitted value of $\lambda_{\rm IOCT}$ corresponds to the calibration results of IOCT.

\begin{figure}[htbp]
\centering
\includegraphics[width=.45\textwidth]{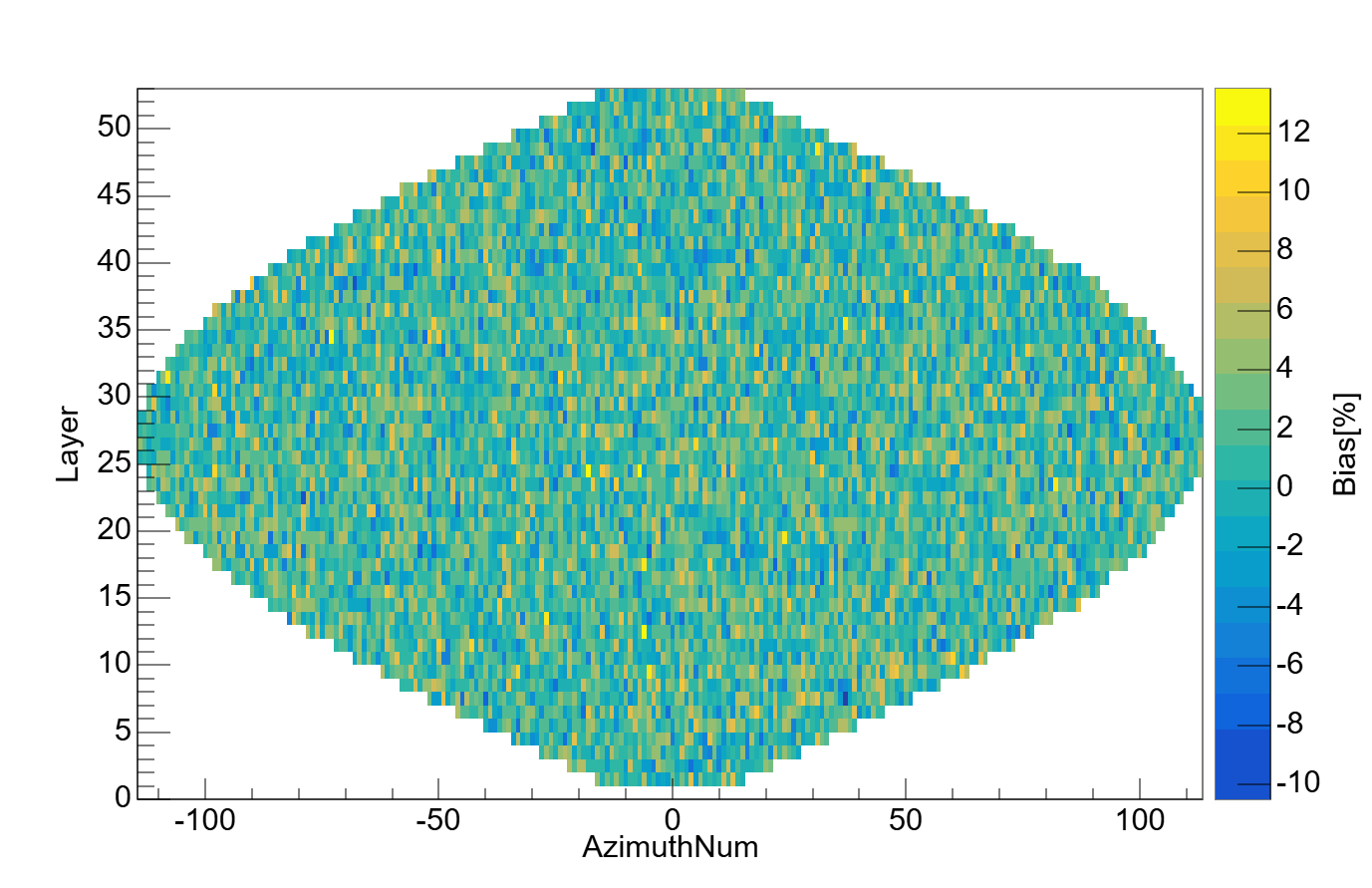}
\qquad
\includegraphics[width=.45\textwidth]{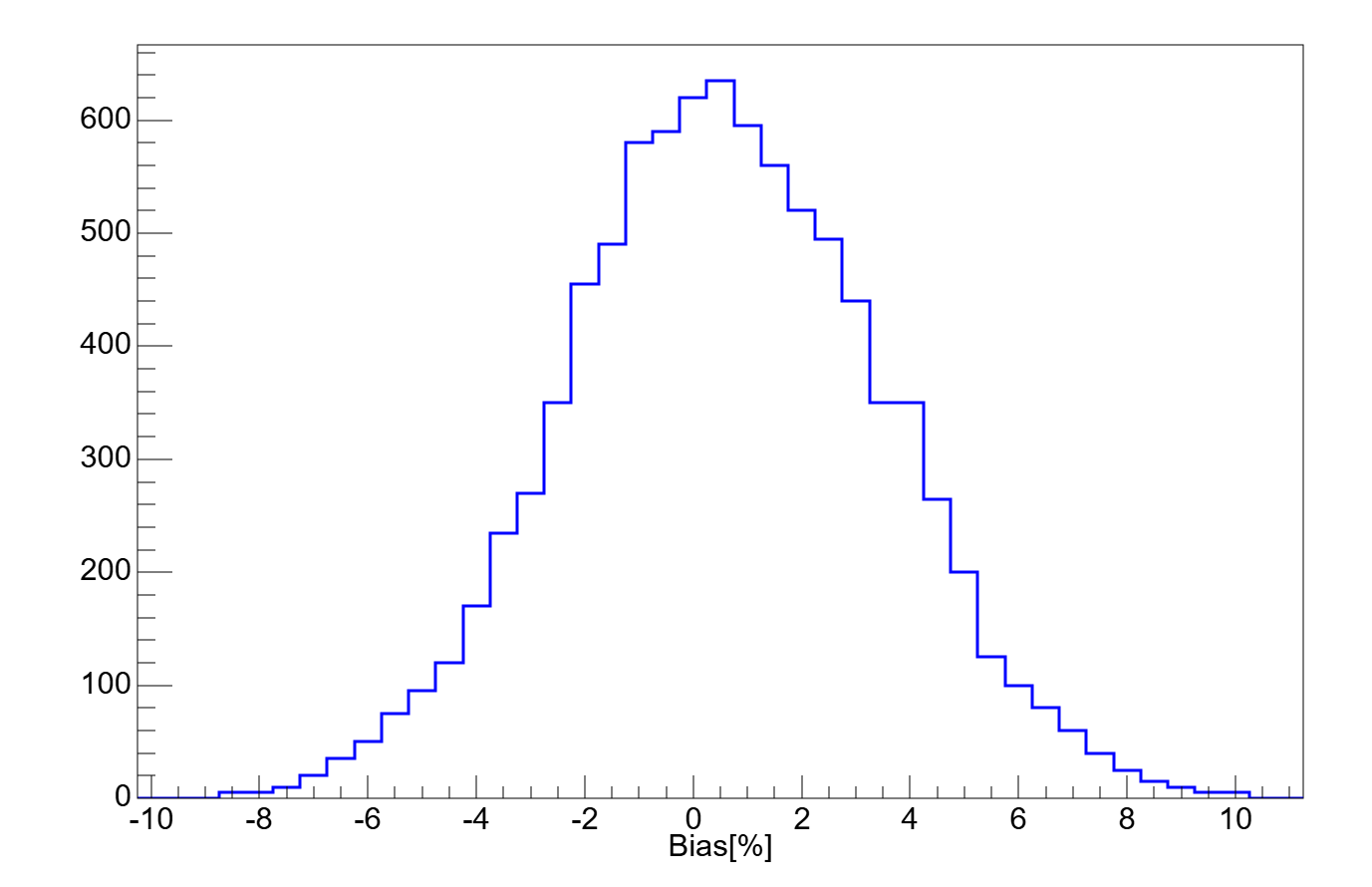}
\caption{IOCT calibration results based on generalized Poisson fitting. The left panel: comparison between the calibration results and the simulation truth in each channel. The right panel: 1D histogram filled with each bin value from the left 2D histogram.}
\label{pic4.5}
\end{figure}

The IOCT calibration results are shown in Figure \ref{pic4.5}; the bias is 1.40\% and the standard deviation is 3.02\%. These results indicate that the bias induced by multiple PEs equivalent dark noise has been effectively eliminated. This method achieves a significant reduction in the IOCT calibration bias without severe degradation of the standard deviation. The rise in standard deviation can be compensated for by increasing the sample size of the calibration dataset. Accordingly, the generalized Poisson fitting method serves as the preferred approach for SiPM IOCT calibration in the TAO detector.

\section{External Optical Crosstalk Calibration}
\label{5}

This paper presents a methodology for calibrating the EOCT rate and the distribution of EOCT emission angle in the TAO CD by switching the SiPM high voltage and using an LED external trigger to acquire data.

As shown in Equation \ref{eq5.1}, when all SiPM channels in the TAO CD are operational, the electrical pulses generated within a SiPM channel include PEs produced by the LED, dark noise, IOCT, AP, and EOCT from other SiPM channels. Therefore, the charge $Q_O$ consists of contributions from the LED photon signal ($Q_{\rm LED}$), dark noise ($Q_{\rm DN}$), IOCT ($Q_{\rm IOCT}$), AP ($Q_{\rm AP}$) and EOCT ($ Q_{\rm EOCT}$). When other SiPM channels are turned off, the charge $Q_C$ only includes LED, dark noise, IOCT, and AP components. Thus, when acquiring "turn on" and "turn off" data over the same time duration, the difference between $Q_O$ and $Q_C$ corresponds to the contribution of EOCT. Here, the EOCT ratio is defined as $Q_{\rm EOCT}$ divided by $Q_C$.

\begin{equation}
\label{eq5.1}
\begin{split}
Q_O &= Q_{\rm LED} + Q_{\rm DN} + Q_{\rm IOCT} + Q_{\rm AP} + Q_{\rm EOCT} \\
Q_C &= Q_{\rm LED} + Q_{\rm DN} + Q_{\rm IOCT} + Q_{\rm AP} \\
\text{Rate}_{\rm EOCT} &= \frac{Q_{\rm EOCT}}{Q_C} \\
&= \frac{Q_O - Q_C}{Q_C}
\end{split}
\end{equation}

In the TAO CD, the operating voltage of each SiPM channel is controlled via the high-voltage system. Specifically, a SiPM channel can be turned off by setting its voltage below the breakdown voltage. Since the number of active SiPM channels in the CD varies before and after SiPM channel switching, data cannot be acquired using the nhit self-trigger. Therefore, an LED external trigger system is employed. Since high voltage is applied to a single SiPM tile, the EOCT calibration in this study can only provide tile-level calibration results rather than channel-level results. (In TAO CD, each tile corresponds to two channels)

\subsection{External Optical Crosstalk Rate Calibration}
\label{5.1}

Three LED external trigger data sets were simulated to study the SiPM EOCT rate calibration. The first data set had all SiPM tiles turned on; the second had all SiPM tiles turned on but with the EOCT effect disabled in the simulation; the third had only one SiPM tile turned on. The EOCT rate calibration results can be derived from the first and third data sets, and the true EOCT rate in the simulation can be obtained from the first and second data sets. By comparing the calibration with the true value, the bias and standard deviation can be determined.

\begin{figure}[htbp]
\centering
\includegraphics[width=0.577\textwidth]{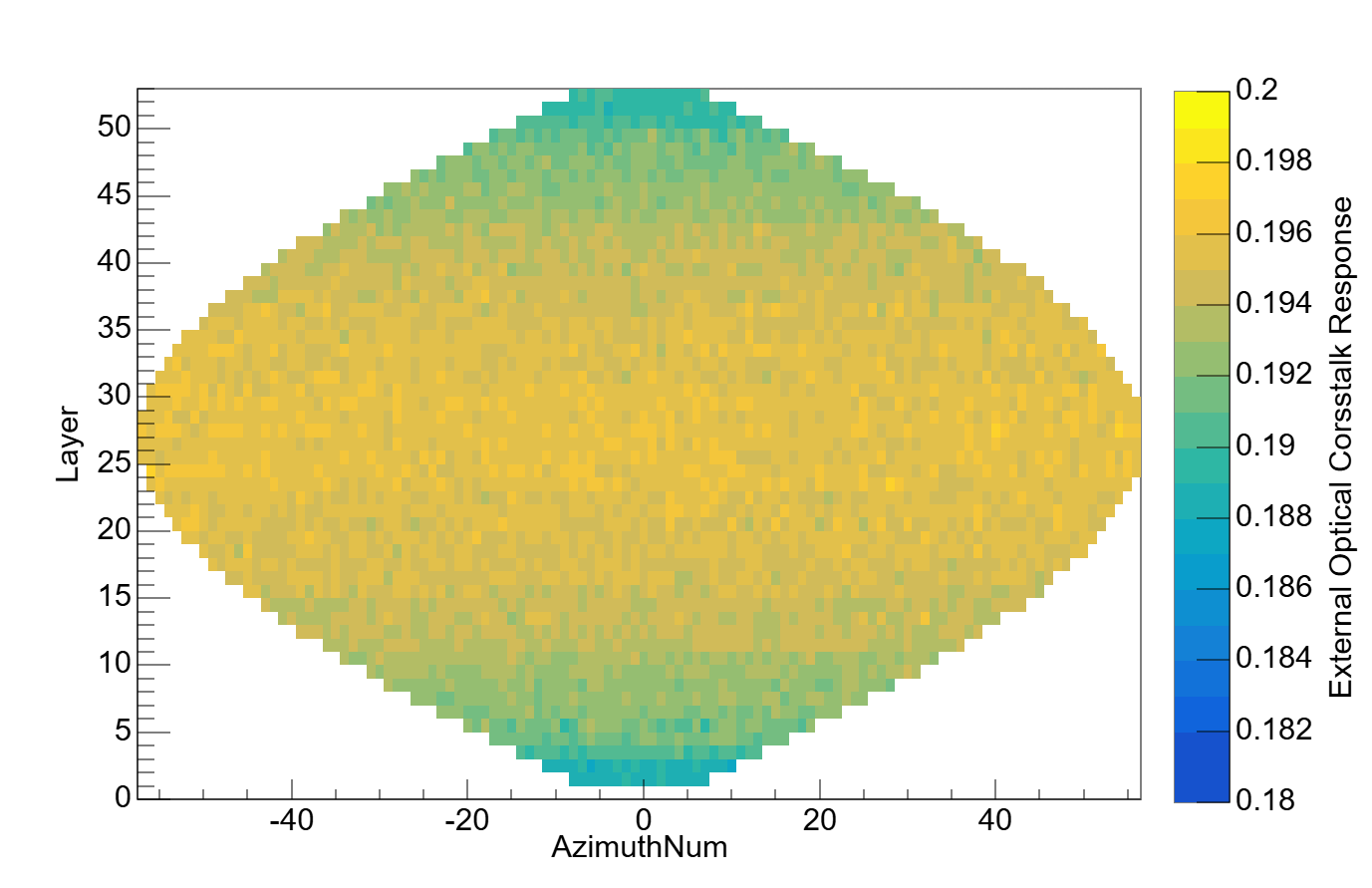}
\\
\includegraphics[width=0.577\textwidth]{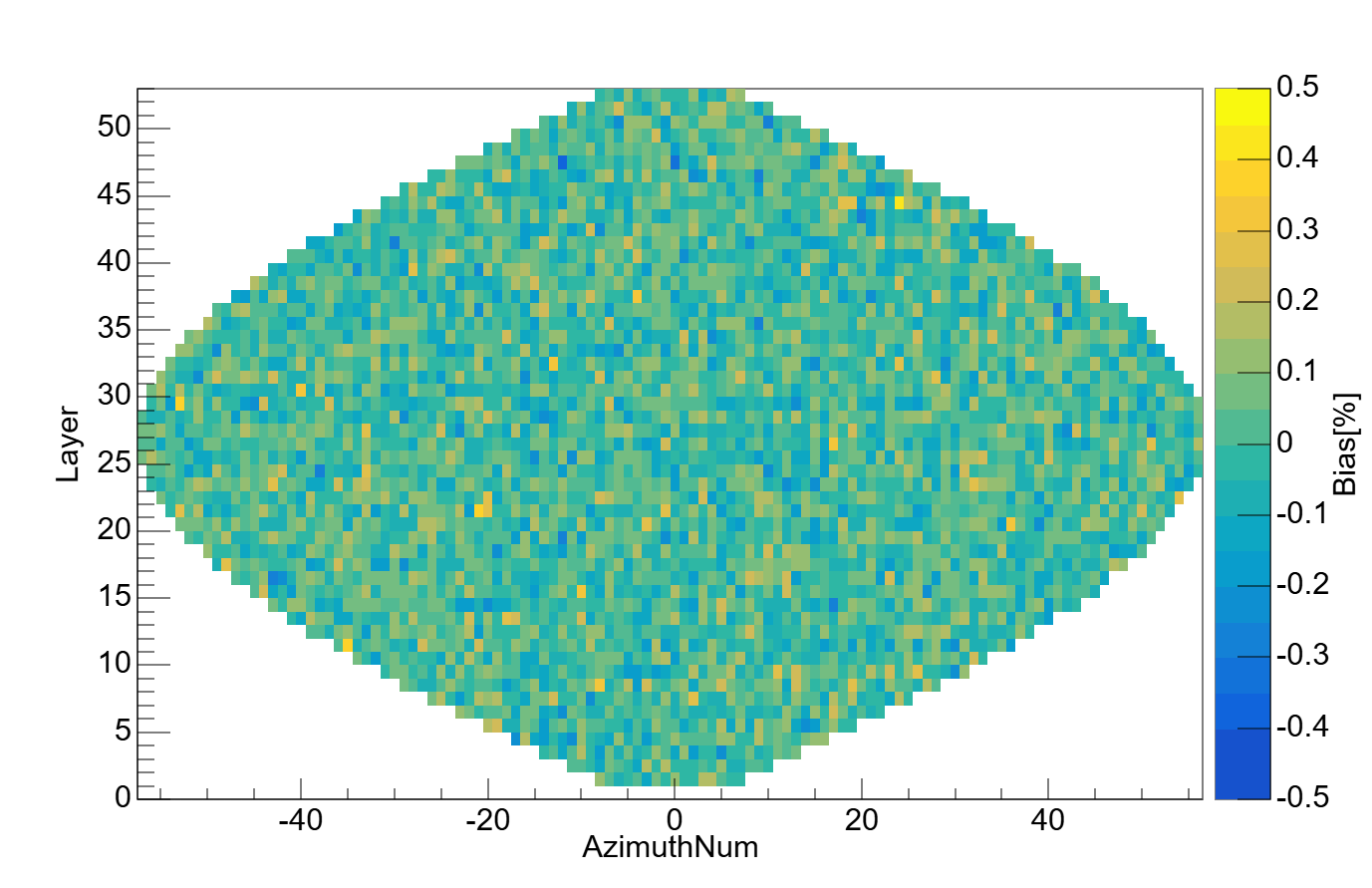}
\\
\includegraphics[width=0.577\textwidth]{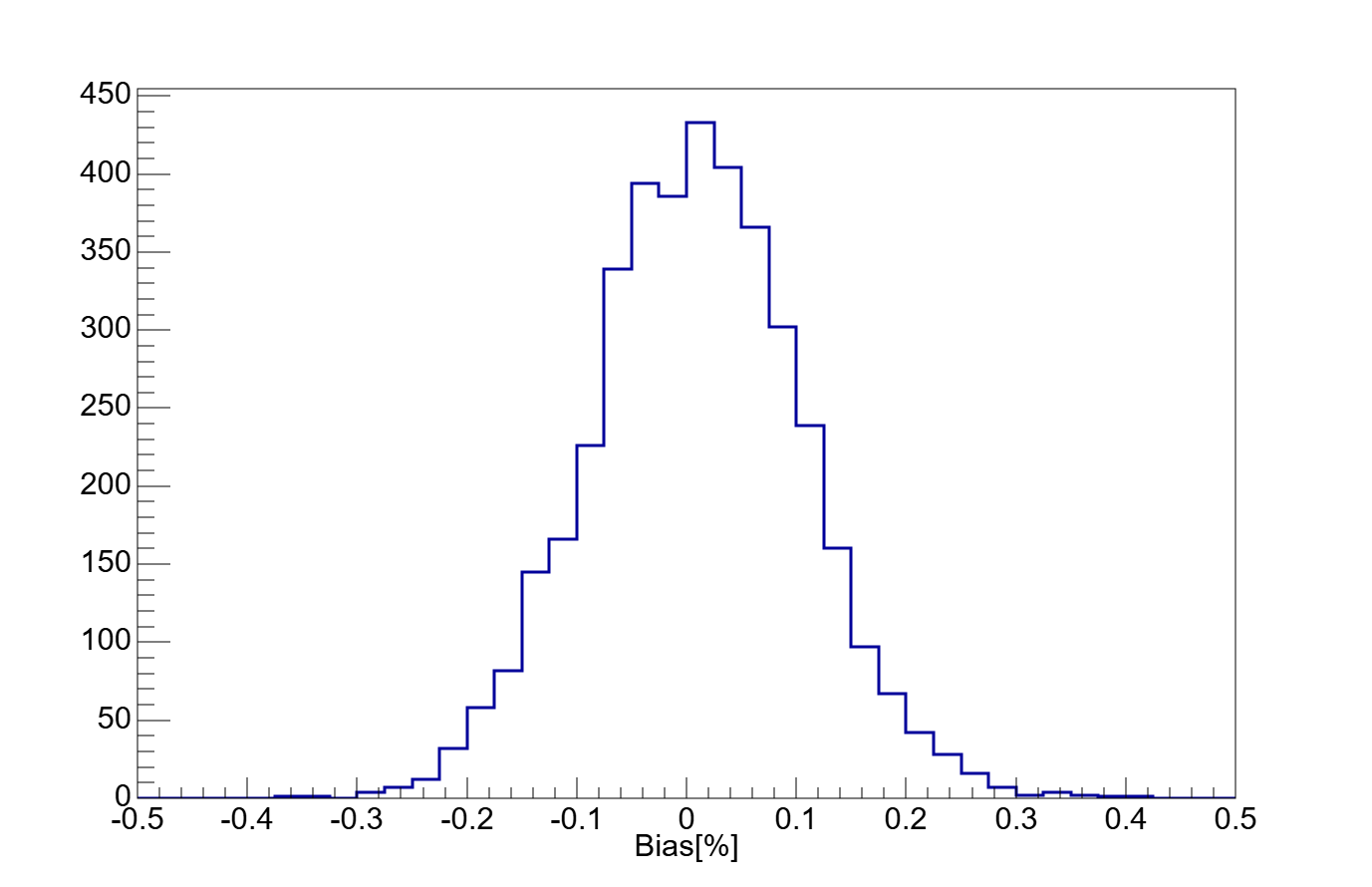}
\caption{EOCT calibration results based on the SiPM on-off switching method. The upper panel: tile-level EOCT calibration results. The middle panel: comparison between the calibration results and the simulation truth in each tile. The lower panel: 1D histogram filled with each bin value from the middle 2D histogram.}
\label{pic5.1}
\end{figure}

The EOCT rate calibration results are shown in Figure \ref{pic5.1}; the bias is 0.009\% and the standard deviation is 0.096\%.

The bias and standard deviation are small because the calibrated EOCT rate (defined in Equation \ref{eq5.1}) is a directly observable quantity rather than a physical parameter such as the EOCT probability for each SiPM tile. The bias is quantified as the relative difference between the rates from Equation \ref{eq5.1} and Equation \ref{eq5.1.1} (In Equation \ref{eq5.1.1}, $\text{Rate}^{\rm truth}_{\rm EOCT}$ is the true EOCT ratio, $Q_{\rm EOCT}$ is the charge from EOCT, $Q_C$ is the charge measured when other SiPM tiles are turned off). Since the LED sources can only be deployed along the central axis, the EOCT emission angle distribution must be calibrated to infer the response off the central axis.

\begin{equation}
\label{eq5.1.1}
\text{Rate}^{\rm truth}_{\rm EOCT} = \frac{Q_{\rm EOCT}}{Q_C}
\end{equation}

\subsection{External Optical Crosstalk Emission Angle Distribution Calibration}
\label{5.2}

In this study, the distribution of the SiPM EOCT emission angle is calibrated by using a reference SiPM tile to detect and quantify EOCT from other SiPM tiles positioned at different 5-degree angular intervals relative to the reference SiPM tile. The procedure is as follows: First, both the reference SiPM tile and the SiPM tiles within each 5-degree angular interval are turned on, and the charge $Q_{Oi}$ on the reference SiPM tile is recorded. Next, only the reference SiPM tile is turned on, and data are acquired over the same duration to measure the charge $Q_{Ci}$ on the reference SiPM tile. Finally, the integrated EOCT response ($E_i(\theta)$) for each angular interval is obtained by subtracting $Q_{Ci}$ from $Q_{Oi}$.

According to the mean value theorem for integrals, the EOCT response $\epsilon_i(\theta)$ in each  5-degree angular interval is given by Equation \ref{eq5.3}:
\begin{equation}
\label{eq5.3}
\begin{split}
\epsilon_i(\theta) &= \frac{E_i(\theta)}{\sum \vec{n}(\theta) \cdot \Delta \vec{S}(\theta)} 
\end{split}
\end{equation}
where $\Delta \vec{S}$ is the area vector of each SiPM tile  (with magnitude equals to the sensitive area and direction normal to the surface), $\vec{n}$ is the unit direction vector pointing from the emitting SiPM tile to the reference SiPM tile. 

After acquiring the $\epsilon_i(\theta)$ values for each 5-degree angular interval, the results are normalized with $\kappa$ as the normalization coefficient:
\begin{equation}
\label{eq5.3.1}
\begin{split}
\kappa\sum_i\epsilon_i(\theta) &= 1
\end{split}
\end{equation}
and the calibrated EOCT emission angle distribution is thus derived.

The reference SiPM tile selected for this study is located at the detector's equator. This choice is made  because apertures are present at the northern and southern poles of the TAO CD. To maximize the collection of EOCT photons, it is essential to position the reference SiPM tile away from these apertures. Therefore, the equatorial location is determined to be the optimal position.

\begin{figure}[htbp]
\centering
\includegraphics[width=.4\textwidth]{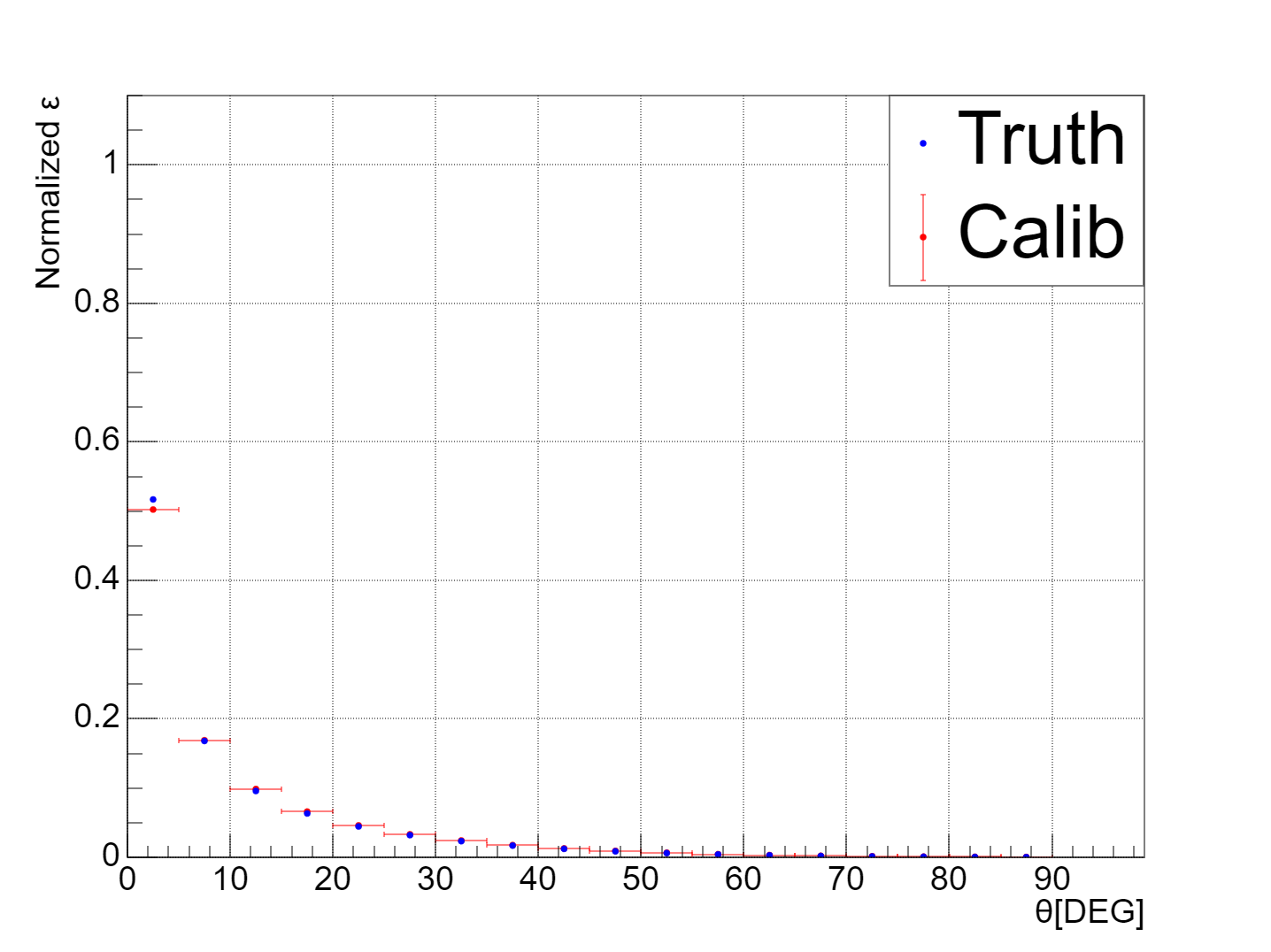}
\qquad
\includegraphics[width=.4\textwidth]{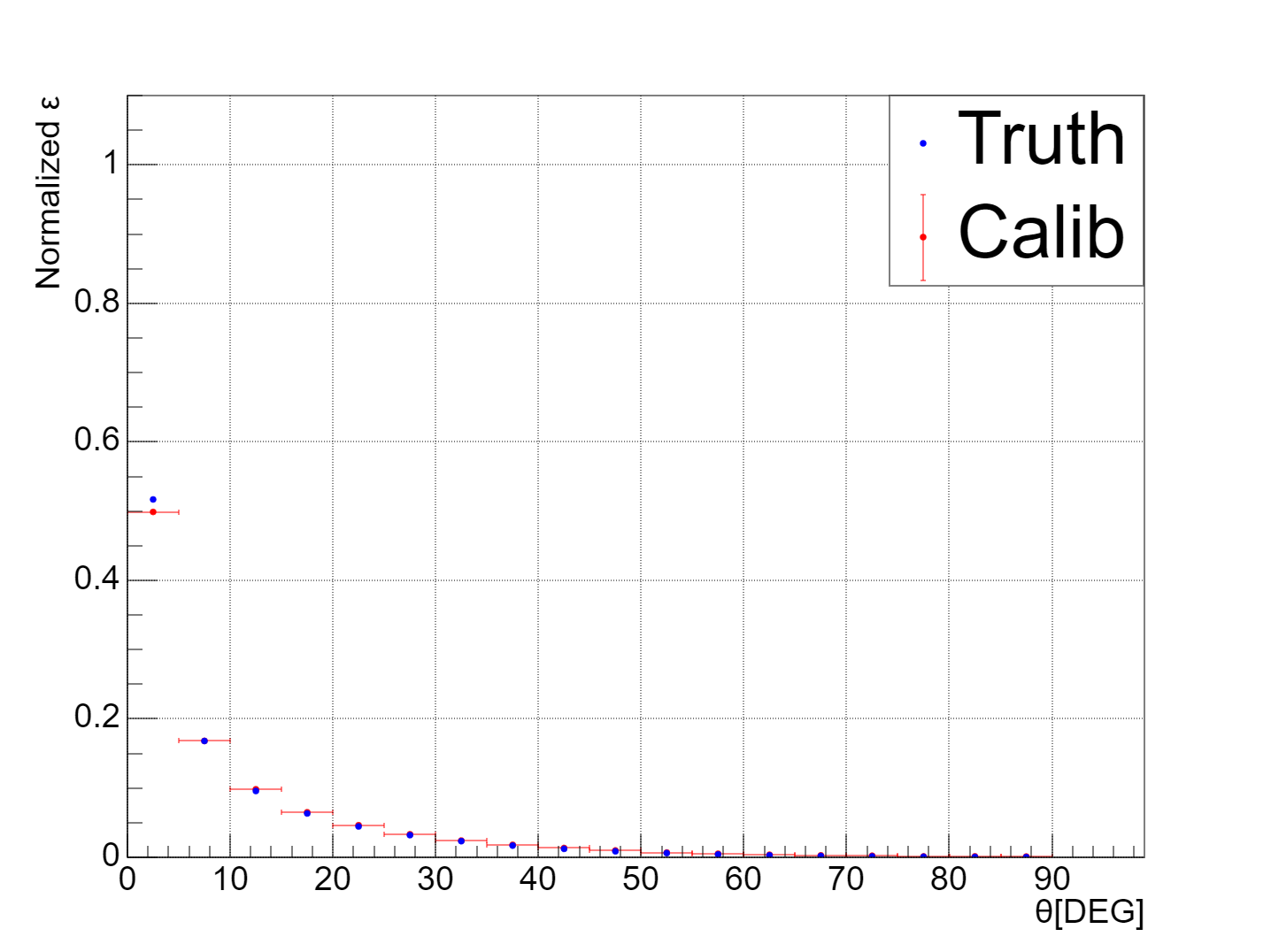}
\\
\includegraphics[width=0.4\textwidth]{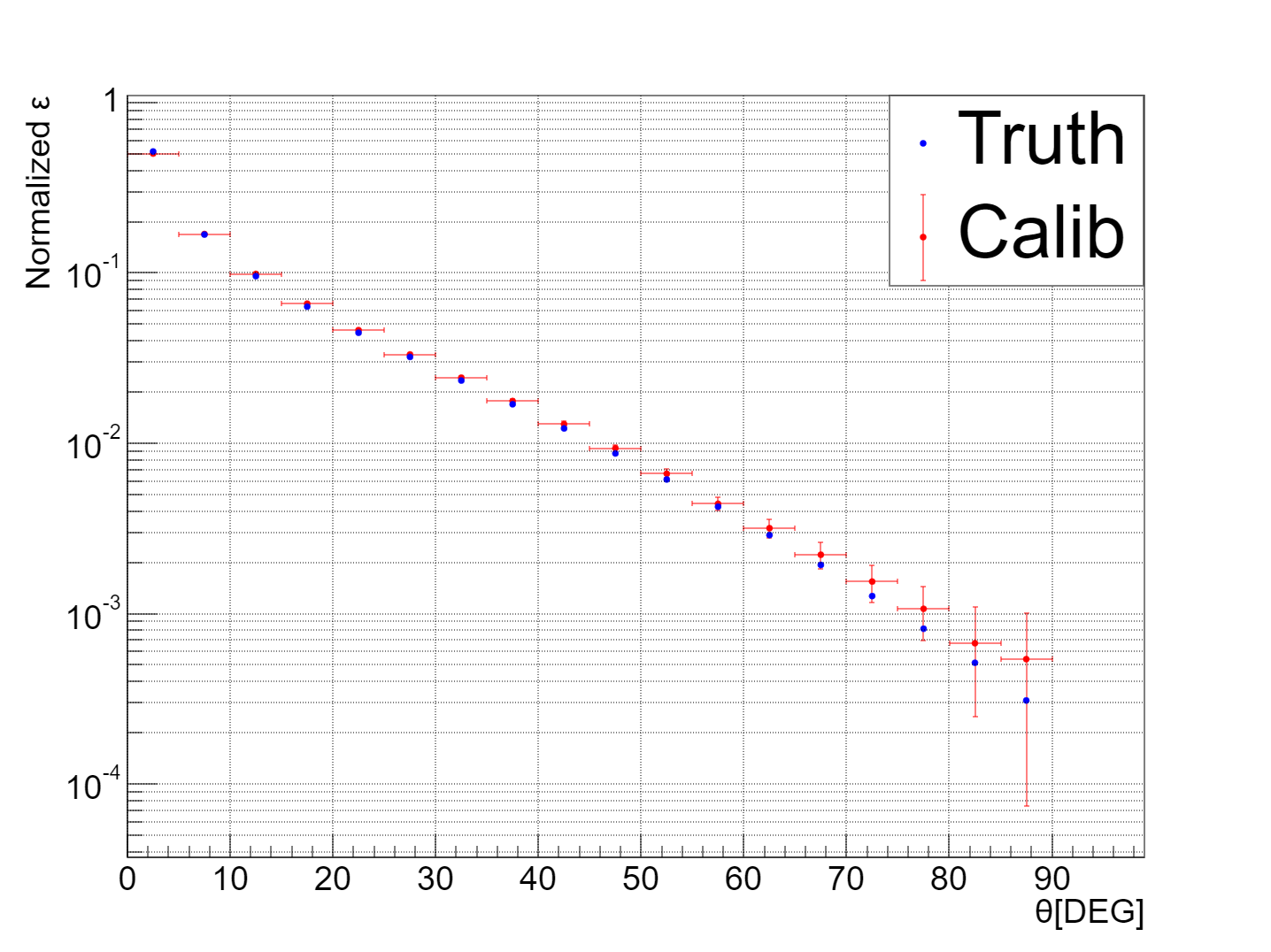}
\qquad
\includegraphics[width=0.4\textwidth]{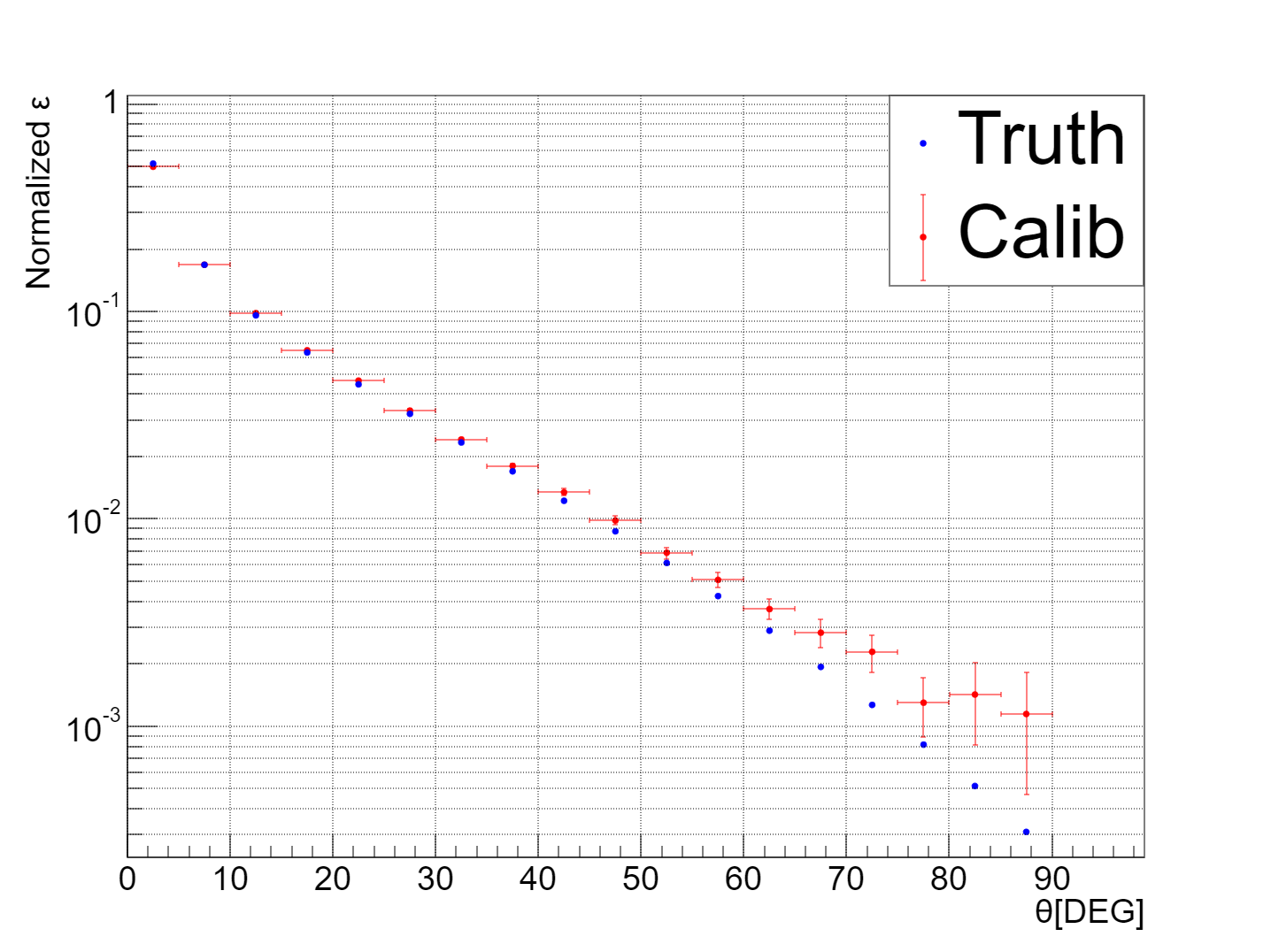}
\caption{The EOCT emission angle distribution calibration results. The left two panels: calibration results with the SiPM tile surface optical reflection effect disabled in simulation, and the numerical values of each data point are presented in Table \ref{tab5.1}. The right two panels: calibration results with the SiPM tile surface optical reflection effect enabled in simulation, and the numerical values of each data point are presented in Table \ref{tab5.2}. The upper two panels: a linear y-axis scale is adopted to display the results in the small-angle range. The lower two panels: a logarithmic y-axis scale is adopted to display the results in the large-angle range.}
\label{pic5.3}
\end{figure}

The EOCT emission angle distribution calibration results are shown in Figure \ref{pic5.3}, Table \ref{tab5.1} and Table \ref{tab5.2}. In this figure, the blue points denote the simulation truth, which is the true EOCT emission angle distribution as configured in the TAO simulation. The red points represent the calibration results, corresponding to the values of $\epsilon_i(\theta)$. The vertical (y-axis) error bars attached to the red points indicate the uncertainty of the calibration results, corresponding to the standard deviation of $\epsilon_i(\theta)$. This standard deviation is derived from the standard deviations of $Q_{Oi}$ and $Q_{Ci}$ via the error propagation formula, with its explicit expression presented in Equation \ref{eq5.3.2},
\begin{equation}
\label{eq5.3.2}
\begin{split}
\sigma_{\epsilon i} = \frac{\sqrt{\sigma_{Q_{Oi}}^{2}+\sigma_{Q_{Ci}}^{2}}}{\kappa \cdot \sum \vec{n}(\theta) \cdot \Delta \vec{S}}
\end{split}
\end{equation}
where $\sigma_{Q_{Ci}}$ is the standard deviation of charge samples acquired by the reference SiPM tile when only the tile itself is enabled, $\sigma_{Q_{Oi}}$ is the standard deviation of charge samples from the reference SiPM tile when both the SiPM tiles in each 5-degree angular interval and the reference SiPM tile are activated. All samples used in this section are obtained from one million LED external trigger events. 

As calculated in Ref. \cite{Ref59} based on a thin-film optical model and Fresnel equations, most EOCT photons emitted from the monocrystalline silicon have their propagation directions confined within $30^\circ$ relative to the surface normal as they travel through the epoxy resin coating on the SiPM tile surface into the LAB medium. Accordingly, the angular interval with EOCT emission angles below $30^\circ$ is defined as the main angular range. The results with the SiPM tile surface optical reflection effect disabled are presented in the left part of Figure \ref{pic5.3} and in Table \ref{tab5.1}. In this case, the emission angle distribution of the EOCT photons could be measured with a bias of less than 4\% and a standard deviation of less than 2\% in the main angular range.

The results with the SiPM tile surface optical reflection effect enabled are presented in the right part of Figure \ref{pic5.3} and in Table \ref{tab5.2}. The results demonstrate that this reflection effect predominantly affects the large-angle intervals, whereas its impact on the calibration results within the main angular range is relatively limited.


\begin{table}[htbp]
  \centering
  \small
  \setlength{\tabcolsep}{3.5pt}
  \caption{The EOCT emission angle distribution calibration results with the SiPM tile surface optical reflection effect disabled in simulation}
  \label{tab5.1}
  \sisetup{
    scientific-notation = true,
    table-format = 1.2e-1,
    group-digits = false
  }
  \begin{tabular}{c S S S c c}
    \toprule
    {Angle ($^\circ$)} & {Simulation Truth} & {Calibration Value} & {Calibration Uncertainty} & {Relative Bias (\%)} & {Relative Uncertainty (\%)} \\
    \midrule
    2.5  & 0.517    & 0.502    & 0.00169  & 3.0  & 0.3  \\
    7.5  & 0.168    & 0.169    & 0.00103  & 0.6  & 0.6  \\
    12.5 & 0.0958   & 0.0987   & 0.000809 & 2.9  & 0.8  \\
    17.5 & 0.0634   & 0.0660    & 0.000681 & 3.9  & 1.0  \\
    22.5 & 0.0446   & 0.0462   & 0.000604 & 3.5  & 1.3  \\
    27.5 & 0.0321   & 0.0332   & 0.000539 & 3.3  & 1.6  \\
    32.5 & 0.0233   & 0.0243   & 0.000513 & 4.1  & 2.1  \\
    37.5 & 0.0170    & 0.0177   & 0.000482 & 4.0  & 2.7  \\
    42.5 & 0.0122   & 0.0130    & 0.000472 & 6.2  & 3.6  \\
    47.5 & 0.00871  & 0.00936  & 0.000449 & 6.9  & 4.8  \\
    52.5 & 0.00613  & 0.00663  & 0.000420  & 7.5  & 6.3  \\
    57.5 & 0.00424  & 0.00441  & 0.000412 & 3.9  & 9.3  \\
    62.5 & 0.00289  & 0.00317  & 0.000389 & 8.8  & 12.3 \\
    67.5 & 0.00193  & 0.00222  & 0.000390  & 13.1 & 17.6 \\
    72.5 & 0.00127  & 0.00154  & 0.000381 & 17.5 & 24.7 \\
    77.5 & 0.000815 & 0.00107  & 0.000372 & 23.8 & 34.8 \\
    82.5 & 0.000515 & 0.000670  & 0.000421 & 23.1 & 62.8 \\
    87.5 & 0.000309 & 0.000540  & 0.000466 & 42.8 & 86.3 \\
    \bottomrule
  \end{tabular}
\end{table}

\begin{table}[htbp]
  \centering
  \small
  \setlength{\tabcolsep}{3.5pt}
  \caption{The EOCT emission angle distribution calibration results with the SiPM tile surface optical reflection effect enabled in simulation}
  \label{tab5.2}
  \sisetup{
    scientific-notation = true,
    table-format = 1.3e-1,
    group-digits = false
  }
  \begin{tabular}{c S S S c c}
    \toprule
    {Angle ($^\circ$)} & {Simulation Truth} & {Calibration Value} & {Calibration Uncertainty} & {Relative Bias (\%)} & {Relative Uncertainty (\%)} \\
    \midrule
    2.5  & 0.517    & 0.499    & 0.00167  & 3.6  & 0.3  \\
    7.5  & 0.168    & 0.168    & 0.000998 & 0.0  & 0.6  \\
    12.5 & 0.0958   & 0.0983   & 0.000807 & 2.5  & 0.8  \\
    17.5 & 0.0634   & 0.0650    & 0.000676 & 2.5  & 1.0  \\
    22.5 & 0.0446   & 0.0464   & 0.000605 & 3.9  & 1.3  \\
    27.5 & 0.0321   & 0.0333   & 0.000541 & 3.6  & 1.6  \\
    32.5 & 0.0233   & 0.0241   & 0.000512 & 3.3  & 2.1  \\
    37.5 & 0.0170    & 0.0180    & 0.000487 & 5.6  & 2.7  \\
    42.5 & 0.0122   & 0.0134   & 0.000480  & 9.0  & 3.6  \\
    47.5 & 0.00871  & 0.00982  & 0.000460  & 11.3 & 4.7  \\
    52.5 & 0.00613  & 0.00687  & 0.000428 & 10.8 & 6.2  \\
    57.5 & 0.00424  & 0.00508  & 0.000442 & 16.5 & 8.7  \\
    62.5 & 0.00289  & 0.00367  & 0.000418 & 21.3 & 11.4 \\
    67.5 & 0.00193  & 0.00283  & 0.000440  & 31.8 & 15.5 \\
    72.5 & 0.00127  & 0.00228  & 0.000464 & 44.3 & 20.4 \\
    77.5 & 0.000815 & 0.00130   & 0.000410  & 37.3 & 31.5 \\
    82.5 & 0.000515 & 0.00142  & 0.000613 & 63.7 & 43.2 \\
    87.5 & 0.000309 & 0.00115  & 0.000679 & 73.1 & 59.0 \\
    \bottomrule
  \end{tabular}
\end{table}

\begin{figure}[htbp]
\centering
\includegraphics[width=.75\textwidth]{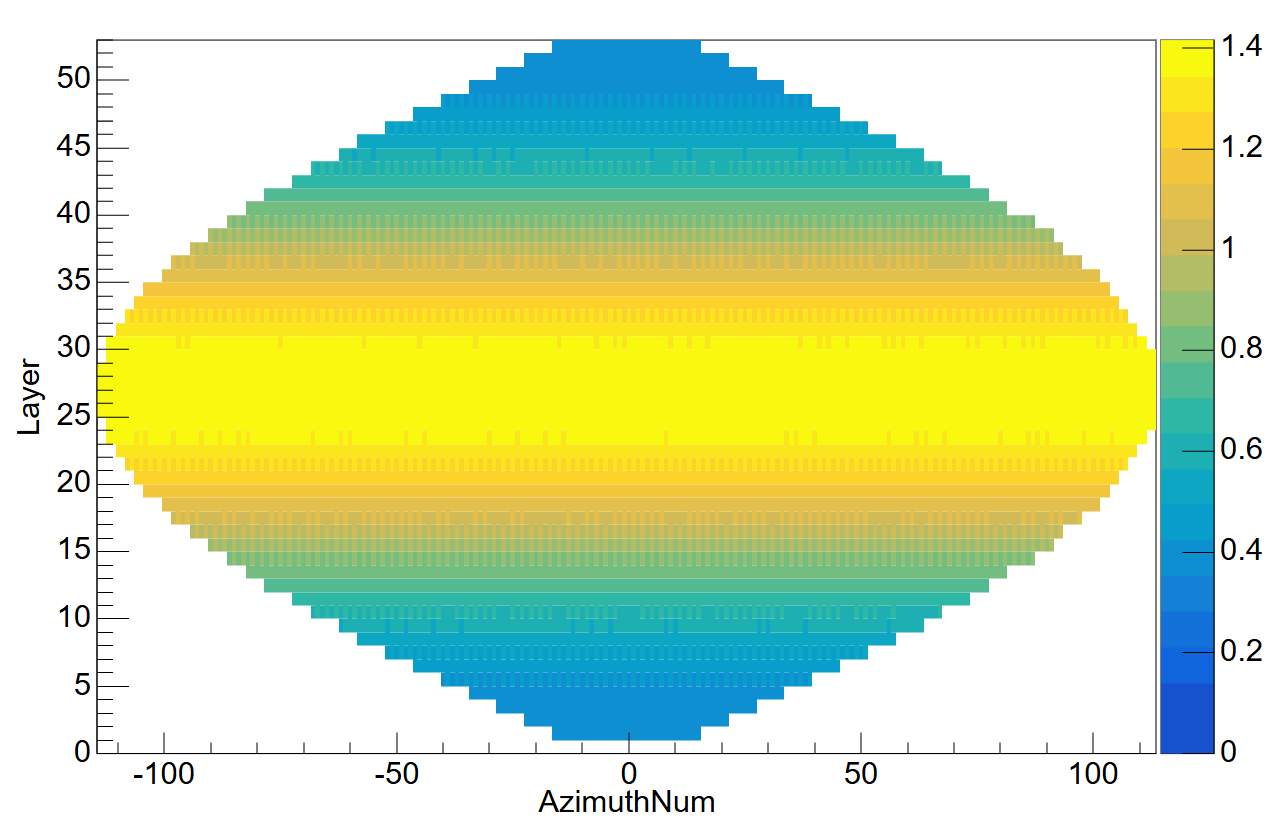}
\caption{LED non-uniform light field distribution configuration in simulation. The distribution follows a Gaussian distribution $\text{Gaus}(\theta,90^\circ,30^\circ)$, where the parameters correspond to the polar angle values in the spherical coordinate system.}
\label{pic5.4}
\end{figure}
\begin{figure}[htbp]
\centering
\includegraphics[width=.4\textwidth]{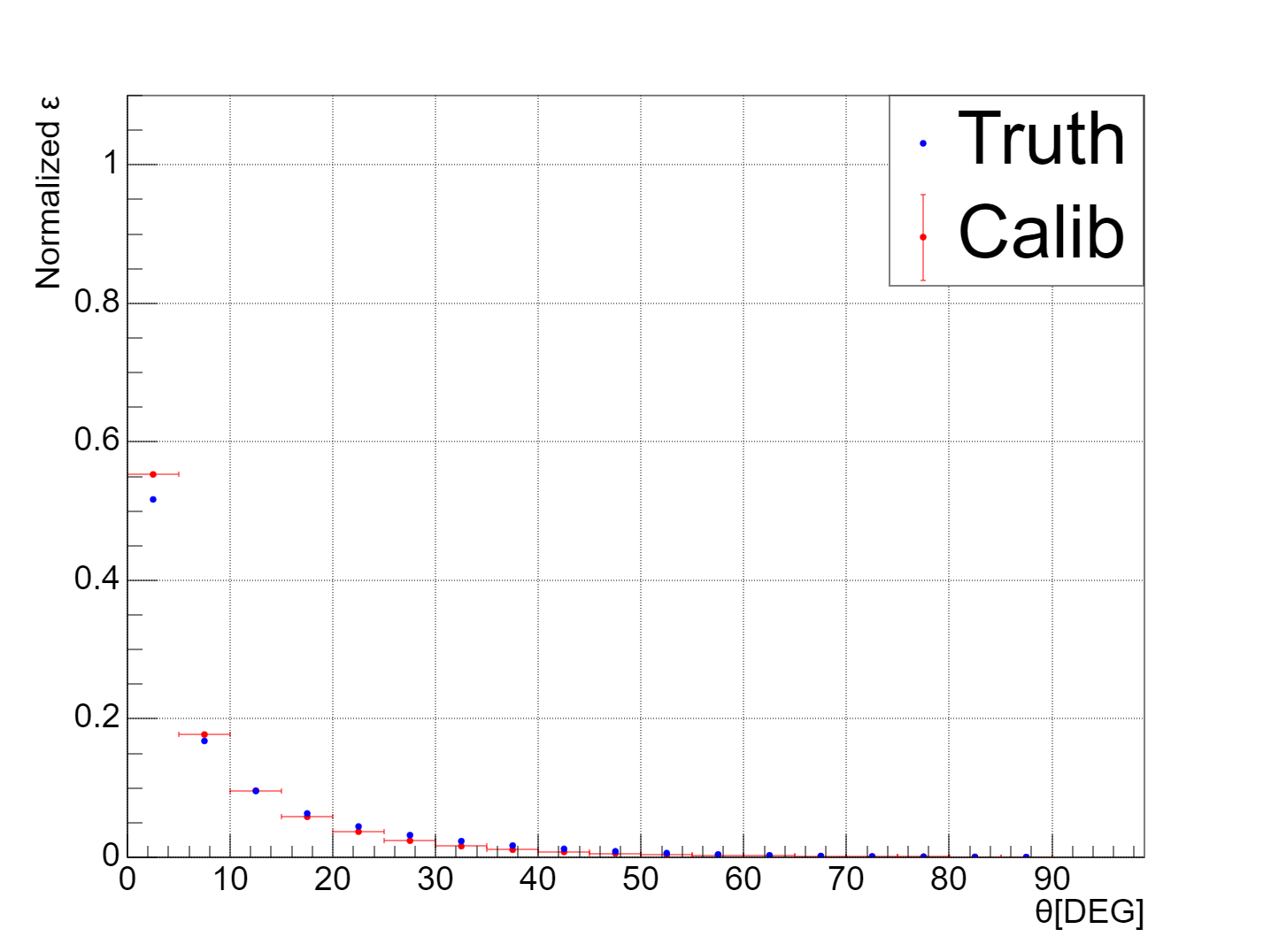}
\qquad
\includegraphics[width=.4\textwidth]{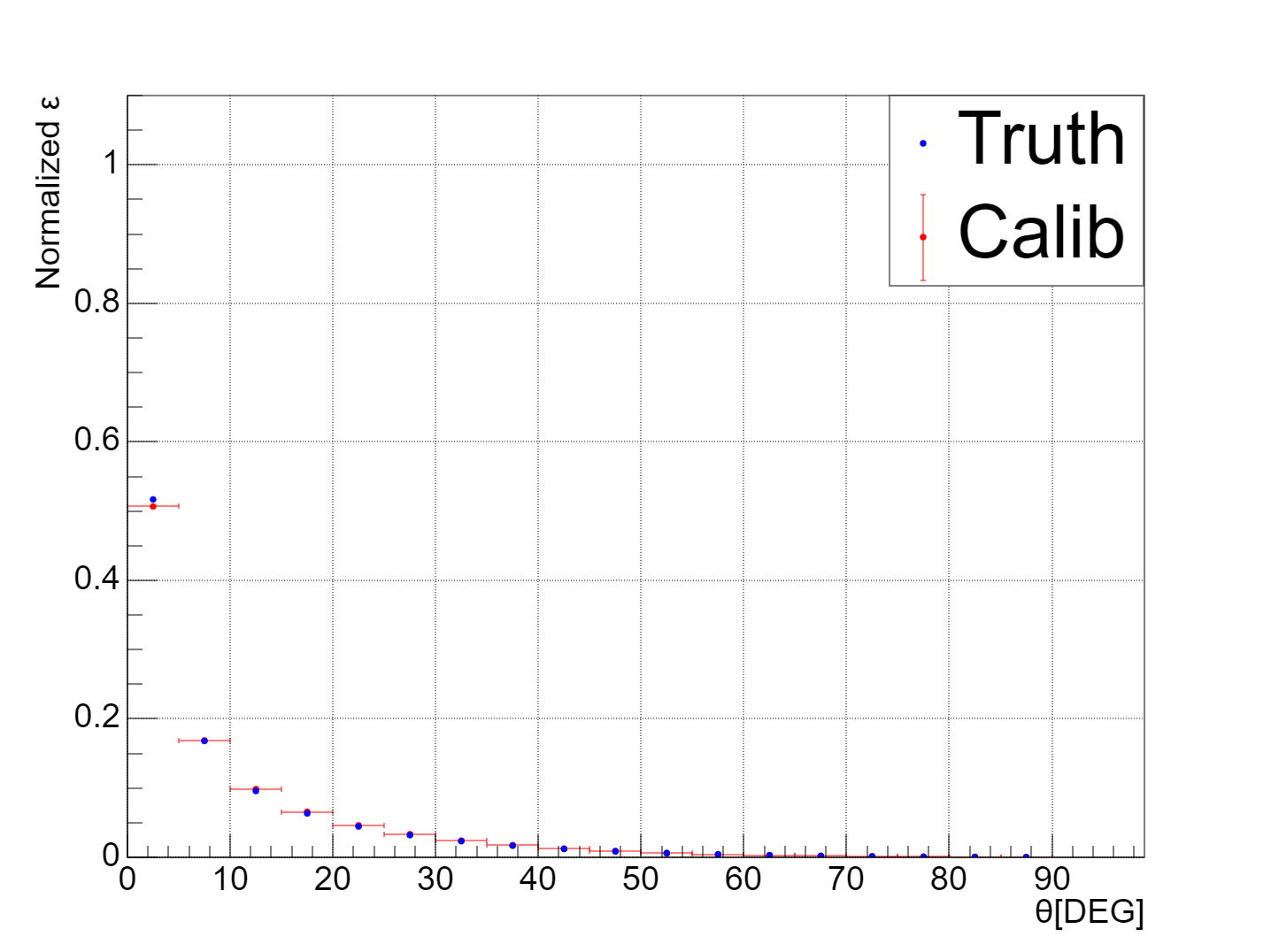}
\\
\includegraphics[width=0.4\textwidth]{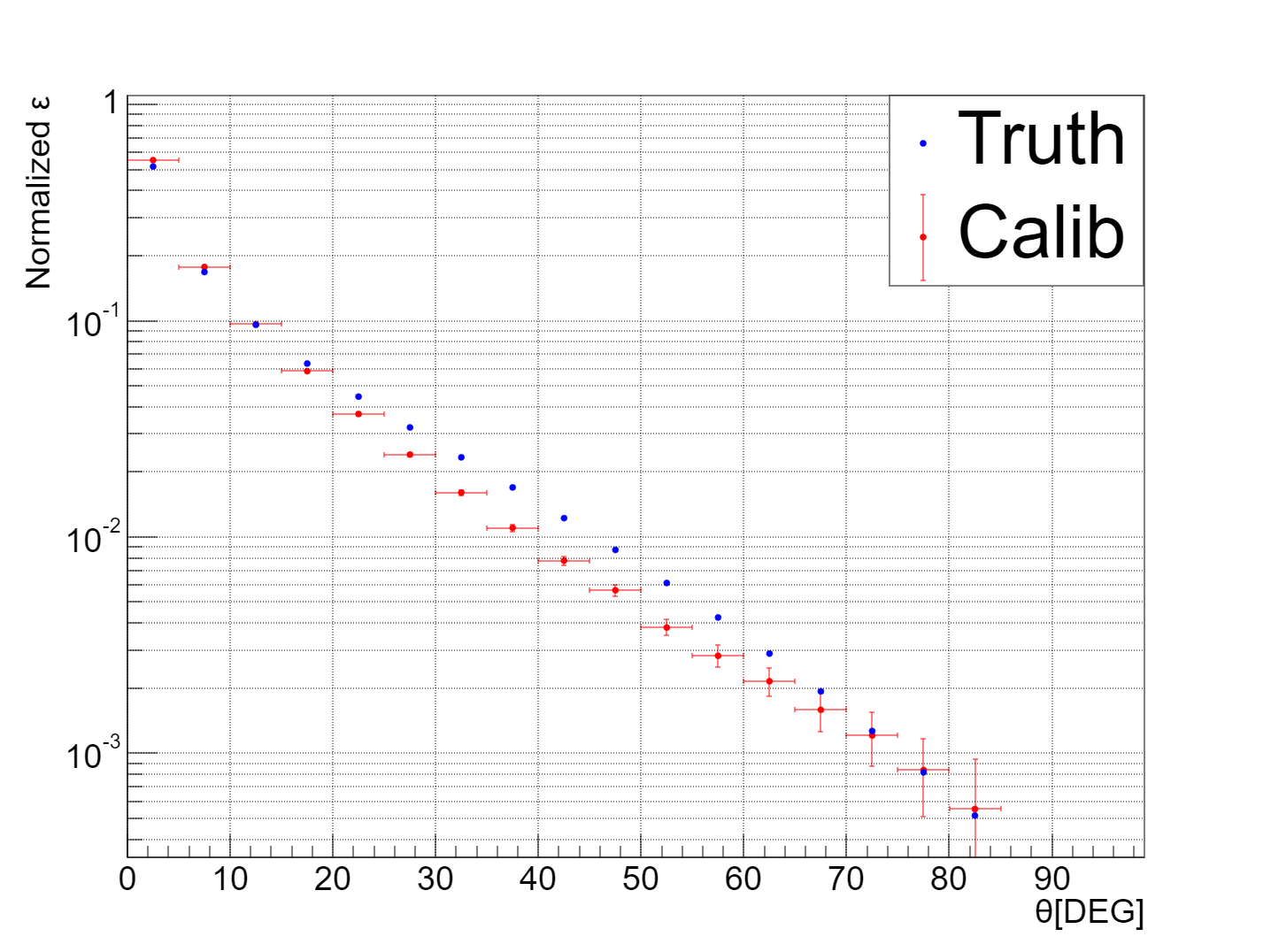}
\qquad
\includegraphics[width=0.4\textwidth]{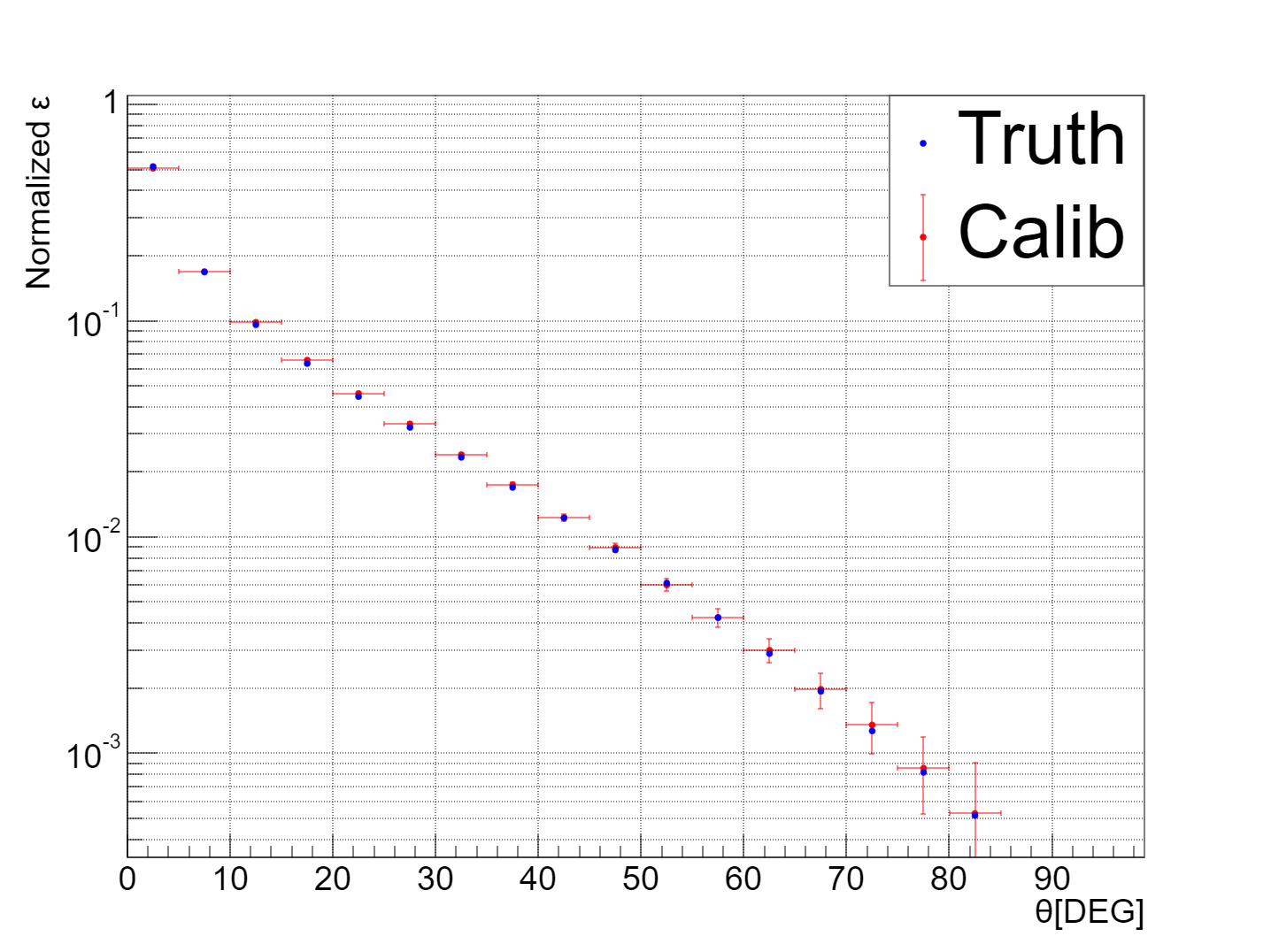}
\caption{The EOCT emission angle distribution calibration results by using a non-uniform light field. The left two panels: calibration results without non-uniform correction, and the numerical values of each data point are presented in Table \ref{tab5.3}. The right two panels: calibration results with non-uniform correction, and the numerical values of each data point are presented in Table \ref{tab5.4}. The upper two panels: a linear y-axis scale is adopted to display the results in the small-angle range. The lower two panels: a logarithmic y-axis scale is adopted to display the results in the large-angle range.}
\label{pic5.5}
\end{figure}

Since the LED light field distribution may be non-uniform, this study simulated a non-uniform LED light field distribution, as illustrated in Figure \ref{pic5.4}. The corresponding EOCT emission angle distribution calibration results are shown in the left part of Figure \ref{pic5.5} and in Table \ref{tab5.3}. These results indicate that the non-uniform LED light field distribution affects the EOCT emission angle distribution calibration within the main angular range.

In the TAO calibration system, the $^{68}{\rm Ge}$ source produces uniform scintillation light in the liquid scintillator, allowing us to correct the non-uniformity of LED light field. The charge response of each SiPM tile for both the $^{68}{\rm Ge}$ source and the LED source are acquired with all SiPM tiles turned on. During the EOCT emission angle distribution calibration, the charge measured by the reference SiPM tile in the "turn on" data is multiplied by a correction factor $\alpha$ defined in Equation \ref{eq5.4}. The summation is performed over all SiPM tiles within the corresponding angular range, $Q_{{\rm Ge}}$ denotes the SiPM tile charge response when using the $^{68}{\rm Ge}$ source, and $Q_{\rm LED}$ denotes the SiPM tile charge response when using the LED light source.


The corrected calibration results are shown in the right panel of Figure \ref{pic5.5} and in Table \ref{tab5.4}. The results confirm that the bias of the calibrated EOCT emission angle distribution within the main angular range, induced by the LED light field non-uniformity, has been successfully corrected.

\begin{table}[htbp]
  \centering
  \small
  \setlength{\tabcolsep}{3.5pt}
  \caption{The EOCT emission angle distribution calibration results by using a non-uniform light field without non-uniform correction}
  \label{tab5.3}
  \sisetup{
    scientific-notation = true,
    table-format = 1.3e-1,
    group-digits = false
  }
  \begin{tabular}{c S S S c c}
    \toprule
    {Angle ($^\circ$)} & {Simulation Truth} & {Calibration Value} & {Calibration Uncertainty} & {Relative Bias (\%)} & {Relative Uncertainty (\%)} \\
    \midrule
    2.5  & 0.517    & 0.553    & 0.00178  & 6.5   & 0.3   \\
    7.5  & 0.168    & 0.177    & 0.00105  & 5.1   & 0.6   \\
    12.5 & 0.0958   & 0.0964   & 0.000799 & 0.6   & 0.8   \\
    17.5 & 0.0634   & 0.0585   & 0.000641 & 8.4   & 1.1   \\
    22.5 & 0.0446   & 0.0371   & 0.000541 & 20.2  & 1.5   \\
    27.5 & 0.0321   & 0.0240    & 0.000458 & 33.8  & 1.9   \\
    32.5 & 0.0233   & 0.0161   & 0.000418 & 44.7  & 2.6   \\
    37.5 & 0.0170    & 0.0110    & 0.000381 & 54.5  & 3.5   \\
    42.5 & 0.0122   & 0.00777  & 0.000365 & 57.0  & 4.7   \\
    47.5 & 0.00871  & 0.00567  & 0.000350  & 53.6  & 6.2   \\
    52.5 & 0.00613  & 0.00381  & 0.000319 & 60.9  & 8.4   \\
    57.5 & 0.00424  & 0.00282  & 0.000329 & 50.4  & 11.7  \\
    62.5 & 0.00289  & 0.00215  & 0.000320  & 34.4  & 14.9  \\
    67.5 & 0.00193  & 0.00159  & 0.000330  & 21.4  & 20.8  \\
    72.5 & 0.00127  & 0.00121  & 0.000337 & 5.0   & 27.9  \\
    77.5 & 0.000815 & 0.000837 & 0.000329 & 2.6   & 39.3  \\
    82.5 & 0.000515 & 0.000552 & 0.000382 & 6.7   & 69.2  \\
    87.5 & 0.000309 & 0.000294 & 0.000344 & 5.1   & 117.0 \\
    \bottomrule
  \end{tabular}
\end{table}

\begin{table}[htbp]
  \centering
  \small
  \setlength{\tabcolsep}{3.5pt}
  \caption{The EOCT emission angle distribution calibration results by using a non-uniform light field with non-uniform correction}
  \label{tab5.4}
  \sisetup{
    scientific-notation = true,
    table-format = 1.3e-1,
    group-digits = false
  }
  \begin{tabular}{c S S S c c}
    \toprule
    {Angle ($^\circ$)} & {Simulation Truth} & {Calibration Value} & {Calibration Uncertainty} & {Relative Bias (\%)} & {Relative Uncertainty (\%)} \\
    \midrule
    2.5  & 0.517    & 0.507    & 0.00170   & 2.0  & 0.3   \\
    7.5  & 0.168    & 0.169    & 0.00103  & 0.6  & 0.6   \\
    12.5 & 0.0958   & 0.0985   & 0.000808 & 2.7  & 0.8   \\
    17.5 & 0.0634   & 0.0657   & 0.000679 & 3.5  & 1.0   \\
    22.5 & 0.0446   & 0.0461   & 0.000603 & 3.3  & 1.3   \\
    27.5 & 0.0321   & 0.0333   & 0.000539 & 3.6  & 1.6   \\
    32.5 & 0.0233   & 0.0240    & 0.000511 & 2.9  & 2.1   \\
    37.5 & 0.0170    & 0.0174   & 0.000479 & 2.3  & 2.8   \\
    42.5 & 0.0122   & 0.0123   & 0.000460  & 0.8  & 3.7   \\
    47.5 & 0.00871  & 0.00895  & 0.000439 & 2.7  & 4.9   \\
    52.5 & 0.00613  & 0.00599  & 0.000399 & 2.3  & 6.7   \\
    57.5 & 0.00424  & 0.00423  & 0.000403 & 0.2  & 9.5   \\
    62.5 & 0.00289  & 0.00299  & 0.000377 & 3.3  & 12.6  \\
    67.5 & 0.00193  & 0.00198  & 0.000368 & 2.5  & 18.6  \\
    72.5 & 0.00127  & 0.00135  & 0.000357 & 5.9  & 26.4  \\
    77.5 & 0.000815 & 0.000852 & 0.000332 & 4.3  & 39.0  \\
    82.5 & 0.000515 & 0.000527 & 0.000373 & 2.3  & 70.8  \\
    87.5 & 0.000309 & 0.000270  & 0.000329 & 14.4 & 121.9 \\
    \bottomrule
  \end{tabular}
\end{table}

\begin{equation}
\label{eq5.4}
\alpha =  \frac{ \sum Q_{{\rm Ge}}}{ \sum Q_{\rm LED}}
\end{equation}

\subsection{Correcting Dark Count Rate Based on the External Optical Crosstalk Rate Calibration Result}
\label{5.3}

The DCR calibrated in Section \ref{3.1} includes contributions from EOCT, which is the equivalent dark count rate (EDCR). Therefore, a correction is necessary. The correction factor $\beta_{\rm DN}$ is defined as:
\begin{equation}
\label{eq5.5.1}
\begin{split}
\text{DCR} &= \text{EDCR} \cdot \beta_{\rm DN} \\
\beta_{\rm DN} &= \frac{Q_{\rm DN}}{Q_{\rm DN}+Q_{\rm DN,\rm EOCT}}
\end{split}
\end{equation}
This correction factor is characterized by the ratio of the dark noise charge response ($Q_{\rm DN}$) to the total charge response from dark noise and dark noise-induced EOCT ($Q_{\rm DN}+Q_{\rm DN,\rm EOCT}$).

Since the AP of the TAO SiPM is very low (<1\%) \cite{Ref32}, $\text{Rate}_{\rm EOCT}$ (from Equation \ref{eq5.1}) can be expressed as:
\begin{equation}
\label{eq5.5.2}
\begin{split}
\text{Rate}_{\rm EOCT} &= \frac{Q_{\rm EOCT}}{Q_{\rm LED}+Q_{\rm DN}+Q_{\rm IOCT}+Q_{\rm AP}} \\
&\approx \frac{Q_{\rm EOCT}}{Q_{\rm LED}+Q_{\rm DN}+Q_{\rm IOCT}} \\
&= \frac{\delta_{\rm EOCT} \cdot (Q_{\rm LED}+Q_{\rm DN})}{Q_{\rm LED}+Q_{\rm DN}+\delta_{\rm IOCT} \cdot (Q_{\rm LED}+Q_{\rm DN})}
\end{split}
\end{equation}
where $\delta_{\text{EOCT}}$ is the mathematical expectation of the total number of EOCT ultimately generated by an initial avalanche signal from a photon hit or dark noise event; the EOCT events stem from two cascading pathways: direct EOCT triggered by the initial avalanche pulse itself, and secondary EOCT further cascaded from IOCT events induced by the same avalanche pulse. $\delta_{\text{EOCT}}$ clearly satisfies:
\begin{equation}
\label{eq5.5.4}
\begin{split}
\delta_{\rm EOCT} &= \frac{Q_{\rm DN,\rm EOCT}}{Q_{\rm DN}}
\end{split}
\end{equation}
and $\delta_{\text{IOCT}}$ is the mathematical expectation of the total number of IOCT ultimately generated by an initial avalanche signal from a photon hit or a dark noise event. For IOCT, the number of IOCT induced by an avalanche signal follows the Borel distribution. Based on the mathematical properties of the Borel distribution, $\delta_{\text{IOCT}}$ satisfies:
\begin{equation}
\label{eq5.5.3}
\begin{split}
\delta_{\rm IOCT} &= \frac{\lambda_{\rm IOCT}}{1-\lambda_{\rm IOCT}}
\end{split}
\end{equation}
According to Equation \ref{eq5.5.1}, Equation \ref{eq5.5.2}, Equation \ref{eq5.5.3} and Equation \ref{eq5.5.4}, $\beta_{\rm DN}$ can be expressed as:
\begin{equation}
\label{eq5.5.5}
\begin{split}
\beta_{\rm DN} &= \frac{1}{1+\frac{\text{Rate}_{\rm EOCT}}{1-\lambda_{\rm IOCT}}}
\end{split}
\end{equation}

\begin{figure}[htbp]
\centering
\includegraphics[width=.45\textwidth]{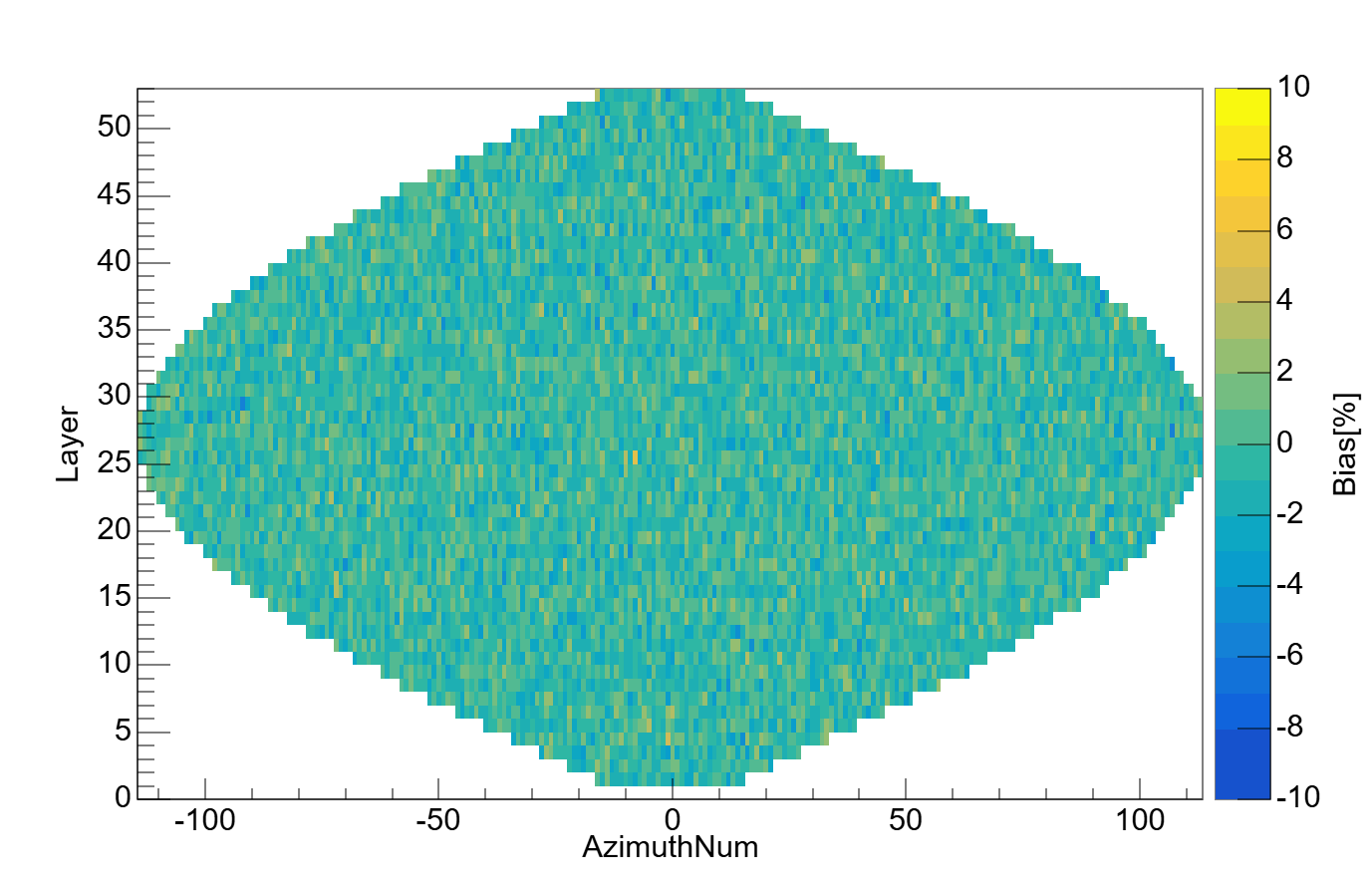}
\qquad
\includegraphics[width=.45\textwidth]{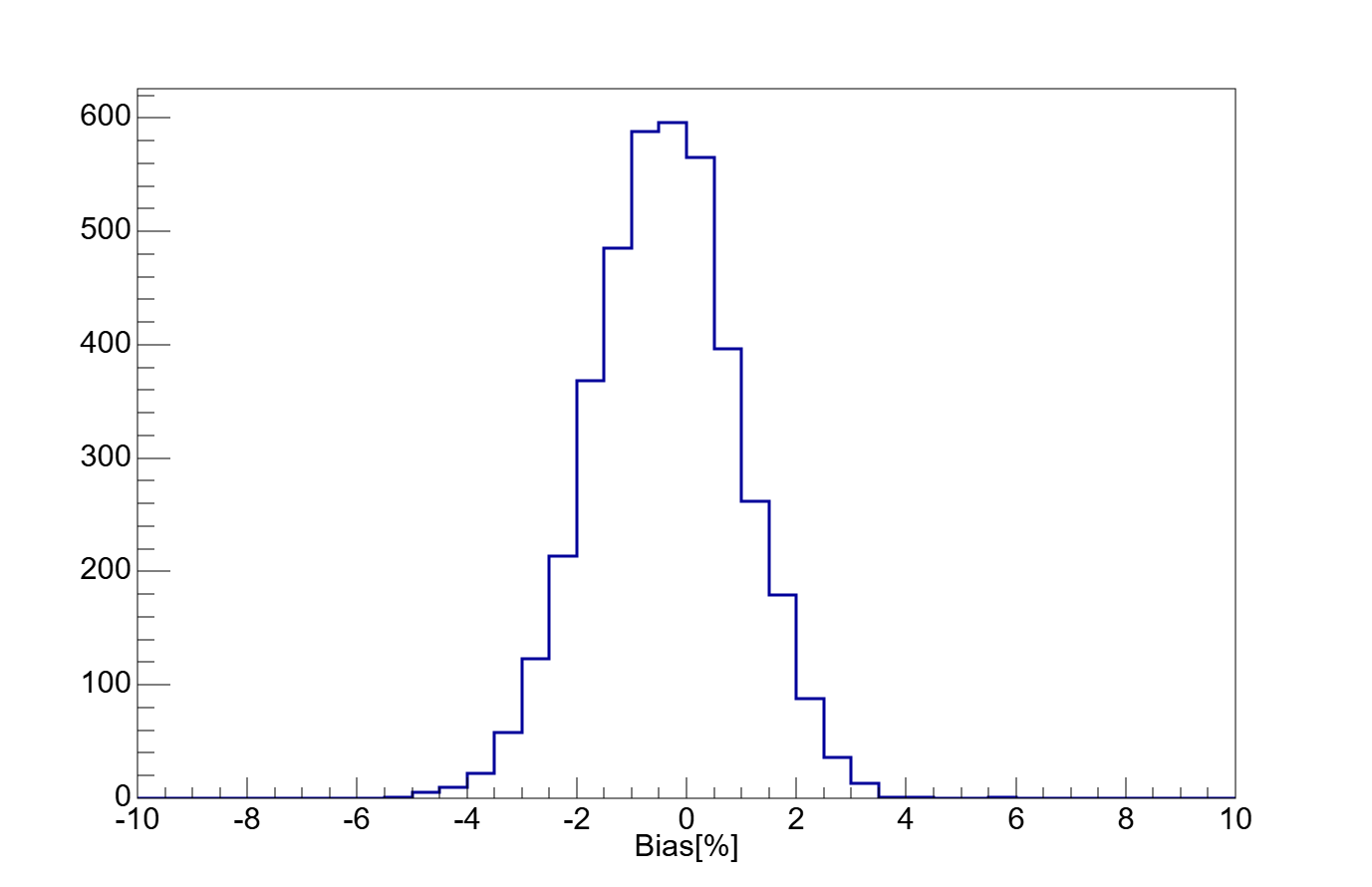}
\caption{DCR calibration results corrected based on EOCT calibration results. The left panel: comparison between calibration results and the simulation truth in each channel. The right panel: 1D histogram filled with each bin value from the left 2D histogram.}
\label{pic5.6}
\end{figure}

The DCR results calibration after correction are shown in Figure \ref{pic5.6}. The bias is -0.40\% and the standard deviation is 1.32\%.

During the EOCT calibration process, some data is acquired with only a single SiPM tile turned on. This data can be used to calibrate the DCR without the influence of the EOCT effect. It provides a reference for independent verification and comparative analysis.


\section{Discussion}
\label{7}
\subsection{Impact of Temperature Fluctuation on the TAO SiPM Parameters}
\label{7.1}
As reported in Ref. \cite{Ref57}, the dependence of SiPM parameters on temperature and overvoltage (OV) can be parameterized as:
\begin{equation}
\label{eq7.1}
\begin{split}
\lambda_{\rm IOCT} &= k_{\rm IOCT} \cdot \text{OV} \,,
\qquad
\lambda_{\rm EOCT} = k_{\rm EOCT} \cdot \text{OV} \,,
\qquad
\text{Gain} = k_{\text{Gain}} \cdot \text{OV}\,
\\
\text{PDE} &= k_{\text{PDE}} \cdot (\text{OV}-\text{OV}_{\text{ref}}) + \text{PDE}(\text{OV}_{\text{ref}})
\\
\text{DCR} &= A_{\text{amp}}\cdot \text{OV} \cdot T^{3/2} \cdot e^{-E_g/kT} \
\end{split}
\end{equation}

According to the TAO SiPM burn-in test results \cite{Ref58}, the relationship between the SiPM overvoltage and temperature follows:
\begin{equation}
\label{eq7.2}
\begin{split}
 \text{OV} = V_{\text{ex}} - V_{\text{bd}} = V_{\text{ex}} - (V_{\text{bd,ref}}+ k_{T}\cdot \Delta T)  
\end{split}
\end{equation}
where $V_{\text{ex}}$ is the externally applied voltage, $V_{\text{bd}}$ is the breakdown voltage, $k_{T}$ = 54.7 $\pm$ 1.1 mV/\si{\degreeCelsius} is the mean value of the temperature coefficient of $V_{\text{bd}}$ for all the SiPM channels measured in the burn-in testing, while $V_{\text{bd,ref}}$ is the $V_{\text{bd}}$ measured for each SiPM channel at -50 \si{\degreeCelsius}. $\Delta T = T - T_{-50\si{\degreeCelsius}}$ denotes the temperature difference.

For the TAO CD, the temperature fluctuation $\sigma_{T}$ is approximately 0.1 \si{\degreeCelsius}. According to Equation \ref{eq7.1} and Equation \ref{eq7.2}, the fluctuation of each SiPM parameter induced by temperature fluctuation is given by:
\begin{equation}
\label{eq7.3}
\begin{split}
\sigma_{T, \rm IOCT} &= k_{\rm IOCT} \cdot k_{T} \cdot \sigma_{T} \,,
\qquad
\sigma_{T, \rm EOCT}  = k_{\rm EOCT} \cdot k_{T} \cdot \sigma_{T} \,,
\\
\sigma_{T, \text{Gain}} &= k_{\text{Gain}} \cdot k_{T} \cdot \sigma_{T}\,
\qquad
\sigma_{T, \text{PDE}} = k_{\text{PDE}} \cdot k_{T} \cdot \sigma_{T}
\\
\sigma_{T, \text{DCR}} &= A_{\text{amp}} \cdot \sigma_{T} [(V_{\text{ex}} -V_{\text{bd,ref}} +  k_{T} \cdot T_{-50\si{\degreeCelsius}}) \cdot \frac{d(T^{3/2} \cdot e^{-E_g/kT})}{dT} - k_{T} \cdot \frac{d(T^{5/2} \cdot e^{-E_g/kT})}{dT}] 
\end{split}
\end{equation}
Among these, the expression for $\sigma_{T, \text{DCR}}$ is relatively complex. Since the TAO CD operates at -50 \si{\degreeCelsius}, the uncertainties of the SiPM parameters induced by temperature fluctuation investigated in this work are all evaluated at this operating temperature. Accordingly, the expression for $\sigma_{T, \text{DCR}}$ can be simplified using the DCR expression at -50 \si{\degreeCelsius}. For reference, at 50 °C, the DCR follows:
\begin{equation}
\label{eq7.4}
\begin{split}
\text{DCR}_{-50\si{\degreeCelsius}} &= A_{\text{amp}}\cdot (V_{\text{ex}}-V_{\text{bd,ref}}) \cdot T_{-50\si{\degreeCelsius}}^{3/2} \cdot e^{-E_g/kT_{-50\si{\degreeCelsius}}}
\end{split}
\end{equation}
Substituting Equation \ref{eq7.4} into the expression for $\sigma_{T, \text{DCR}}$ in Equation \ref{eq7.3}, the following expression can be derived:
\begin{equation}
\label{eq7.5}
\begin{split}
\sigma_{T_{-50\si{\degreeCelsius}}, \text{DCR}} = \left( \frac{3}{2} + \frac{E_g}{k T_{-50\si{\degreeCelsius}}} \right) \frac{\text{DCR}_{-50\si{\degreeCelsius}} \cdot \sigma_{T}}{T_{-50\si{\degreeCelsius}}}  - \frac{\text{DCR}_{-50\si{\degreeCelsius}} \cdot k_T \cdot \sigma_{T}}{V_{\text{ex}} - V_{\text{bd, ref}}}
\end{split}
\end{equation}

According to the TAO SiPM mass and burn-in test results \cite{Ref32,Ref58}, Equation \ref{eq7.3} and Equation \ref{eq7.5} we calculated the relative uncertainties induced by a 0.1 \si{\degreeCelsius} temperature fluctuation at an operating temperature of -50 \si{\degreeCelsius} and a SiPM OV of 3.2 V. The results are summarized in Table \ref{t7.1}.

\begin{table}[htbp]
  \centering
  \caption{Relative uncertainties of SiPM parameters induced by a 0.1 \si{\degreeCelsius} temperature fluctuation}
  \begin{tabular}{ccc}
    \hline
     Parameter & Relative Uncertainty \\
    \hline
     DCR & 0.75\% \\
     Gain & 0.17\%  \\
     PDE & 0.058\%  \\
     IOCT & 0.17\% \\
     EOCT Rate & 0.17\% \\
    \hline
  \end{tabular}
  \label{t7.1}
  
\end{table}

\subsection{Impact of SiPM Parameter Uncertainties and Calibration Biases on the Reconstructed Vertex and Energy}
\label{7.2}

In the TAO simulation, we individually add the uncertainties of each SiPM parameter, which originate from temperature fluctuation and the calibration method (the standard deviation derived from calibration). Meanwhile, we introduce the corresponding calibration biases between the SiPM parameters used in reconstruction and those configured in the TAO simulation. It is worth noting that the relative PDE calibration bias features a spatial distribution, and accordingly the PDE bias applied in our analysis adopts the spatially distributed pattern shown in Figure \ref{pic3.4}. We simulate 1 MeV kinetic energy electrons at the center of the liquid scintillator in the TAO CD, reconstruct the vertex and energy of each electron via the charge center method \cite{Ref16}, and derive the corresponding reconstructed vertex uncertainty and energy resolution for both scenarios with and without the inclusion of SiPM parameter uncertainties and calibration biases. The results are summarized in Table \ref{t7.2}. Notably, the IOCT rate implemented in the TAO simulation is the in-air value for SiPMs (15\%), whereas the IOCT rate of SiPMs in the LAB environment is lower. Accordingly, the impact of IOCT on reconstruction performance presented in Table \ref{t7.2} is a conservative upper bound.

\begin{table}[htbp]
  \centering
  \small
  \caption{Impact of SiPM parameter uncertainties and calibration biases on reconstructed vertex uncertainty and energy resolution}
  \label{t7.2}
  \begin{tabular}{ll
    S[table-format=2.3]
    S[table-format=+1.3]
    S[table-format=1.4]
    S[table-format=+1.4]
  }
    \toprule
    \multirow{2}{*}{SiPM Parameter} & \multirow{2}{*}{Uncertainty / Bias Source} & \multicolumn{2}{c}{Vertex Uncertainty} & \multicolumn{2}{c}{Energy Resolution} \\
    \cmidrule(lr){3-4} \cmidrule(lr){5-6}
    & & {Absolute (\si{mm})} & {Relative (\si{\%})} & {Absolute (\si{\%})} & {Relative (\si{\%})} \\
    \midrule
    Baseline & No Uncertainty / Bias & 13.080 & 0.000 & 1.8081 & 0.0000 \\
    \midrule
    DCR & \multirow{6}{*}{Temperature} & 13.082 & +0.015 & 1.8084 & +0.0166 \\
    PDE & & 13.082 & +0.015 & 1.8084 & +0.0166 \\
    Gain & & 13.082 & +0.015 & 1.8104 & +0.1272 \\
    IOCT & & 13.081 & +0.008 & 1.8083 & +0.0111 \\
    EOCT & & 13.082 & +0.015 & 1.8086 & +0.0277 \\
    All Parameters & & 13.089 & +0.069 & 1.8117 & +0.1991 \\
    \midrule
    DCR & \multirow{6}{*}{Calibration Uncertainty} & 13.083 & +0.023 & 1.8085 & +0.0221 \\
    PDE & & 13.099 & +0.145 & 1.8197 & +0.6416 \\
    Gain & & 13.081 & +0.008 & 1.8089 & +0.0442 \\
    IOCT & & 13.097 & +0.130 & 1.8134 & +0.2931 \\
    EOCT & & 13.082 & +0.015 & 1.8084 & +0.0166 \\
    All Parameters & & 13.121 & +0.313 & 1.8270 & +1.0453 \\
    \midrule
    DCR & \multirow{6}{*}{Calibration Bias} & 13.084 & +0.031 & 1.8088 & +0.0387 \\
    PDE & & 13.214 & +1.024 & 1.8225 & +0.7964 \\
    Gain & & 13.081 & +0.008 & 1.8081 & <0.0053 \\
    IOCT & & 13.087 & +0.054 & 1.8101 & +0.1106 \\
    EOCT & & 13.080 & <0.008 & 1.8081 & <0.0053 \\
    All Parameters & & 13.226 & +1.116 & 1.8251 & +0.9402 \\
    \midrule
    All Parameters & All Uncertainties / Biases & 13.276 & +1.498 & 1.8469 & +2.1459 \\
    \bottomrule
  \end{tabular}
\end{table}

According to the results in Table \ref{t7.2}, it can be seen that the SiPM parameter uncertainties and biases have a limited impact on the vertex and energy reconstruction precision obtained via the charge center method \cite{Ref16}. The additional vertex smearing and extra degradation of energy resolution introduced by these uncertainties and biases are small, which directly confirms that the calibration precision fully meets the performance requirements of the TAO detector.

\subsection{Impact of Tile-Level External Optical Crosstalk Calibration on Reconstructed Vertex Uncertainty and Energy Resolution}
\label{7.3}

The high-voltage system for the TAO SiPMs can only regulate the voltage for each SiPM tile, and cannot provide independent voltage control for the two readout channels within a single SiPM tile. For this reason, the SiPM on-off switching method can only achieve tile-level EOCT calibration. If the EOCT values of the two channels within the same tile differ, this discrepancy will affect the final energy and vertex reconstruction performance. 

\begin{figure}[htbp]
\centering
\includegraphics[width=.45\textwidth]{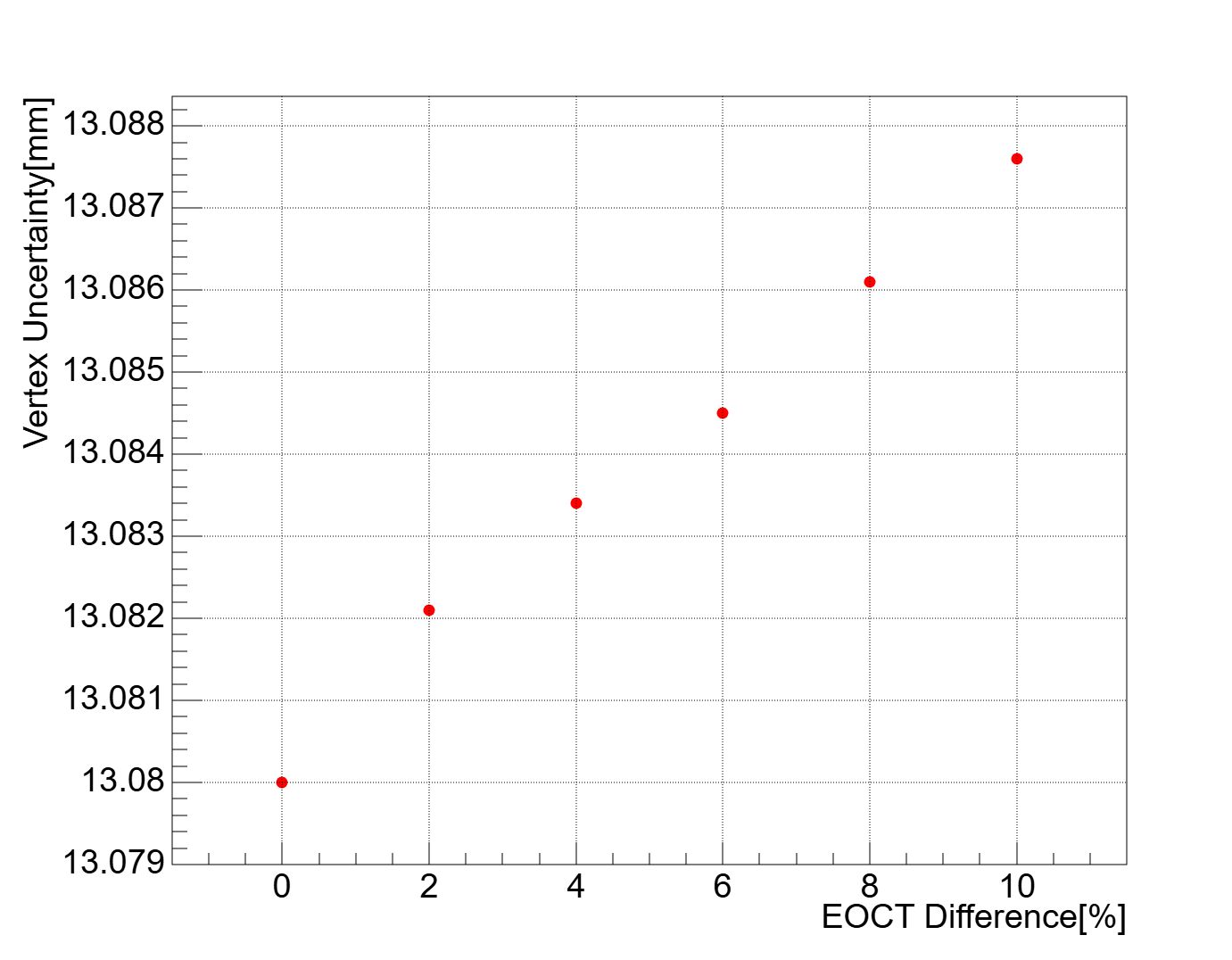}
\qquad
\includegraphics[width=.45\textwidth]{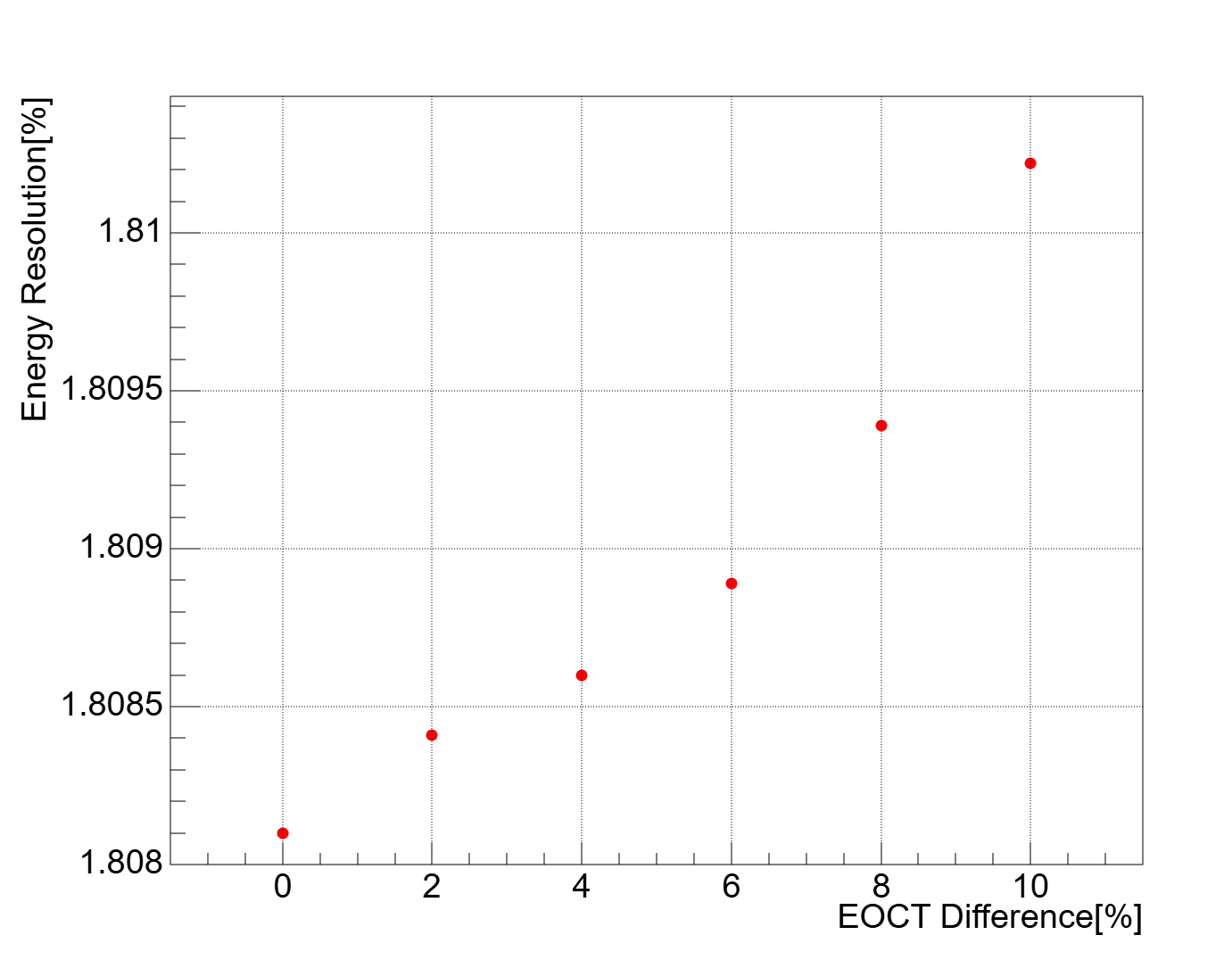}
\caption{Vertex uncertainty and energy resolution under varying EOCT differences between the two channels within the same tile.}
\label{pic7.1}
\end{figure}

To quantitatively evaluate this effect, we assign distinct EOCT values to the two channels within the same tile in the TAO simulation, simulate 1 MeV kinetic energy electrons at the center of the liquid scintillator in the TAO CD, and perform event reconstruction via the charge center method. We derive the corresponding reconstructed vertex uncertainty and energy resolution under different magnitudes of EOCT difference between the two channels within the same tile, with the results shown in Figure \ref{pic7.1}. As described by Equation \ref{eq7.1}, the EOCT parameter scales linearly with the SiPM overvoltage. According to the TAO Conceptual Design Report \cite{Ref15}, the overvoltage non-uniformity within a single tile is required to be below 10\%. As a result, the channel-to-channel variation of the EOCT parameter within one tile is generally bounded by 10\%. The simulation results show that the degradation of vertex uncertainty and  energy resolution caused by the EOCT difference between the two channels within the same tile is limited, which confirms that tile-level EOCT calibration fully satisfies the performance requirements of the TAO detector.

\section{Conclusion}
\label{6}

This work presents a systematic and comprehensive calibration strategy for the SiPMs in the TAO CD,  covering the core performance parameters including DCR, time offset, relative PDE, gain, IOCT rate and EOCT. Furthermore, we propose a novel method based on the SiPM on-off switching scheme, which is specifically designed for the calibration of the EOCT rate and its emission angular distribution in detectors with high dark noise levels. The feasibility of all proposed calibration methods is fully verified based on the TAO offline software simulation framework, with the bias, standard deviation, and underlying physical sources of bias for each method systematically quantified.

For the key SiPM parameters, the core calibration results are summarized as follows: the DCR calibration achieves a bias of -0.4\% and a standard deviation of 1.32\% after correction with the EOCT rate calibration results, which effectively suppresses the 23.6\% large bias caused by EOCT; the time offset calibration yields both bias and standard deviation less than 0.2 ns, where the uncertainty is dominated by the approximately 1 cm positional deviation of the deployed source inside the detector; for the relative PDE calibration, after correcting for the dark noise contribution, the method yields a bias of 0.011\% and a standard deviation of 0.17\% when the SiPM tile surface optical reflection effect is disabled, and when the reflection effect is enabled, the degradation of light field uniformity induced by SiPM tile surface reflection introduces a maximum bias of approximately 3\% between the pole and equator regions; the IOCT rate calibration using the generalized Poisson distribution fitting method achieves a bias of 1.40\% and a standard deviation of 3.02\%, and this method corrects the contribution of dark noise-induced multiple PEs hits to the calibration bias obtained from the multiple PEs hit analysis method, and the residual bias is mainly attributed to the AP effect.

For the proposed novel EOCT calibration method, the SiPM on-off switching scheme enables EOCT rate calibration with both bias and standard deviation less than 0.1\%. Meanwhile, we realize the calibration of the EOCT emission angle distribution by controlling the SiPMs on-off in each angular interval, achieving a bias of less than 4\% and a standard deviation of less than 2\% in the main angular range.


In addition, this work investigates the SiPM parameter uncertainties induced by temperature fluctuation in the TAO CD, and systematically analyzes the impacts of SiPM parameter uncertainties, calibration biases, and the adoption of tile-level rather than independent channel-level EOCT calibration on the detector's vertex reconstruction uncertainty and energy resolution. The results demonstrate that the corresponding impact is limited, which confirms that the proposed calibration scheme fully meets the stringent performance requirements of the TAO detector.

This work provides solid theoretical guidance for the calibration of individual SiPM performance parameters in the future TAO real detector. Accurate calibration of SiPM parameters will facilitate an in-depth understanding of the TAO detector response and the origins of its energy resolution, and further support the optimization of particle event reconstruction algorithms and the overall performance of the detector.

\FloatBarrier

\acknowledgments
We are grateful to the JUNO-TAO and the technical staff of the participating institutions. This work is supported by the National Key Research and Development Project of China, Grant No. 2022YFA1602002, National Natural Science Foundation of China under Grant Number 12275281.

\bibliographystyle{JHEP}
\bibliography{refer.bib}


\end{document}